\documentclass{article}
\usepackage{arxiv}
\usepackage{graphicx,color,amsmath,amsfonts,amscd,amssymb,pstricks}
\usepackage{dcolumn}
\usepackage{bm}
\usepackage{longtable}
\usepackage{amsmath}
\usepackage{mathrsfs}
\usepackage{amsfonts}
\usepackage{textcomp}
\usepackage{esvect}
\usepackage{floatrow}
\usepackage[USenglish,british]{babel}
\usepackage{environ}
\usepackage{xifthen}
\usepackage{xargs}			
\usepackage{hyperref}
\usepackage{physics}
\usepackage{multirow,here}
\usepackage{blindtext}
\usepackage{subcaption,tabularx,caption}
\newfloatcommand{capbtabbox}{table}[][\FBwidth]
\title{Split or unsplit electron beams?}

\author{A. Franchi\\ 
ESRF, 71 Avenue des Martyrs, 38000 Grenoble, France \\
\And 
M. Giovannozzi\thanks{Corresponding author: massimo.giovannozzi@cern.ch}\\
Beams Department, CERN, Esplanade des Particules 1, 1211 Meyrin, Switzerland}
\begin{document}
\maketitle
\begin{abstract}%
Non-linear effects have become increasingly relevant in modern circular particle accelerators, and in recent years a change of paradigm has appeared, the attitude towards nonlinear effects having shifted from fighting them to exploiting them with the goal of devising new beam manipulations, such as the splitting of the beam in the transverse phase space by crossing a stable resonance. In the field of hadron accelerators, well-established operational techniques based on nonlinear effects exist, whereas for the case of synchrotron light sources these new techniques are only beginning their way into the field. In this paper, we discuss novel techniques aimed at providing split beams in synchrotron light sources that are obtained by using stable islands in the transverse phase space or unsplit beams with AC dipoles to generate periodic closed orbits. The results of detailed numerical simulations, which support the proposed methods, are presented and discussed in this paper, together with possible applications.
\end{abstract}
%
%

\section{Introduction} \label{sec:intro}
Splitting bunches in the longitudinal plane to lower their intensity or the value of the longitudinal emittance by means of Radio-Frequency (RF) gymnastics has been in routine operation for decades at the CERN Proton Synchrotron (PS)~\cite{Garoby:1998kea,Garoby:EPAC00-WEOAF102,Burnet:1359959}, mainly for the generation of beams for the Large Hadron Collider (LHC), as well as in other circular hadron accelerators~\cite{Laxdal:EPAC92,PhysRevE.49.2484}. At the CERN PS, splitting bunches in the horizontal plane has been in operation for several years~\cite{Borburgh:2137954,PhysRevAccelBeams.20.014001,PhysRevAccelBeams.22.104002}, with the goal of performing multiturn extraction to fill the Super Proton Synchrotron (SPS) with high-intensity proton beams~\cite{Giovannozzi:987493}. In this manipulation, five beamlets, four stable islands plus the central core, are generated when the horizontal tune crosses the fourth-order resonance adiabatically, while the sextupole and octupole magnets are adjusted to generate stable fixed points around which phase-space islands are populated with protons~\cite{PhysRevLett.88.104801,PhysRevSTAB.7.024001}. A series of further applications has since been proposed, making use of variations of this principle~\cite{PhysRevSTAB.18.074001,GiovannozziEPJPlusTransition}. 

Until a few years ago, the beam split in the horizontal phase space belonged to the exclusive domain of hadron circular accelerators. Lepton machines, owing to their non-Hamiltonian feature of radiation damping and quantum diffusion, were thought not to be suitable for such manipulations, whose theoretical framework has its foundations in Hamiltonian, hence conservative, classical mechanics. Nevertheless, some years ago the synchrotron-light-source community received from Berlin first notifications of experimental evidence and possible applications of transversely split beams. Multiple beamlets were already observed since decades in several synchrotron light sources when operating in a quasi-isochronous mode (i.e.\ with low momentum compaction $\alpha_\mathrm{c}$) to deliver short bunches, and then voluntarily created and manipulated via sextupole and octupole magnets at the Metrology Light Source (MLS) (see Ref.~\cite{MarkusRies:phd} and references therein). These were the result of stable fixed points and islands in the longitudinal phase space, called $\alpha$-buckets, resulting in beams split horizontally along the dispersive regions. Still in the MLS, islands could also be created by a purely horizontal gymnastics of setting the tune close to a resonance and by exciting the beam with an AC exciter whose frequency was set at $1/N$ in units of the tune, where $N$ is the number of islands to be generated~\cite{MLS-BESSY1:ipac15}. By varying the amount of betatron coupling in the ring, those beamlets could be transferred from the horizontal plane to the vertical one and thus made visible on standard imaging systems. This idea has found an operational application in the BESSY~II ring, under the project named TRIBs, where some bunches have been put into the islands and used to generate synchrotron light for some users (see Refs.~\cite{Kramer:IPAC18-TUPML052,TRIBs-BESSY1:ipac19} and references therein). First tests of the same scheme were also recently reported at the MAX~IV light source~\cite{TRIBs-MAXIV:ipac19}. 

These intriguing experimental observations were reported as being the result of stable islands of phase space, similar to those created in the PS (a hadron machine), populated by displacing horizontally the electrons with a harmonic excitation of a beam shaker (or AC dipole) allowing the particles to abandon the core (i.e.\ the stable area around the nominal central orbit), to cross the separatrix (which defines the frontier between core and stable islands around the off-axis fixed points), and eventually to fill the stable islands. In early experiments, the existence of split beams was conditioned by the activation of the beam shaker: whenever this would be switched off, electrons would \emph{abandon the islands} and \emph{merge} into the central core.

The aim of this paper is to scrutinise this interpretation and help provide a more comprehensive, although not yet complete, picture. First of all, the theoretical literature, to the best of our knowledge, has never investigated the possibility of having electrons populating stable islands of the horizontal phase space, owing probably to the assumption that radiation damping would make the transverse momentum (and thus position) to converge towards the centre of the phase space. A previous work~\cite{NOCE17} presented numerical simulations in which the survival of electron beams within such stable islands was possible, although not always granted.\footnote{It has been pointed out to us that trapping of electrons in stable islands was also studied in Ref.~\cite{Piminov:PAC09-TH6PFP093}, in the context of the analysis of the strong non-linear effects in beam dynamics in the CLIC damping rings.} In the same paper, it was discussed how to correctly evaluate the equilibrium emittance and the optical parameters (among which the beta and dispersion functions) of the beam trapped in the islands. These indeed cannot be the same as those of the nominal on-axis beam, as the electrons trapped in the islands cross all magnets off-axis, thus experiencing additional dipolar feed-down fields from the quadrupoles, and quadrupolar fields from the sextupoles. Moreover, the standard use of numerical codes such as MAD-X~\cite{madx,Persson:IPAC21-WEPAB028} or Accelerator Toolbox (AT)~\cite{Terebilo:PAC01-RPAH314,Nash:IPAC15-MOPWA014} can lead to errors if the correct off-axis closed orbit is not properly accounted for in the evaluation of the optical parameters and of the quantum diffusion. 

This paper represents a step forward, trying to offer a wider understanding of the experimental observations discussed above. To this end, two scenarios are presented and investigated by means of multi-particle simulations, all carried out in 6D, of the original ESRF lattice (in use from 1994 to 2018), based on a Double Bend Achromat (DBA) design~\cite{4327987}. 

In the first scenario, the nonlinear optics is adjusted so as not to feature stable Hamiltonian islands. By introducing an AC dipole operating at a frequency of $1/3$ in units of the horizontal tune, we will show how three separate beam images can be generated by a diagnostic device. However, these are not related to the nonlinear setting nor to the Hamiltonian islands, and, more importantly, they do not represent split beams. The three images that may appear on a radiation monitor stem instead from the periodic distortion of the closed orbit induced by the AC dipole, which closes after three turns, hence providing three different values of the closed orbit as seen on a turn-by-turn basis. Typical cameras are not capable of turn-by-turn resolution, and the signal integrated over several turns produces the illusion of having three separate beamlets. 

In the second scenario, the nonlinear optics is adjusted so as to exhibit stable Hamiltonian fixed points in the horizontal phase space and stable islands around them. The existence of a configuration of equilibrium within their surface is investigated numerically. Different ways to populate the Hamiltonian islands with electrons, thus creating real split beams, are also presented. 
 
The paper is structured as follows: in Section~\ref{sec:scenario2}, the generation of a resonant orbit creating a beam which is only apparently split is presented and discussed with supporting simulation results. In Section~\ref{sec:scenario1}, the concept of stable Hamiltonian islands is briefly reviewed, and its interplay with lepton beams experiencing radiation damping and quantum diffusion is discussed (Sections~\ref{sec:Equilbrium} to~\ref{sec:inter}), and four possible schemes for populating resonance islands with electrons are presented (Section~\ref{sec:Populate}). Section~\ref{sec:comparing} provides a summary of the comparison (differences and similarities) in terms of electron and x-ray beams between the resonant orbit scheme and those based on resonance islands. Possible applications and benefits to synchrotron-radiation users of electron bunches split by resonance islands are outlined in Section~\ref{sec:applications}. Finally, some conclusions are drawn in Section~\ref{sec:conc}.

\section{Use of a resonant closed orbit in synchrotron light sources} \label{sec:scenario2}

Throughout the paper, the term AC dipole will refer to any type of programmable harmonic device (beam shaker, RF exciter, dipole, or feedback system with AC features) capable of generating an oscillating dipolar field $\theta_n=\bar{k}_0\sin{(2\pi n f_\mathrm{AC}+\phi)}$, $n$ being a generic turn number, $\bar{k}_0$ the maximum deflecting strength (i.e.\ angle), $\phi$ an initial arbitrary phase, and the frequency $f_\mathrm{AC}$ shall be comparable to that of the betatron motion. 

If the excitation frequency is set to a rational value in units of the horizontal tune, for instance, $f_\mathrm{AC}=1/N$, the beam experiences different $N$ dipolar kicks during the first $N$ turns. However, each kick is replicated identically every $N$ turns, namely
\begin{equation}\label{eq:ACdip1}
\left\{\begin{array}{l}
\quad\theta_{n}=k_0\sin{(2\pi\frac{n}{N}+\phi)} \vspace{2mm}\\
\theta_{n+1}=k_0\sin{(2\pi\frac{n+1}{N}+\phi)} \\
...
\end{array}\right. \ , \qquad
\left\{\begin{array}{l}
\quad\theta_{n+N}=k_0\sin{(2\pi\frac{n+N}{N}+\phi)}=\theta_{n}  \vspace{2mm} \\
\theta_{n+N+1}=k_0\sin{(2\pi\frac{n+N+1}{N}+\phi)}=\theta_{n+1}\\
...
\end{array} \right. \, . 
\end{equation} 
This implies that in a ring $R_N$, consisting of $N$ times the original lattice $R_1$, the AC dipole will be seen as $N$ DC dipole errors located at $N$ locations, evenly distributed along $R_N$, resulting in a standard closed-orbit distortion described by textbook formulae, periodic over one turn of the ring $R_N$. When looking at the original single-turn accelerator $R_1$, the orbit distortion is time-dependent, resulting in $N$ different closed orbits, each one occupied by the beam on a turn-by-turn basis. 
 
\begin{figure}[htb]
  \begin{center}
  \includegraphics[trim=2truemm 2truemm 1truemm 1truemm, width=0.49\linewidth,angle=0,clip=]{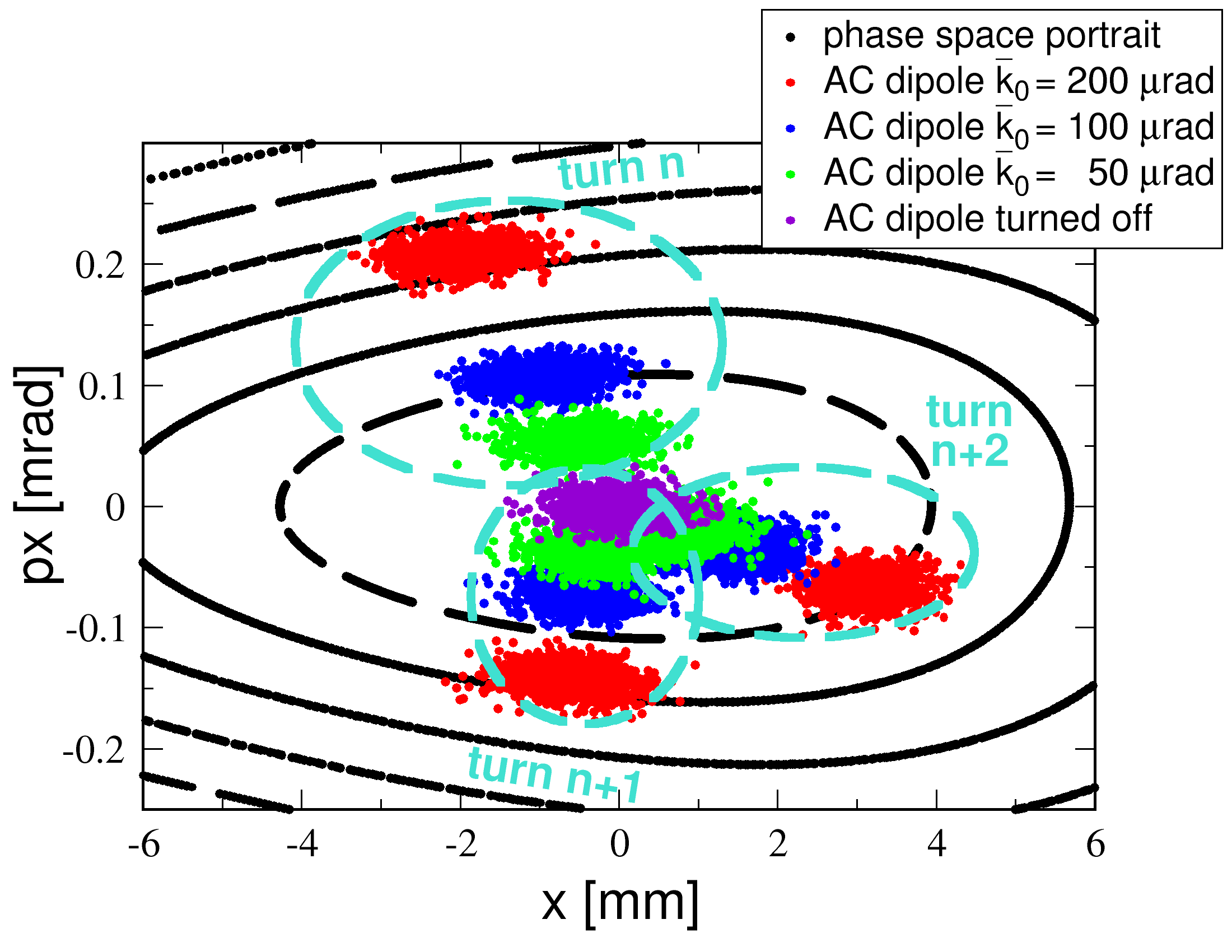}\ \ \ 
  \includegraphics[trim=2truemm 2truemm 1truemm 1truemm, width=0.49\linewidth,angle=0,clip=]{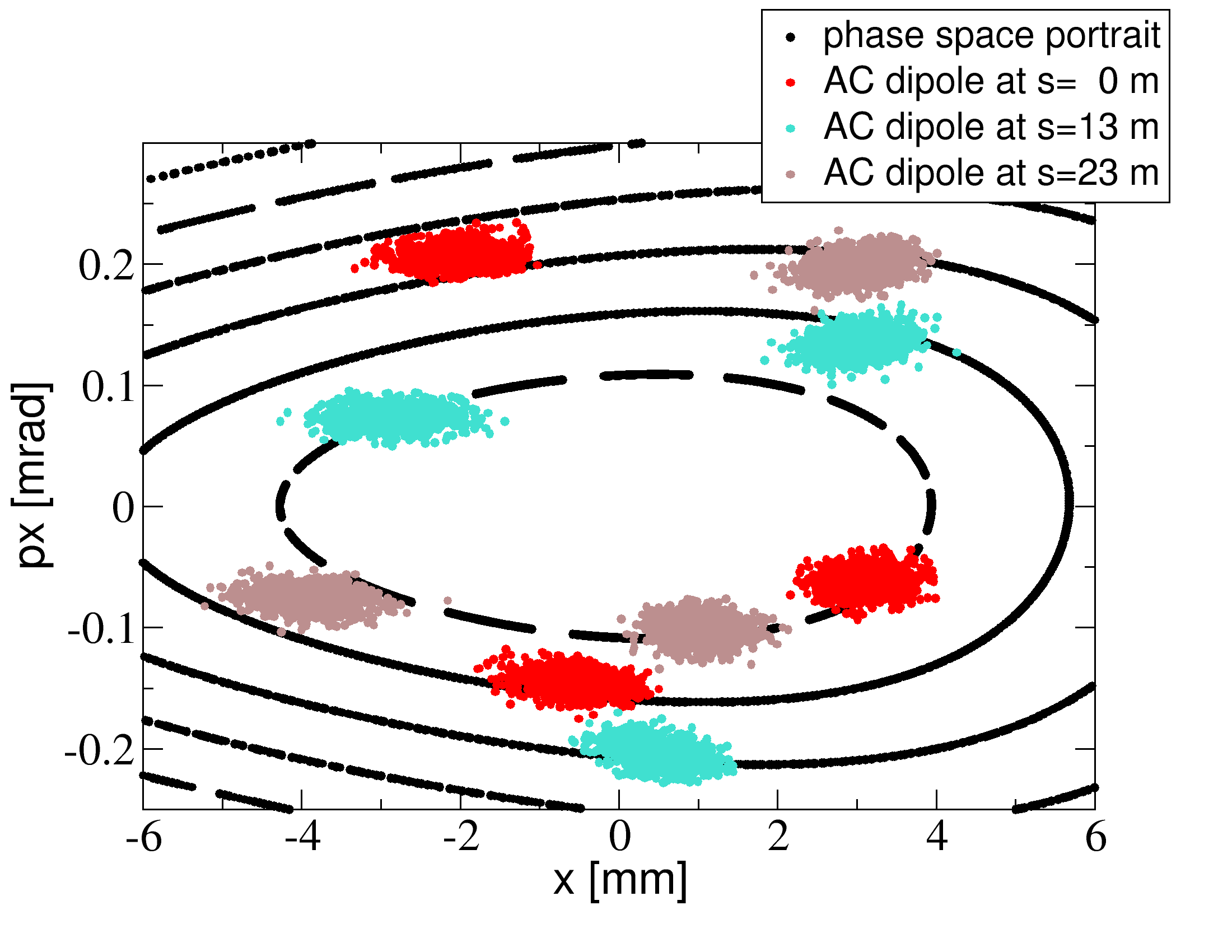}
  \caption{Phase-space portrait of the original ESRF lattice without resonance islands (black contour lines) and particle distributions obtained by simulating an AC dipole excitation with $f_\mathrm{AC}=1/3$ (in tune units) for different strength values (left) and different locations in the ring (with $\bar{k}_0=200\ \mu$rad, right). Each particle distribution is displayed stroboscopically, projecting onto the same phase space $(x,p_x)$ data from three consecutive turns, represented by the turquoise dashed ellipses. When the AC dipole is turned off, the three spots converge to the centre restoring the standard on-axis equilibrium distribution (violet distribution in the left plot). Position and angle of the three beamlets depend on the phase advance between the observation point and the location of the AC dipole (right). Note that the red distributions in the left plot correspond to the case shown in Fig.~\ref{ResOrb02}}
  \label{ResOrb01}
  \end{center}
\end{figure}

\begin{figure}[htb]
  \begin{center}
  \includegraphics[trim=2truemm 2truemm 2truemm 25truemm, width=0.45\linewidth,angle=0,clip=]{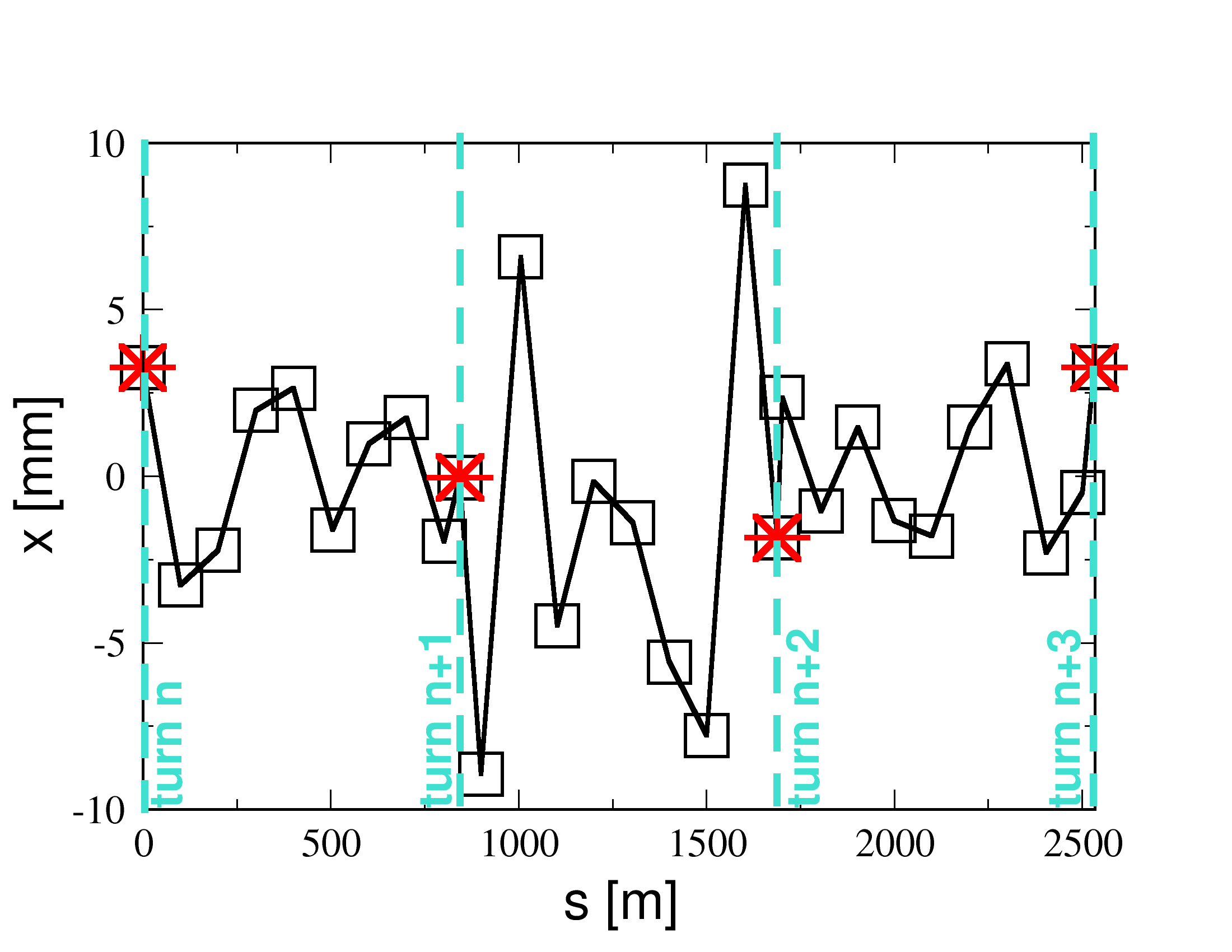}
  \includegraphics[trim=2truemm 2truemm 2truemm 25truemm, width=0.45\linewidth,angle=0,clip=]{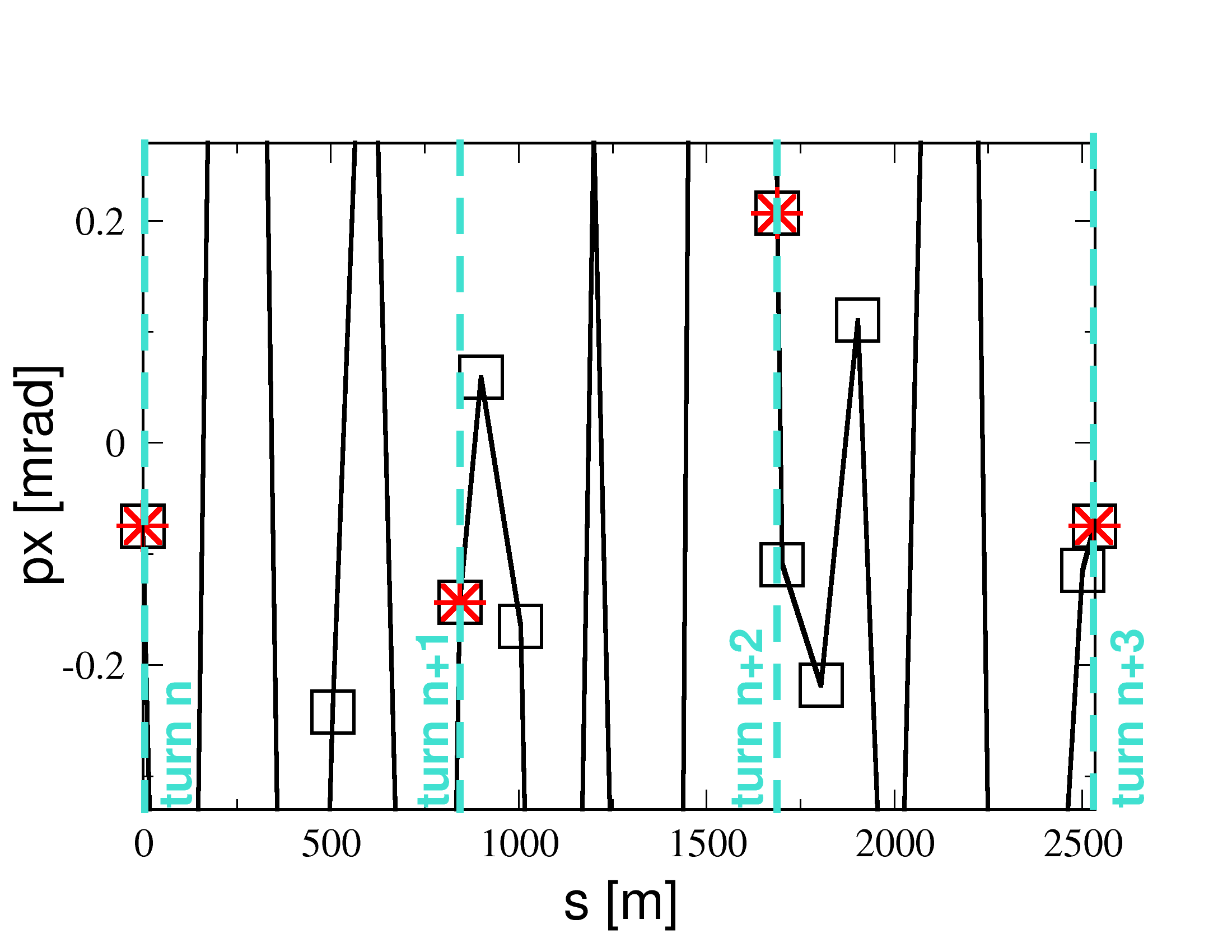}
  \caption{Left: Closed-orbit position over three turns for the ring $R_3$ (only a limited number of locations is shown for the sake of clarity) generated by an AC dipole at s=0 m, $f_\mathrm{AC}=1/3$ and $\bar{k}_0=200\ \mu$rad as computed by MAD-X (black curve). The centroids of the three orbit beamlets corresponding to the red data of Fig.~\ref{ResOrb01} are denoted by the red stars. Right: same plot for the horizontal closed-orbit angles.}
  \label{ResOrb02}
  \end{center}
\end{figure}

If the orbit distortion is sufficiently large, $N$ images on x-ray cameras, or any beam-imaging device, may be generated depending on the camera's position relative to that of the AC dipole, i.e.\ depending on the $N$ consecutive (in time) values of the closed orbit at the camera's location. Two values of the closed orbit with similar position $x_{1,2}$, though different divergence $x'_{1,2}$, would be indistinguishable, whereas whenever $x_{1,2}$ are separated by a distance much greater than the beam size or envelope $3\sigma_x$, i.e.\ $|x_1-x_2| >> 3\sigma_x$, two spots would be visible on the cameras. The integration time of these devices is typically much longer than the revolution time, hence preventing the camera from recording images on a turn-by-turn basis. This implies that the displayed $N$ spots\footnote{In this case these spots are called orbit beamlets to distinguish them from the beamlets used to describe particles tapped in stable islands.} do not correspond to $N$ separate orbit beamlets, but rather to the same original beam which on a turn-by-turn basis moves from one value of the closed orbit to another one. An example of this configuration with $N=3$ is displayed in Fig.~\ref{ResOrb01}. The nominal lattice of the original ESRF storage ring features no stable resonance islands within the region of interest, as shown by the black contour lines. By introducing an AC dipole at the beginning of the ring ($s=0$~m) and $f_\mathrm{AC}=1/3$ (in tune units), the particle distribution will jump from one phase space position to the next one on a turn-by-turn basis. Distributions are displayed stroboscopically, projecting onto the same phase space $(x,p_x)$ data from three consecutive turns, represented by the turquoise dashed ellipses. The distance of each of the three islands from the centre scales linearly with the AC dipole peak strength $\bar{k}_0$. The AC-dipole strength is also increased from zero to the maximum over several ms, so as to preserve the equilibrium between radiation damping and quantum diffusion (both included in the 6D simulations). When eventually the AC dipole is adiabatically turned off, the standard on-axis closed orbit is recovered and the particles move back to the centre of the phase space. 
To further validate the interpretation of the three spots as a mere resonant orbit distortion, the turn-by-turn centroid position and angle of the orbit beamlets are compared in Fig.~\ref{ResOrb02} with the closed orbit computed by MAD-X, after setting up a lattice $R_3$ comprising three identical lattices of the ESRF storage ring plus three DC dipole kicks evaluated as in Eq.~\eqref{eq:ACdip1}. The agreement is evident and perfect up to numerical precision. 

This use of an AC dipole in synchrotron light sources has several implications. Canted beamlines aligned to receive x-rays from a single orbit beamlet will be illuminated only every $N$ turns, with no photon crossing their front-end during $(N-1)$ turns. The total photon flux from the beamlet will be $1/N^{\rm th}$ of that originating from the electron beam on the nominal central orbit, although this reduction does not come from a reduced electron beam intensity (the beam is not split in $N$ islands), but rather from the discontinuity of the x-ray pulses turn by turn, as photons are only collected on 1 turn every $N$ revolutions. This can be an interesting feature for experiments requiring no primary photons from the ring during a certain time (the time can be controlled easily by choosing $N$ appropriately), without conflicting with users requiring continuous photon beams. 

The AC excitation can be performed on selected bunches along the train by making use of the electronics of the bunch-by-bunch orbit feedback system. This allows canted beamlines that receive x-rays from the orbit beamlet not to be disturbed by the photons originated by the main train of electron bunches. 

The orbit beamlets exist only in the presence of the AC excitation. By turning this off, the induced orbit distortion fades away and the beam returns to the nominal on-axis closed orbit. Multi-particle simulations confirm the reversibility of the process, as shown by the violet particle distribution in the left plot of Fig~\ref{ResOrb01}, which goes back to the standard central orbit after turning off the AC dipole. Moreover, in the exotic case of a vertically canted beamline, the desired vertical angle and offset can be tailored by a vertical AC dipole, without the need of adding linear betatron coupling to the storage ring optics to generate orbit beamlets in the vertical phase space, with an overall beneficial impact on the accelerator operation, given that the linear betatron coupling impacts the performance of the whole machine. 

The evaluation of the parameters of the orbit beamlets, such as the optical functions and the equilibrium emittance, requires some care. Since the beam crosses off-axis many quadrupoles and sextupoles, additional dipolar and quadrupolar feed-down fields are experienced by the electrons, thus impacting both the dispersion functions and the Courant-Snyder parameters ($\beta,\ \alpha$ and $\gamma$), and eventually the equilibrium emittance and the beam size. As a general consideration, the stronger is the AC excitation, the larger will be the orbit distortion and thus the difference between the optics of the nominal on-axis beam and the one of the orbit beamlets. The computation of the latter requires some preliminary steps. First of all, a model lattice $R_N$ comprising $N$ sequences of the same ring $R_1$ needs to be defined and used as a baseline structure. Then, the strength of the AC dipole needs to be inserted as $N$ DC dipole errors, one per ring, representing the harmonic dipolar kick at the frequency of $1/N$, as in Eq.~\eqref{eq:ACdip1}. Any optics code such as MAD-X or AT will then automatically compute the distorted closed orbit and the corresponding optics parameters. These will be of course periodic for $R_N$, but not for the original single-turn structure $R_1$. This is a natural consequence of the fact that the orbit beamlet sits on $N$ different closed orbits, each with different linear optics. The equilibrium emittance is evaluated by the standard integrals computed for $R_N$, and it is the same for the $N$ orbit beamlets. The linear optics of the orbit beamlet will instead be defined on a single-turn basis and will be different for each single orbit beamlet. Such optics calculations can be always cross-checked by means of multi-particle simulations with a proper description of the AC dipole and correct computation of the quantum diffusion term. There are different ways to evaluate this term which is fundamental, along with the damping term, to obtain the correct final equilibrium emittance. It is worth stressing that some software models use a one-turn diffusion matrix computed from the initial baseline lattice: this option is clearly not suitable to describe the orbit beamlets, because the dipolar feed-down fields from the off-axis orbit across quadrupoles vary on a turn-by-turn basis and depends on the strength of the AC dipole. A more realistic computation of the diffusion and damping terms on an element-by-element basis, as a function of the electron position, such as the one implemented in MAD-X~\cite{Helmut-diffusion} is then needed. 

From the considerations made above, it should be clear that both the existence and the nature of orbit beamlets is unrelated to the nonlinear optics and to the resonance islands of Refs.~\cite{Borburgh:2137954,PhysRevSTAB.7.024001}. These phase space structures applied to electron beams will be discussed in the next section. 

\section{Use of stable islands in synchrotron light sources} \label{sec:scenario1}

\begin{figure}[htb]
  \begin{center}
  \includegraphics[trim=2truemm 2truemm 2truemm 2truemm, width=0.3\linewidth,angle=0,clip=]{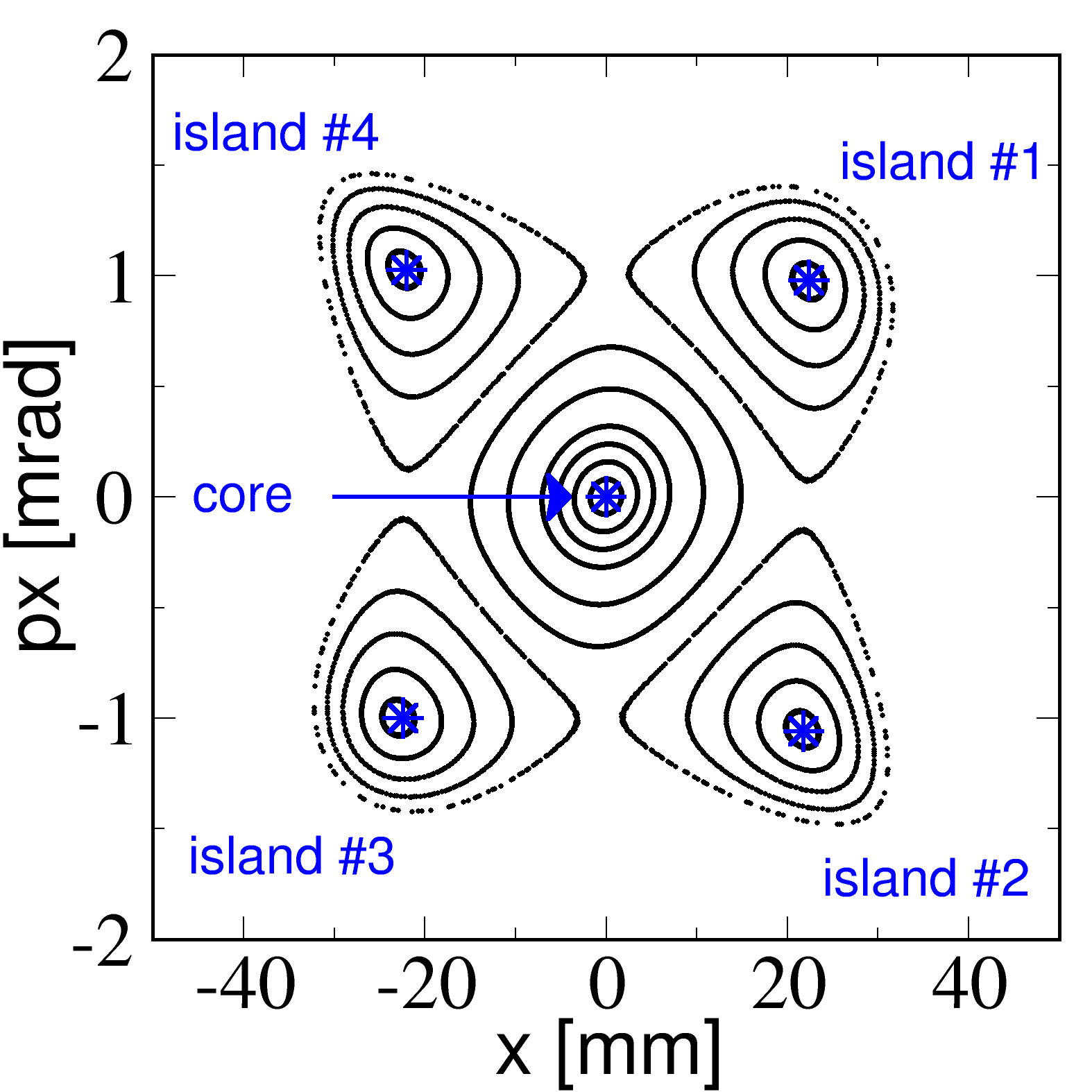}
  \includegraphics[trim=2truemm 2truemm 2truemm 2truemm, width=0.3\linewidth,angle=0,clip=]{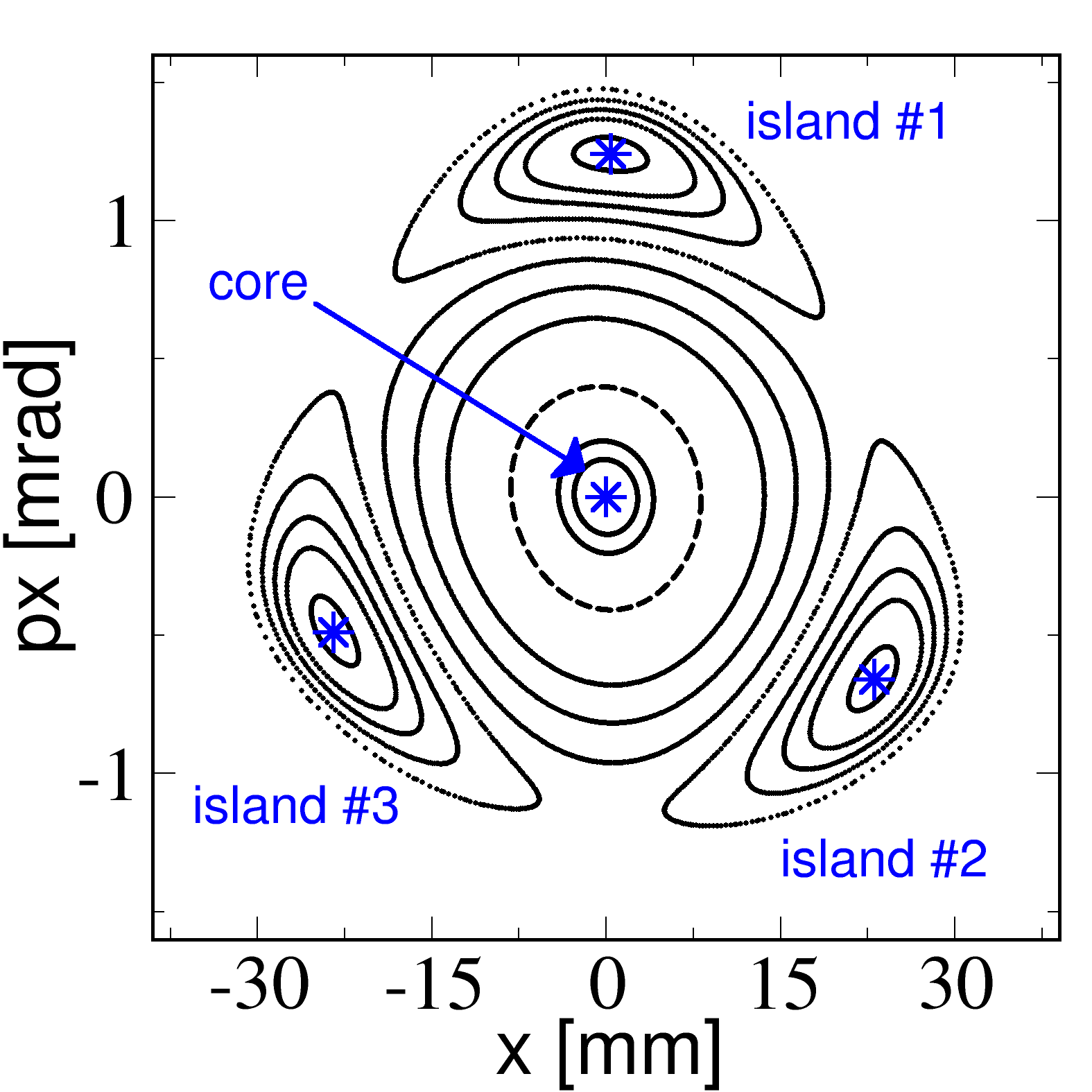}
  \includegraphics[trim=2truemm 2truemm 2truemm 2truemm, width=0.3\linewidth,angle=0,clip=]{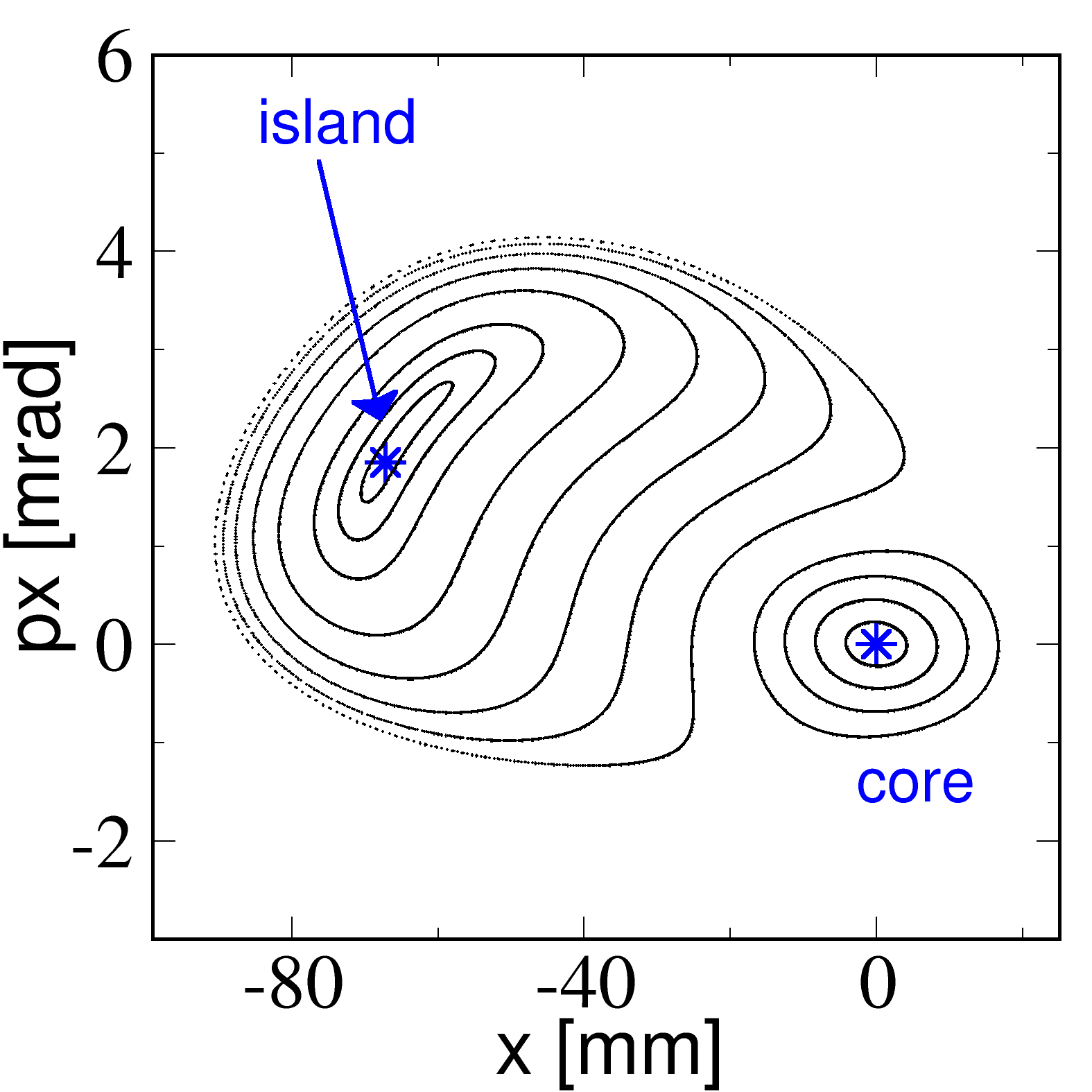}
  \caption{Examples of horizontal phase-space portraits exhibiting stable nonlinear fixed points and resonance islands of the CERN PS lattice. These configurations are obtained with the same setting of nonlinear magnets by simply changing the betatron tune: $Q_x=0.237$ (left plot, generating 4 stable islands), $Q_x=0.320$ (centre plot, with 3 stable islands) and $Q_x=0.05$ (right plot, with one stable island). The positions of the corresponding fixed points as computed by MAD-X-PTC are denoted by the blue stars (including the trivial on-axis fixed point).}
  \label{PS_PhSp}
  \end{center}
\end{figure}

In Ref.~\cite{giovannozzi2021novel}, the basic elements for the existence and determination of the positions of the fixed points are provided. An accelerator lattice with a given amplitude-dependent detuning $\Omega_2$ generated by sextupole and octupole magnets will make the frequency of the betatron oscillations in phase space vary with amplitude, represented by the value of the nonlinear invariant $\rho$ associated to this oscillation. In the horizontal plane, for example, the amplitude-dependent tune $\nu_x$ reads 
\begin{equation}\label{eq:detuning1}
\nu_x(\rho_x)\simeq Q_x+\frac{\Omega_2}{2\pi}\rho_x \, ,
\end{equation} 
where $Q_x$ is the betatron (i.e.\ amplitude-independent) tune and the $2\pi$ factor stems from the definition of $\Omega_2$ in Ref.~\cite{giovannozzi2021novel}. The approximation originates from possible additional resonant terms scaling with $\rho$, that are ignored here, and from higher-order terms scaling with $\rho^2$. Equation~\eqref{eq:detuning1} suggests the possible existence of a special amplitude $\rho_x^*$ such that $\nu_x(\rho_x^*)=\bar{Q}_x=1/N$, i.e.\ that the tune satisfies a resonant condition:
\begin{equation}\label{eq:fixedpoint1}
\rho_x^*(\Delta,\ \rho_x)\simeq -\frac{2\pi\Delta}{\Omega_2}\, ,\qquad \Delta=Q_x-\bar{Q}_x\, ,
\end{equation} 
where $\Delta$ is the distance of the linear betatron tune from the resonant condition. In phase space, one or more chains of $N$ points may correspond to $\rho_x^*$: if particles oscillate with this amplitude (and with a suitable phase that is not discussed here), they will jump turn by turn from one point to the other, to come back to the initial point after $N$ turns. Such special points are called fixed points. The action $\rho$ being by definition a positive quantity, fixed points may exist only if the detuning $\Omega$ and the distance to the resonant tune $\Delta$ have opposite signs, as indicated in Eq.~\eqref{eq:fixedpoint1}. Three examples of horizontal resonance islands of the CERN PS obtained with different tunes are shown in the phase-space portraits of Fig.~\ref{PS_PhSp}. The corresponding fixed points are represented by blue stars inside the islands. In the left plot, the fractional part of the tune is set at $Q_x=0.237$, i.e.\ close to the $1/4$ resonance. The tune of the fixed points is indeed $\bar{Q}_x=1/4$, so that the fixed point returns to its initial position after 4 turns. The central phase-space portrait corresponds to the same lattice with the horizontal tune moved to $Q_x=0.320$, close to $\bar{Q}_x=1/3$. The case of the stable integer resonance with $Q_x=0.05$ and $\bar{Q}_x=1$ is depicted in the right picture. It is worth noting that these three configurations are obtained with the same configuration of the sextupole and octupole magnets and by simply varying the linear tune. 

As for the orbit beamlets, the evaluation of the optical parameters, such as the dispersion functions and the Courant-Snyder parameters, by the optics codes requires a preliminary setup. If a resonance of order $N$ is to be studied, a lattice ($R_N$) comprising $N$ sequences of the original ring lattice ($R_1$) needs to be built. Unlike the orbit beamlet case, these sequences will be strictly identical, i.e.\ $R_N=R_1^N$, since there is no need to define different strengths, e.g.\ of the AC dipole, along the $N$ turns. A fundamental input for the optics codes is the initial condition in phase space around which the program has to look for an orbit closing over $N$ turns, to properly account for all feed-down effects. Indeed, it is customary for the codes to assume that the closed orbit search is to be carried out around the origin of the phase space, thus computing the standard optical parameters. Initial guesses for the fixed-point position can be obtained from single-particle tracking of several initial conditions and by plotting the phase space portraits, such as those of Fig.~\ref{PS_PhSp}. For the examples of Fig.~\ref{PS_PhSp}, the corresponding horizontal optical parameters for the standard optics and fixed points are reported in Table~\ref{tab:optics-functions-PS}. 

\begin{table}[htb]
\centering
\begin{tabular}{l |l |l |l |l} 
\hline\hline\vspace{-0.25cm} \\ \vspace{0.10 cm}
Optical   & nominal \ \ \ , [4 islands] & nominal\ \ \ , [3 islands] &  nominal\ \ \ , 1 island      \\
parameter &                             & 
         &                               \\
        \vspace{-2 mm} \\
\hline
\hline\vspace{-2 mm} \\ \vspace{1 mm}
$\beta_x$ [m] & 22.13 \ \ ,  [22.52, 22.00, 22.25, 22.33]       & 20.15  , [ 22.73, 21.65, 52.86]  & 18.63 , 24.49  \\
\hline\vspace{-2 mm} \\ \vspace{1 mm}
$\alpha_x$  [\ ] & -0.0735 ,  [0.219, 0.0976, 0.2187, 0.1932]\ \ \ \ & 0.060 , [-0.773, 0.890, 0.192] & 0.073 , -2.752\\
\hline\vspace{-2 mm} \\ \vspace{1 mm}
$D_x$  [m]    &3.06 \ \ \ \ , [8.35, -2.72, -3.80, 7.27]                 & 2.98 \  , [10.82, -4.84, 2.97]         & 2.43\ \  , -0.48\\
\hline
\end{tabular}
\caption{Horizontal optical parameters at the beginning of the PS lattice corresponding to the three phase-space portraits of Fig.~\ref{PS_PhSp}. For each case, the standard values, i.e.\ those for nominal closed orbit, are given along with the ones for the fixed points, as computed by MAD-X-PTC.}
\label{tab:optics-functions-PS}
\end{table}
It shall be noticed that the PS lattice provides a wide linear phase space area, the nonlinear effects being negligible within a phase-space area of $\pm10$~mm and $\pm0.5$~mrad. 
\begin{figure*}[b]
\begin{floatrow}
\ffigbox{
    \includegraphics[trim=2truemm 2truemm 2truemm 22truemm, width=0.9\linewidth,angle=0,clip=]{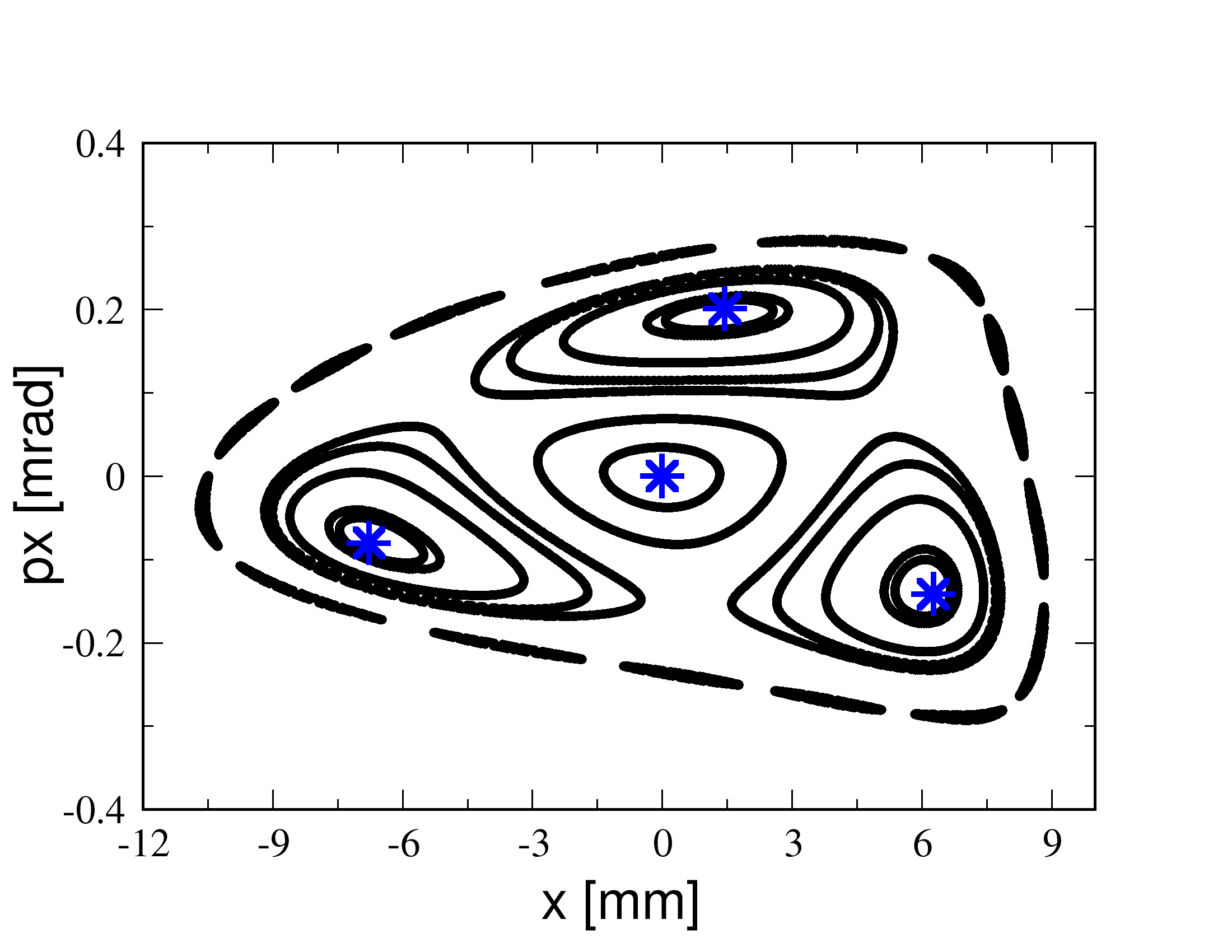}
    }{ \caption[]{Examples of horizontal phase-space portrait exhibiting three stable nonlinear fixed points and resonance islands of the original ESRF storage ring lattice.   \label{functions-ESRF}}}
\capbtabbox{
        \begin{tabular}{l l l}  
        \hline\hline\vspace{-0.25cm} \\ \vspace{1 mm}
        Optical     & nominal & [3 islands] \\
        parameter   &         &             \\
        \vspace{-2 mm} \\
        \hline \hline \vspace{-2 mm} \\ \vspace{1 mm}
        $\beta_x$ [m] & 37.08   & [ 43.75, 73.50, 18.17]  \\
        \hline\vspace{-2 mm} \\ \vspace{1 mm}
        $\alpha_x$  [\ ] & 0.045 & [0.702, -0.354, 0.004] \\
        \hline\vspace{-2 mm} \\ \vspace{1 mm}
        $D_x$  [m]    &0.1344  & [-0.002, 0.171, 0.264]         \\
        \hline\vspace{1.10 cm} 
        \end{tabular}  
        }{\caption{Optical parameters around the nominal closed orbit and the three fixed points of Fig.~\ref{functions-ESRF} as computed by MAD-X-PTC. \label{tab:optics-functions-ESRF}}}
\end{floatrow}
\end{figure*}

Synchrotron light sources, with their stronger focusing and thus larger chromatic and transverse aberrations compared to hadron machines, may show a much bigger difference in the linear optics of the fixed points. In Table~\ref{tab:optics-functions-ESRF}, the horizontal optical functions at the beginning of the original ESRF storage ring lattice for the nonlinear fixed points are compared with the ones for the standard closed orbit for a nonlinear setting generating the three stable islands displayed in the phase space portrait of Fig.~\ref{functions-ESRF}. The differences are indeed striking. 

\subsection{Is the equilibrium possible inside the resonance islands?} \label{sec:Equilbrium}
\begin{figure}
  \begin{center}
  \includegraphics[trim=2truemm 2truemm 2truemm 20truemm, width=0.45\linewidth,angle=0,clip=]{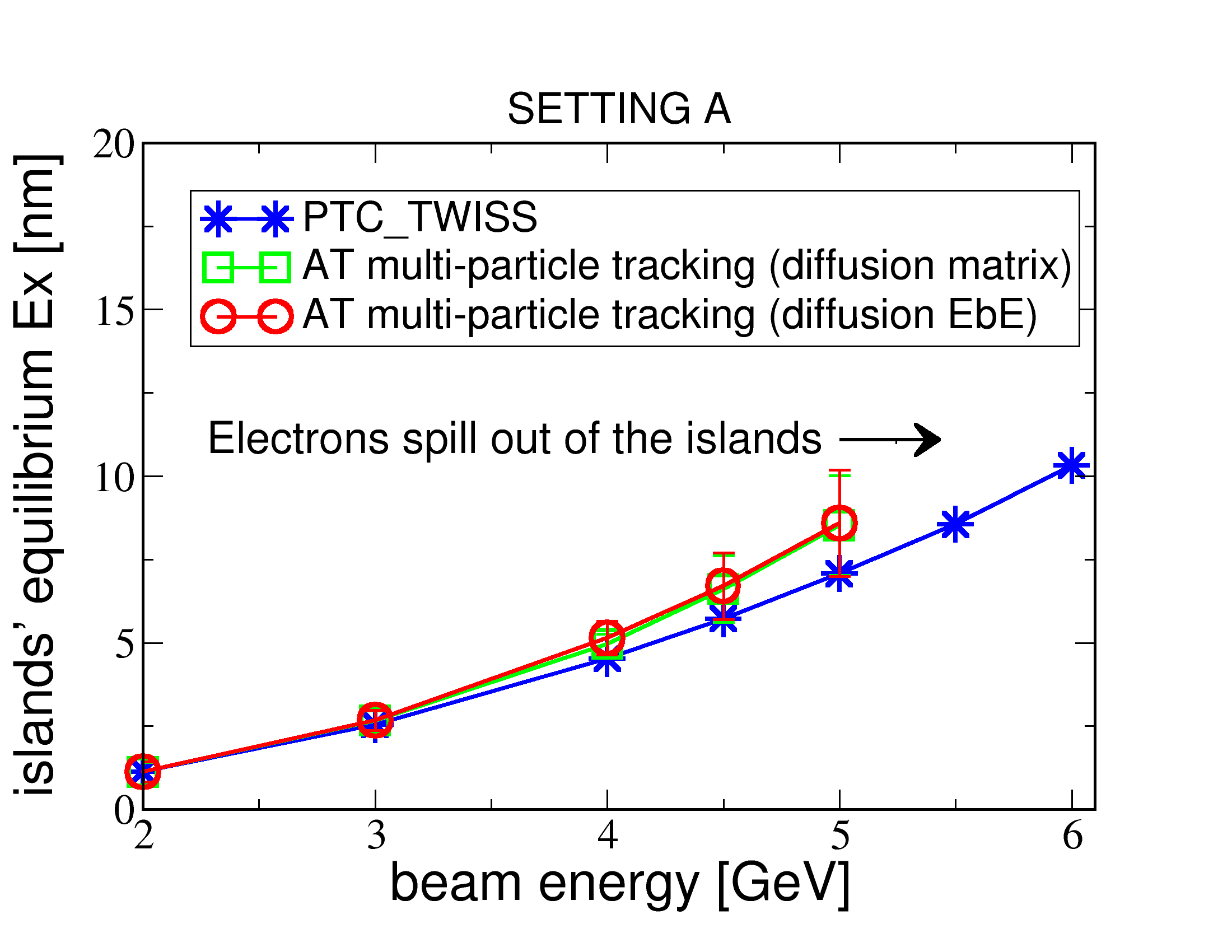}
  \includegraphics[trim=2truemm 2truemm 2truemm 20truemm, width=0.45\linewidth,angle=0,clip=]{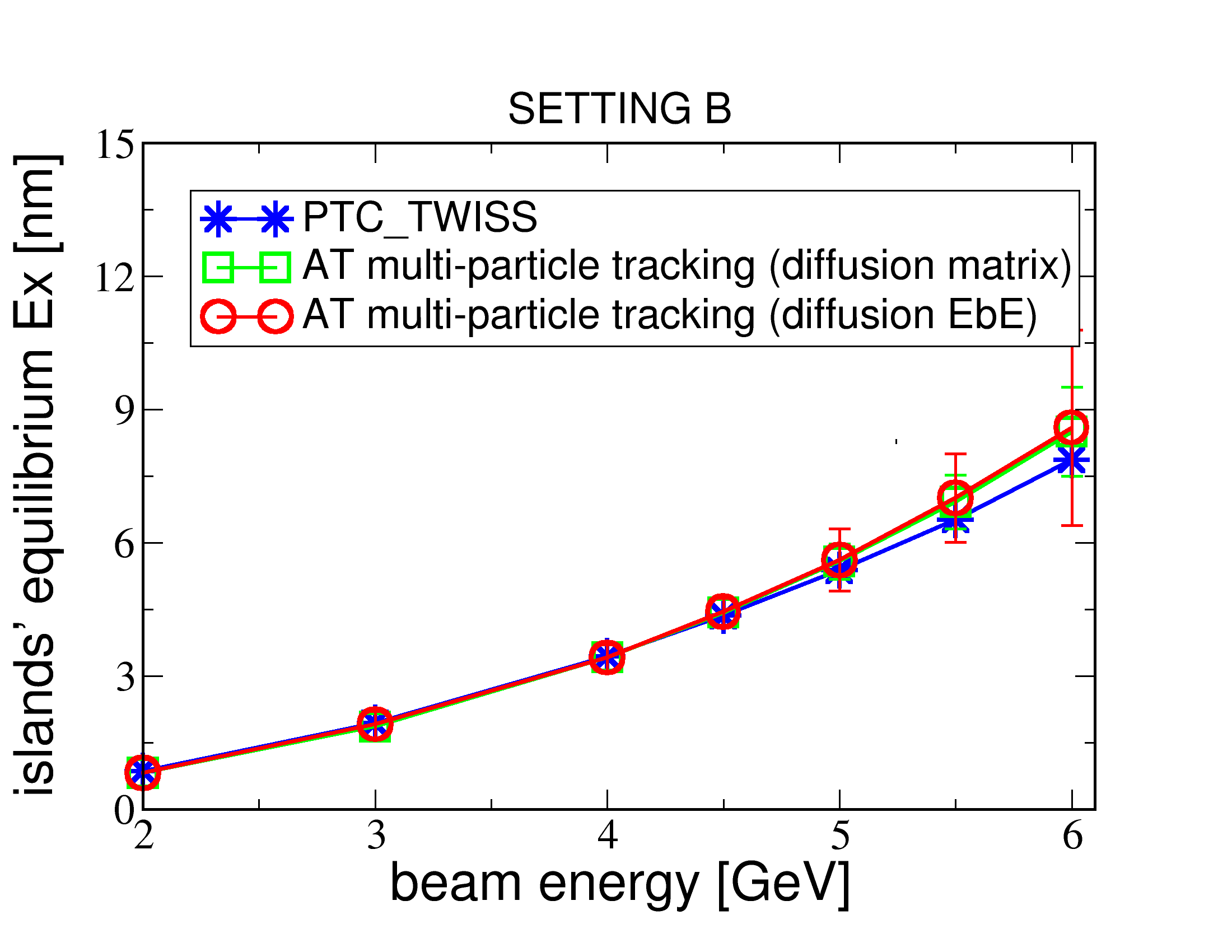}\\ \ \\
  \includegraphics[trim=2truemm 2truemm 2truemm 2truemm, width=0.3\linewidth,angle=0,clip=]{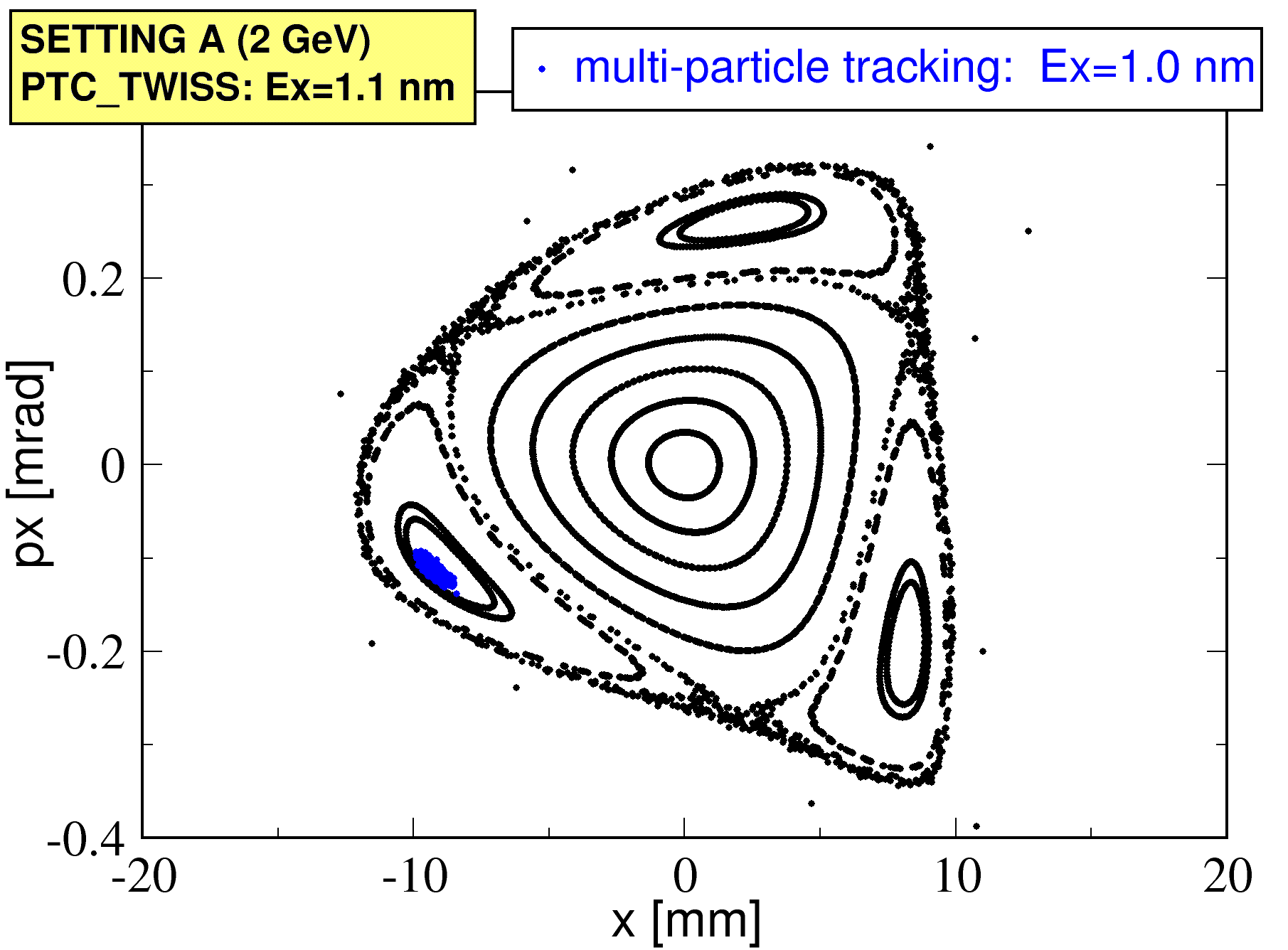}
  \includegraphics[trim=2truemm 2truemm 2truemm 2truemm, width=0.3\linewidth,angle=0,clip=]{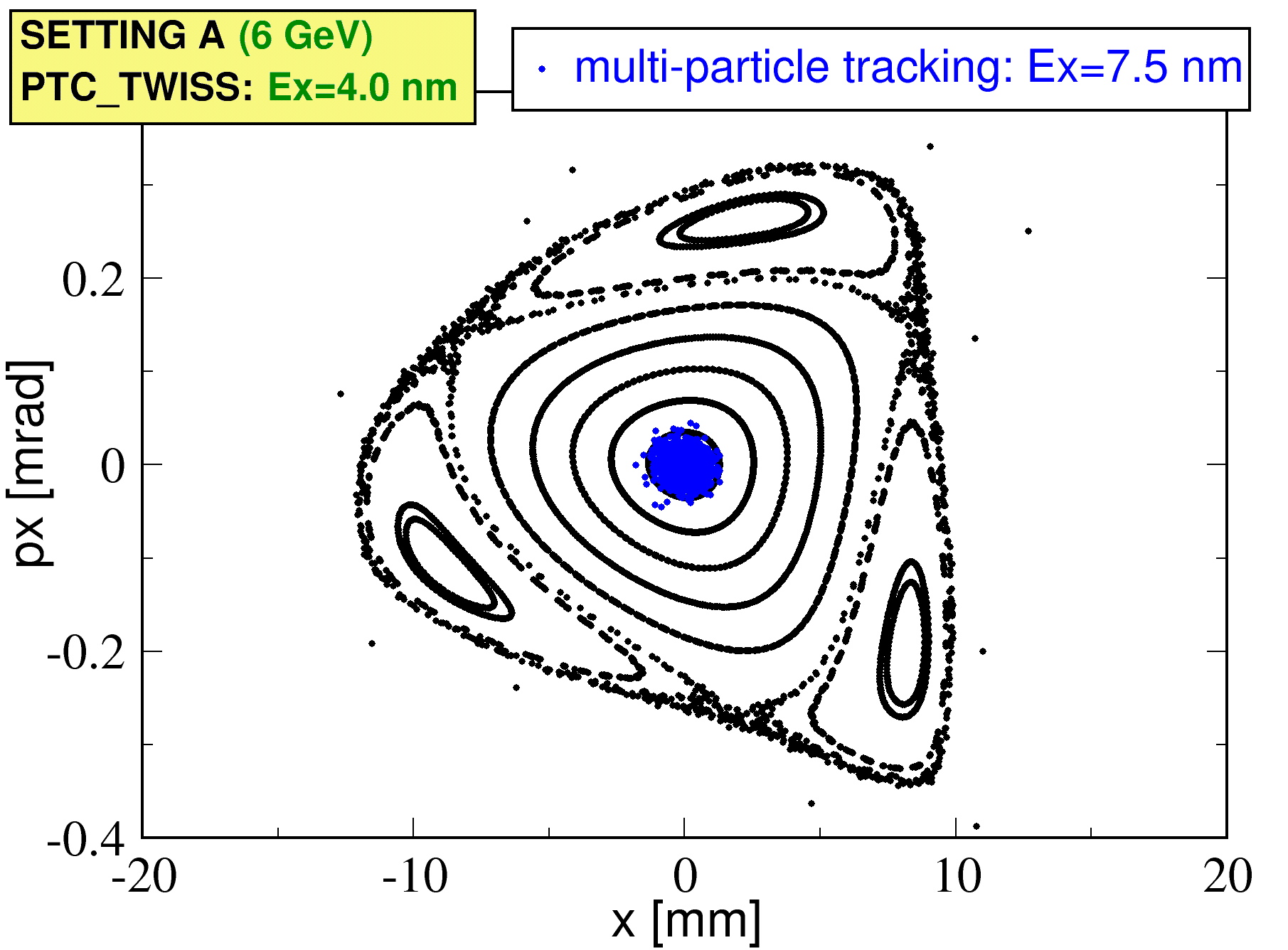}
  \includegraphics[trim=2truemm 2truemm 2truemm 2truemm, width=0.3\linewidth,angle=0,clip=]{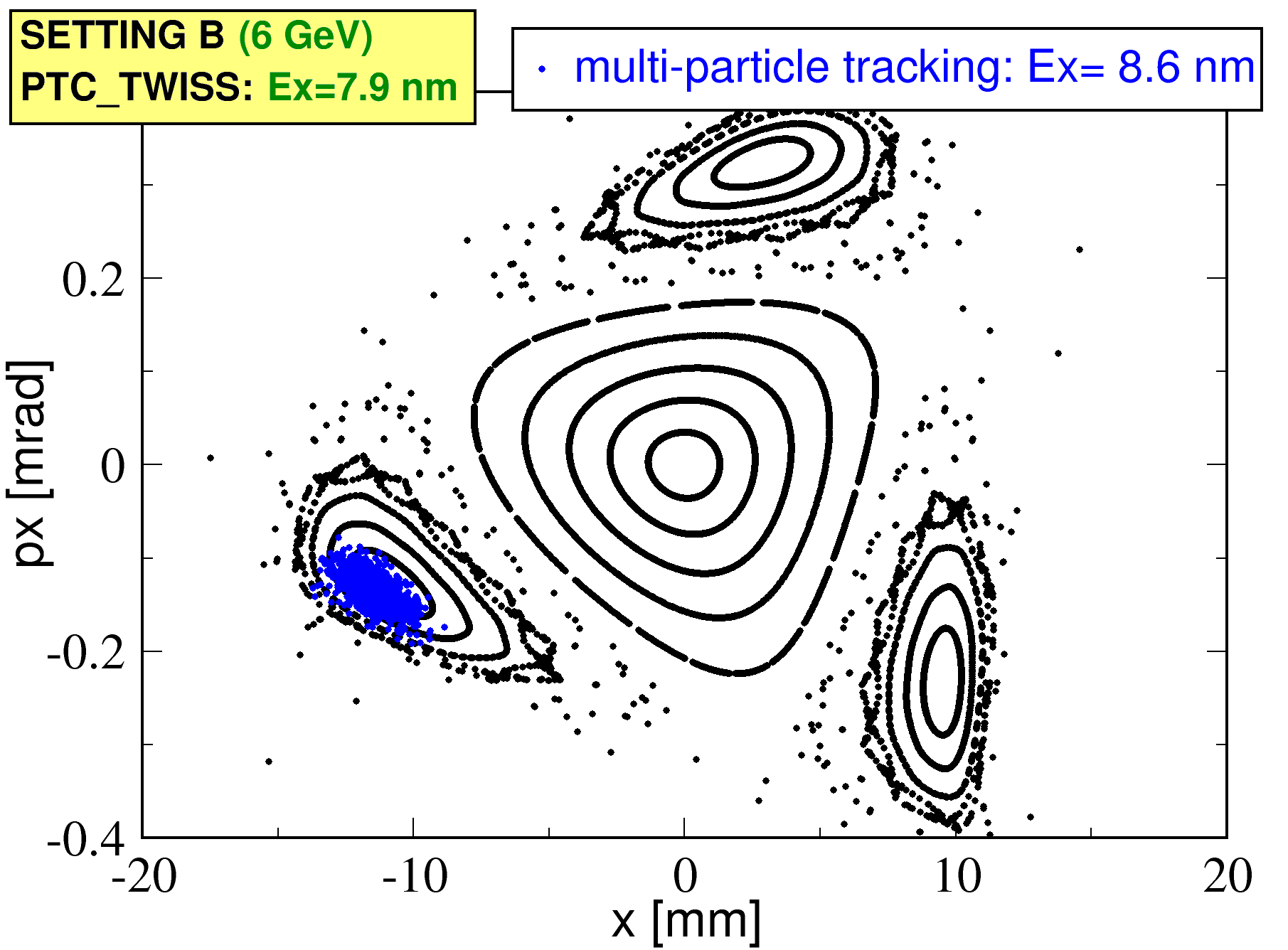}
  \caption{Top: The equilibrium horizontal emittance of the islands computed by MAD-PTC and from AT multi-particle tracking data (with two different algorithms describing the quantum diffusion) for two sextupole settings of the original ESRF storage ring. Bottom: Final phase-space multi-particle distribution from AT and portraits computed by MAD-X-PTC for some selected simulations (from Ref.~\cite{NOCE17}).}
  \label{Fig:Ex-SET6-9}
  \end{center}
\end{figure}

\begin{figure}
  \begin{center}
  \includegraphics[trim=2truemm 2truemm 2truemm 0truemm, width=0.45\linewidth,angle=0,clip=]{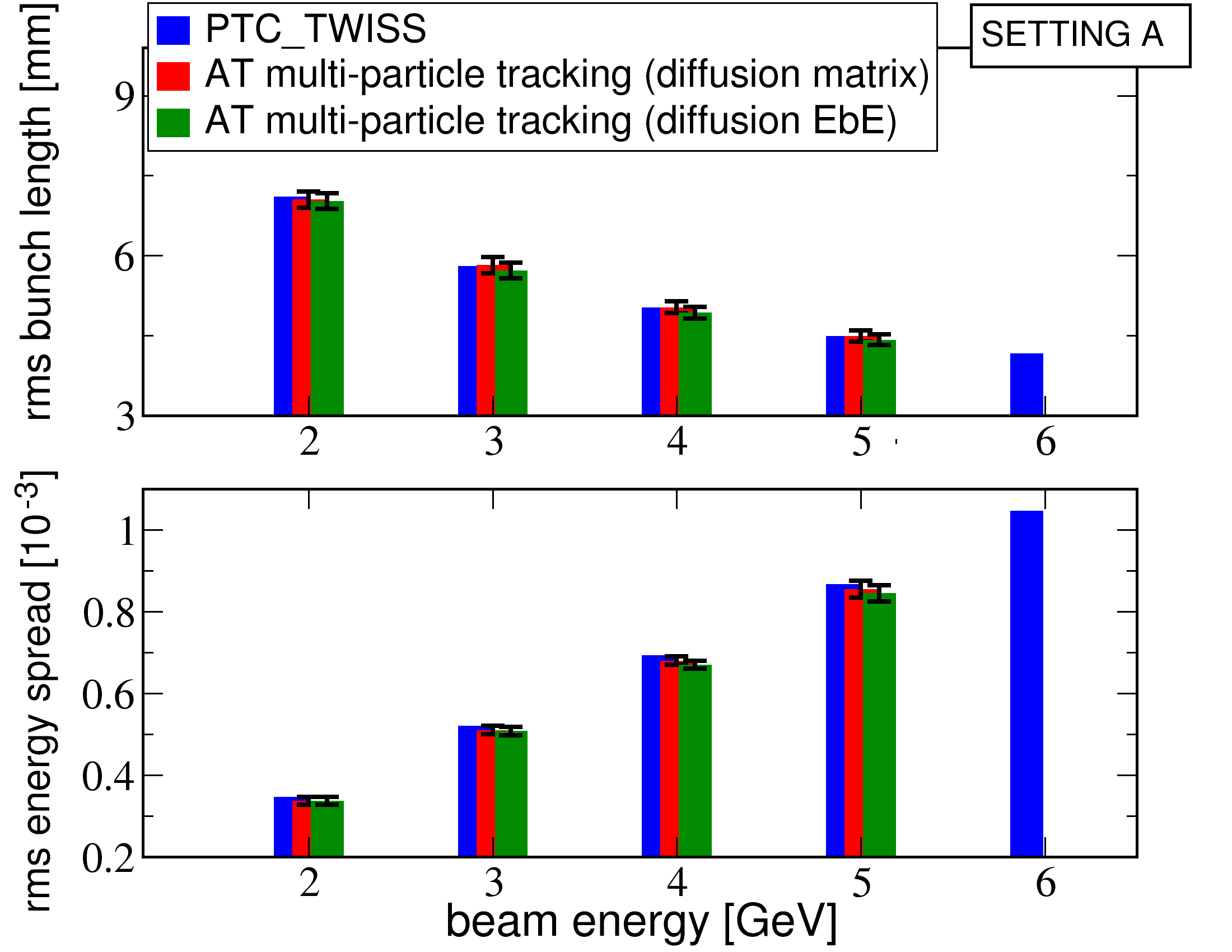}
  \includegraphics[trim=2truemm 2truemm 2truemm 0truemm, width=0.45\linewidth,angle=0,clip=]{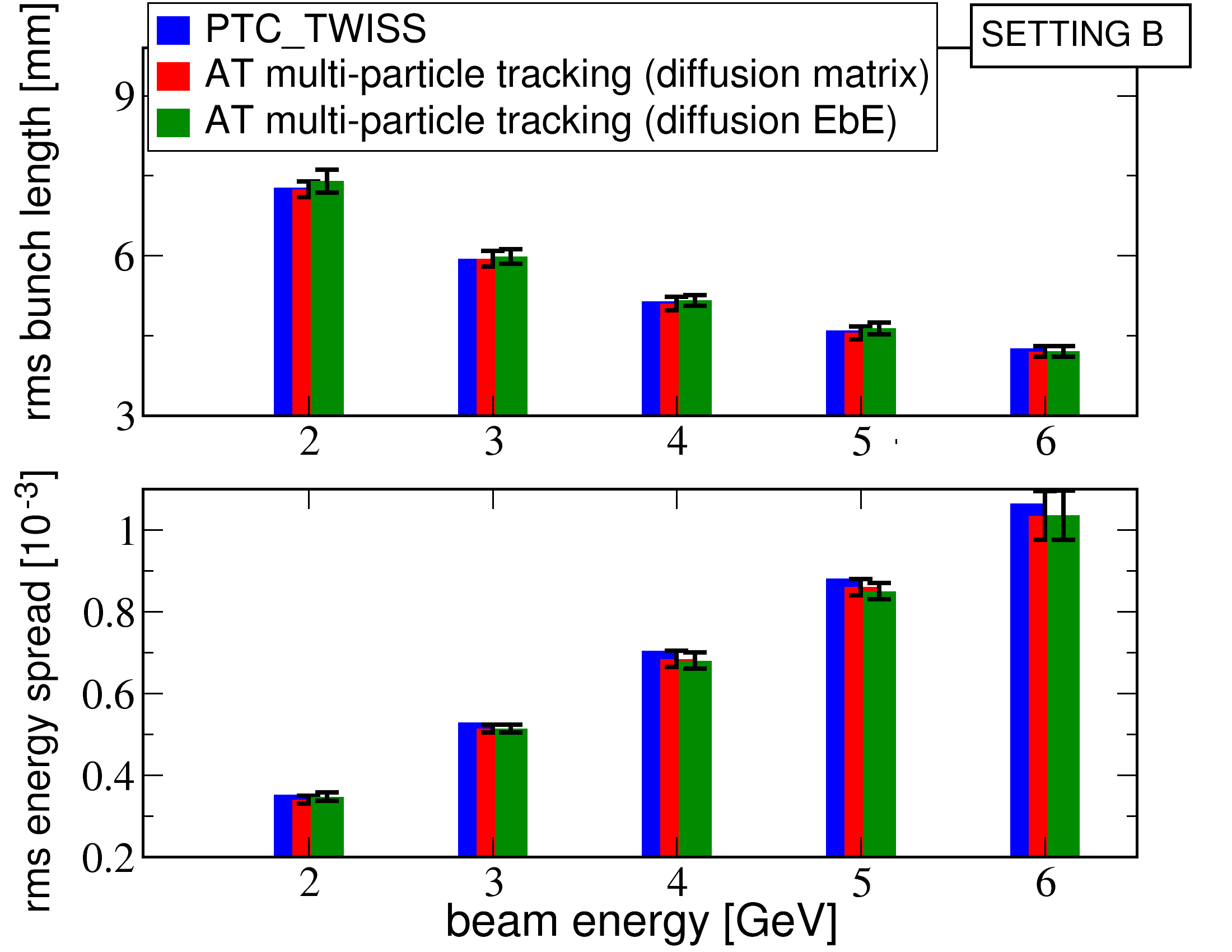}
  \caption{Final RMS bunch length (top) and energy spread (bottom) corresponding to the simulations of Fig.~\ref{Fig:Ex-SET6-9} (top) for Setting~A and B. The agreement between AT multi-particle tracking and MAD-X-PTC, when possible, is remarkable (from Ref.~\cite{NOCE17}).}
  \label{Fig:LongEquil}
  \end{center}
\end{figure}

MAD-X-PTC predictions for the longitudinal equilibrium parameters for both sets agree very well with the values obtained with multi-particle tracking, as shown in Fig.~\ref{Fig:LongEquil}.

The obvious question is whether lepton beams initially placed inside stable resonance islands will remain confined there, as is the case of hadron beams, which are governed by a Hamiltonian, or whether the radiation damping and diffusion processes will force leptons out of the islands and back to the nominal closed orbit. Early numerical studies~\cite{NOCE17} showed that electrons may indeed remain confined inside the stable islands. Furthermore, the numerical prediction of the equilibrium emittance for a beam in the stable islands carried out by MAD-X-PTC can agree with the one obtained through multi-particle tracking by AT, although these observations are lattice-dependent. 

To explore the different regimes of equilibrium emittances, the lattice of the original ESRF storage ring was used for numerical simulations with electrons at different energies, thus at different equilibrium emittances, without changing the nonlinear configuration of the lattice. This implies that the phase-space topology remains unchanged, in particular the position and surface of the islands. Indeed, an equilibrium emittance comparable or larger than the island's surface would make electrons spill out of the islands and reach the core. To answer the question about the existence of an equilibrium within the islands, several ensembles of $10^3$ macro-particles were generated. These distributions represent beams of different initial emittance values, smaller and larger than the equilibrium emittance estimated by MAD-X-PTC, and with optical mismatch with respect to the Courant-Snyder parameters of the islands. The different ensembles of macroparticles were placed around one fixed point and tracked with AT for several thousands turns. In all cases, the same equilibrium situation was reached, either inside the islands or around the core. It is worth mentioning that the AT multi-particle tracking simulations were carried out with two different algorithms for the evaluation of the quantum diffusion term. The first one is based on the evaluation of a global one-turn diffusion matrix, derived from the lattice properties and the closed orbit, to be applied turn by turn. The second algorithm is based on the random evaluation of the diffusion terms element by element, derived from the angular kick imparted by each magnet to every electron, thus accounting for the closed orbit of the fixed points~\cite{Helmut-diffusion}. The final equilibrium emittances from AT multi-particle tracking were then compared to the analytical predictions performed by the \verb PTC_Twiss module of MAD-X-PTC, which are based on the evaluation of the radiation integrals.

Results for two sextupole settings, hence with different horizontal phase-space topology and island properties, are shown in the two upper plots of Fig.~\ref{Fig:Ex-SET6-9}. The first observation is that equilibrium inside the islands is possible and that the predictions by MAD-X-PTC are consistent with the evaluation of the equilibrium emittance from multi-particle simulations. The lower the equilibrium emittance, corresponding to lower beam energy in these examples, the better the agreement. For the Setting~B (top-right plot), the agreement is good even at higher energy, with $E_x\sim8$~nm~rad (note that $E_x=4$~nm~rad for the standard beam at the nominal closed orbit). The corresponding final (projected) distribution in the horizontal phase space is displayed in the bottom-right plot. The situation is less clear for the other sextupole configuration (Setting~A), whose results are shown in the upper-left plot of Fig.~\ref{Fig:Ex-SET6-9}. Equilibrium is possible at low energies, i.e.\ at emittance $E_x < 2$~nm~rad, whereas for larger values a discrepancy between the MAD-X-PTC predictions and results from multi-particle tracking appears. Starting from an energy of $5.5$~GeV ($E_x\sim7.5$~nm~rad) electrons start to spill out of the islands and reach the nominal closed orbit, thus suggesting that in these regimes equilibrium within the islands cannot be reached. 

The origin of the conditional equilibrium is not yet fully understood. As far as the discrepancy in the equilibrium emittance between MAD-X-PTC and AT multi-particle tracking is concerned, the reason may be found in the limited region of linear dynamics around the fixed points, which are denoted by distorted ellipses. If the beam distribution covers this nonlinear region, additional diffusive terms may act, which are not contemplated by the classical textbook formulae or algorithms for the evaluation of the equilibrium emittance implemented in MAD-X-PTC. It has been observed that electrons leaving the islands to reach the nominal closed orbit are those with the largest energy deviation resulting from quantum diffusion: the fact that for Setting~B none leave the islands suggests that the corresponding momentum acceptance of the islands is greater than that of Setting~A. It is interesting to note that the final equilibrium emittance of Setting A at $6$~GeV once all electrons abandon the islands to populate the core is wrong when the diffusion matrix is used (bottom-centre phase-space portrait of Fig.~\ref{Fig:Ex-SET6-9}): it is of about $7.5$~nm~rad instead of the true $4$~nm~rad. This is a consequence of using the diffusion matrix initially evaluated from the closed orbit of the islands for electrons jumping to the standard on-axis closed orbit. The more realistic element-by-element diffusion algorithm provides the correct final equilibrium emittance. 
 
\subsection{Do electrons in resonance islands experience synchro-betatron coupling?} 
By looking at the $N-$turn ring $R_N$, the tune around the fixed point is close to $0$ or to $1$. Therefore, it is natural to suspect a possible effect of synchro-betatron coupling in the electron dynamics inside the islands, with the horizontal and synchrotron tunes being much closer than for the standard optics around the nominal closed orbit. To explore this effect, the equilibrium emittance can be computed from three different observables, similarly to what can be done to evaluate the betatron coupling between the transverse planes~\cite{Andrea-CouplingESRF}. From the covariance matrix of the final 6D particle distribution, the following equilibrium emittances can be computed: $(i)$ the eigen-emittance, i.e.\ the matrix eigenvalue corresponding to the horizontal phase space; $(ii)$ the projected emittance, i.e.\ the determinant of the $2\times2$ projection of the original $6 \times 6$ covariant matrix onto the horizontal phase space, $\sigma_x\sigma_p-\sigma_{xp}^2$; $(iii)$ the apparent emittance, i.e.\ the horizontal betatron rms beam size divided by the island's beta function, $\sigma_x^2/\beta_x$, which is the only observable quantity. In the absence of any coupling, these three definitions are equivalent and yield the same result. If coupling is present, either horizontal-vertical or synchro-betatron, differences between the three definitions appear: the larger the coupling, the larger the differences~\cite{Andrea-CouplingESRF}. No significant differences above the numerical noise levels were found in any of the simulated lattices and configurations, thus suggesting that the synchro-betatron coupling plays no role in the electron dynamics inside the islands.

\subsection{Do the electrons in resonance islands have the nominal energy?} 
\begin{figure}
  \begin{center}
  \includegraphics[trim=2truemm 2truemm 2truemm 2truemm, width=0.4\linewidth,angle=0,clip=]{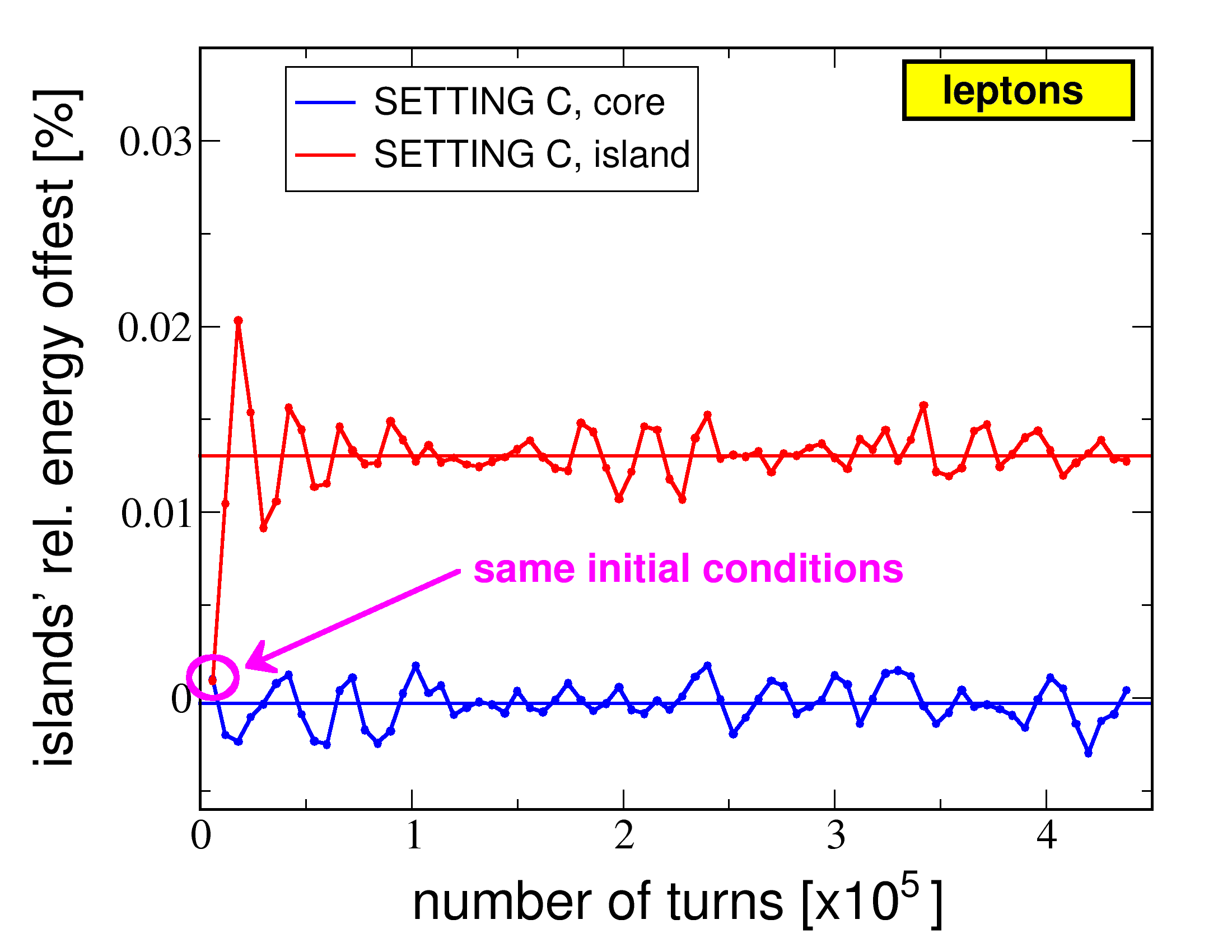}\ \ \ \ 
  \includegraphics[trim=2truemm 2truemm 2truemm 2truemm, width=0.4\linewidth,angle=0,clip=]{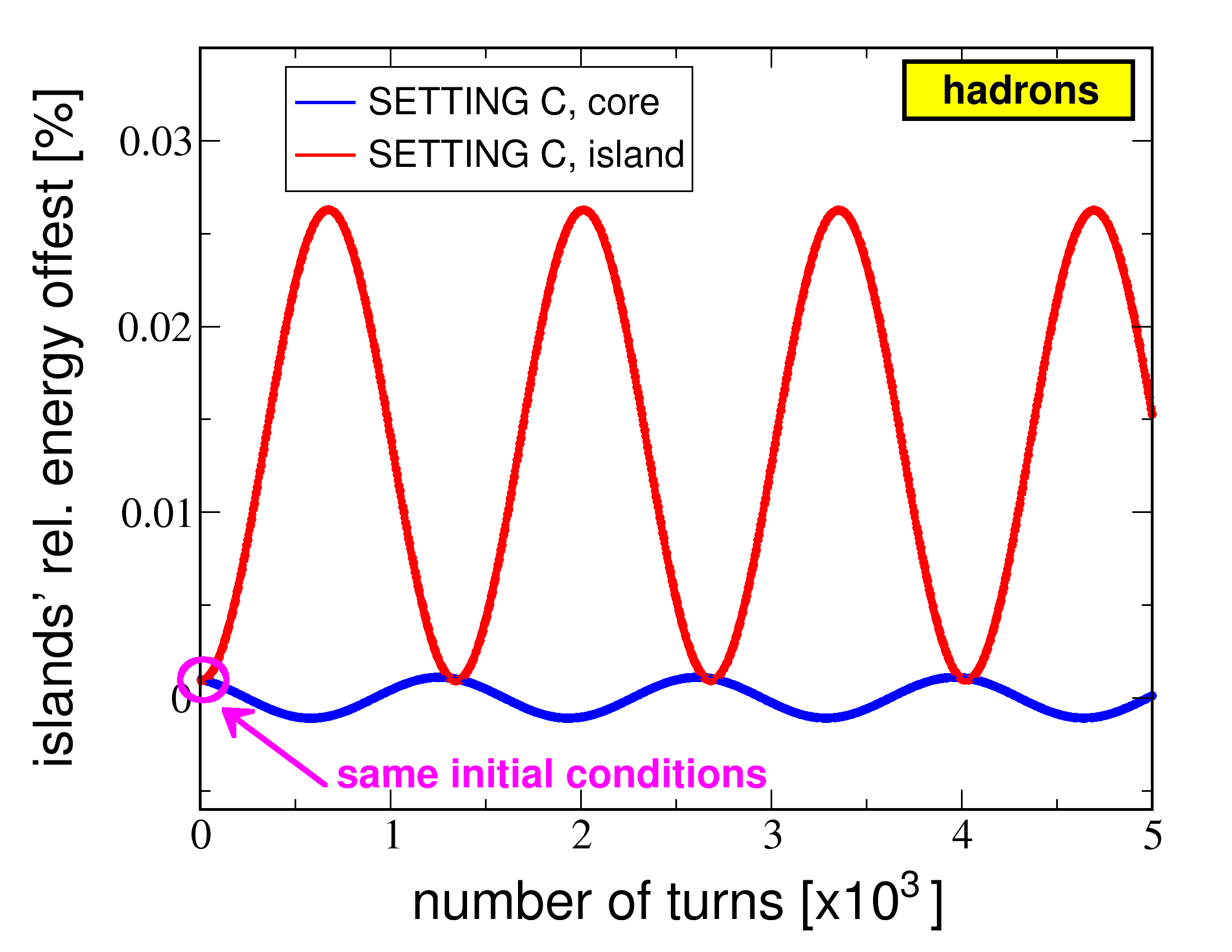} \\ \ \\
  \includegraphics[trim=2truemm 2truemm 2truemm 2truemm, width=0.4\linewidth,angle=0,clip=]{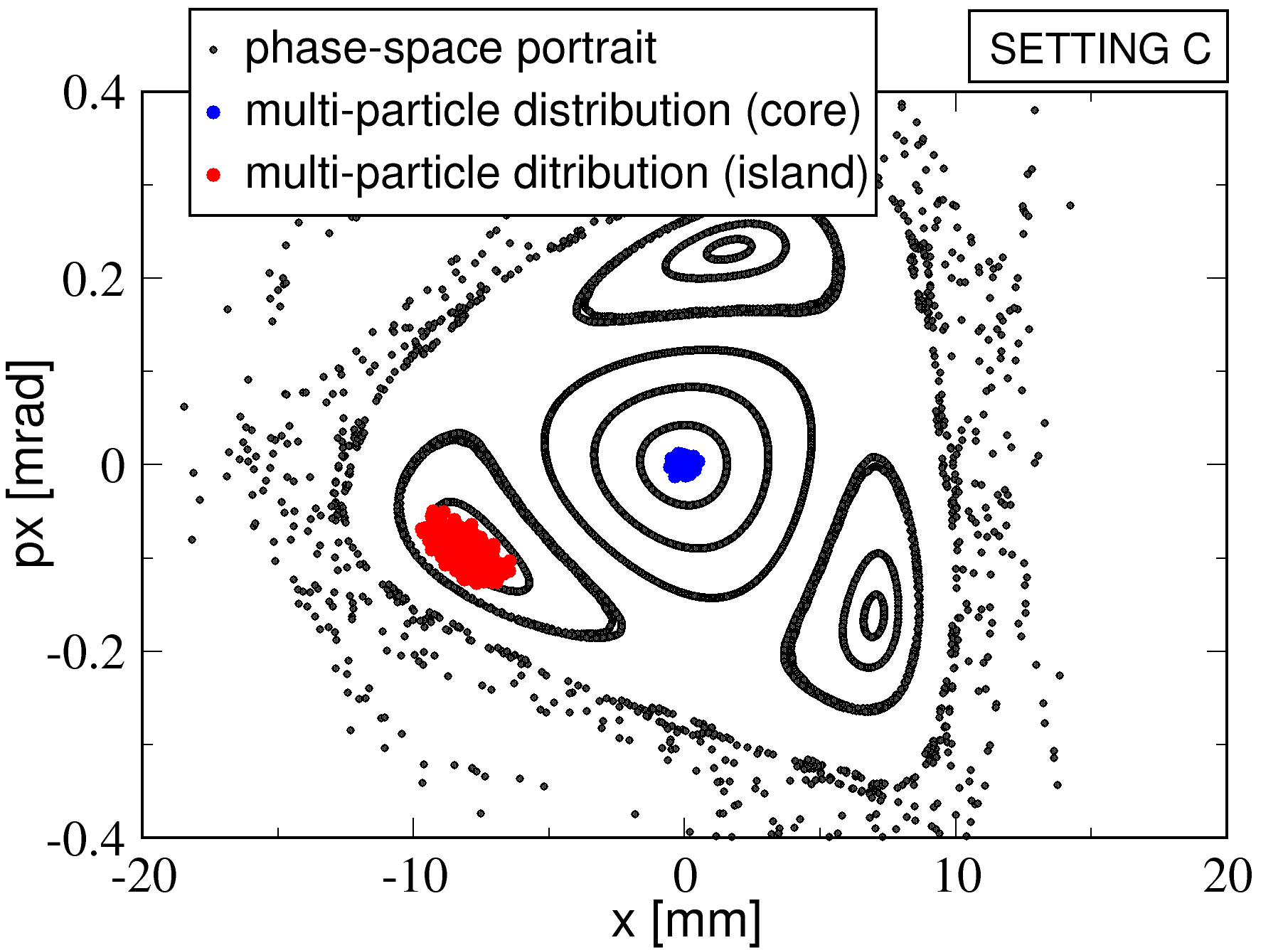}\ \ \ \
  \includegraphics[trim=2truemm 2truemm 2truemm 2truemm, width=0.4\linewidth,angle=0,clip=]{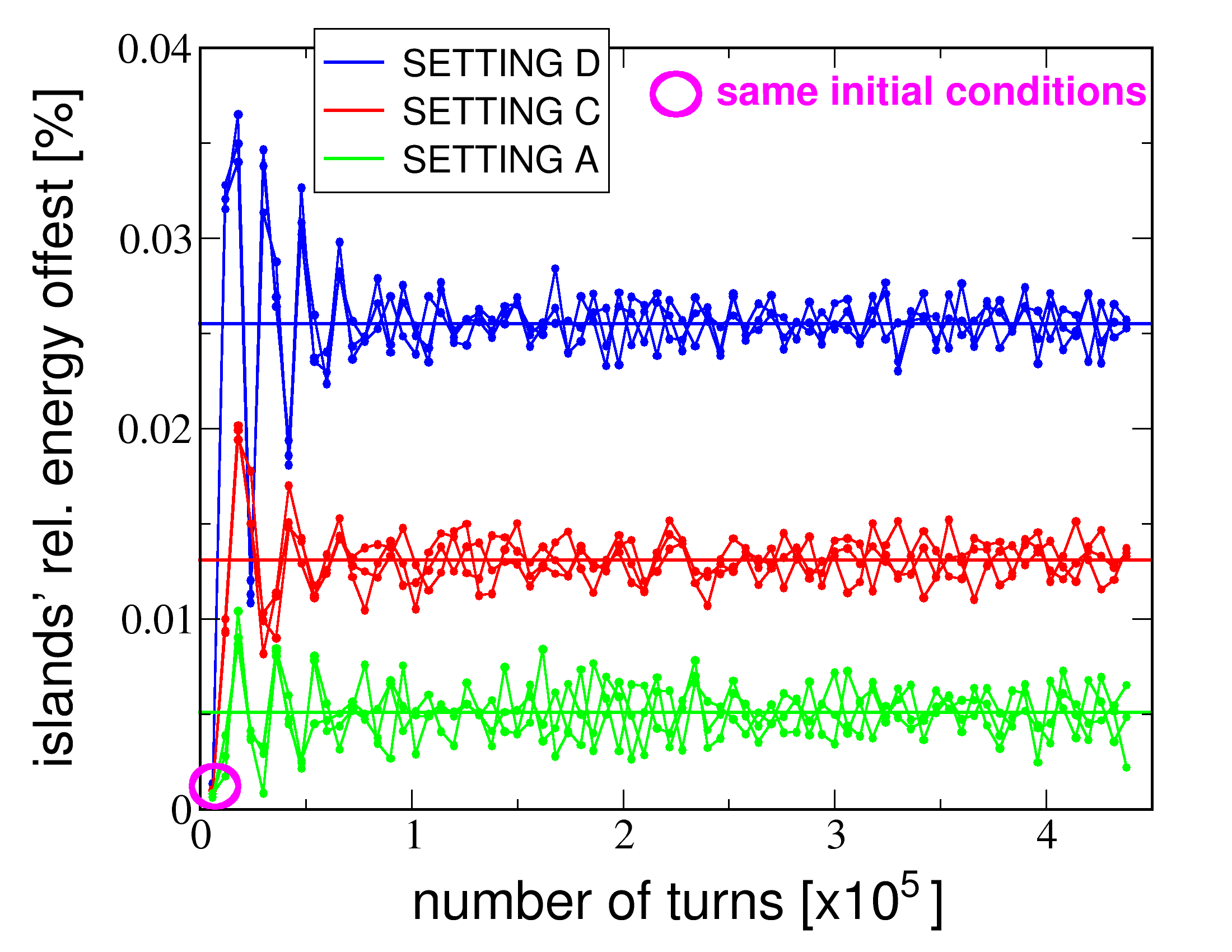}
  \caption{Top: Temporal evolution of the energy offset of particles trapped inside the islands (red curves) and around the nominal closed orbit (blue curves) inferred from AT multi-particle simulations of $10^3$ electrons (left) and protons (right): the absence of radiation damping and diffusion makes the protons trapped inside the islands oscillate continuously in energy at the synchrotron frequency. Bottom: Initial horizontal phase-space distribution corresponding to the top plots (left) and temporal evolution of the energy offset for electrons trapped inside the islands of three sextupole settings (right).}
  \label{Fig:IslandEnergyOffset}
  \end{center}
\end{figure}
Islands are expected to have a tiny energy offset compared to the beam around the nominal closed orbit, because of their longer path over one turn. This is confirmed by multi-particle simulations. In the upper-left plot of Fig.~\ref{Fig:IslandEnergyOffset} the energy offset of a beam of $10^3$ electrons placed either around the nominal closed orbit or in a stable resonance island (see bottom-left phase-space image in the same figure), initially with the same nominal energy, is plotted over several hundreds of thousands of turns. The energy of the electrons in the islands quickly jumps and reaches equilibrium in about one thousand turns. The energy deviation is tiny ($\sim1.3\times10^{-4})$ but visible. The damping of the energy offset is a peculiarity of lepton beams. By simulating the same lattice with a hadron beam (i.e.\ without radiative effects), the energy offset of the beam in the islands oscillates around the same value as the energy offset of the electrons at the synchrotron frequency, never reaching equilibrium, as shown in the top-right plot of Fig.~\ref{Fig:IslandEnergyOffset}. This small energy offset depends on several parameters, mainly the distance of the island from the central orbit (for a given lattice setting, the larger the distance, the greater the offset), the horizontal dispersion and the island's momentum compaction, which can be different from the standard one~\cite{giovannozzi2021novel}. This is shown in the bottom-right plot of the same figure, where the energy offset of the island is displayed for three different sextupole settings.
\begin{figure}[htb]
  \begin{center}
  \includegraphics[trim=2truemm 2truemm 2truemm 2truemm, width=0.49\linewidth,angle=0,clip=]{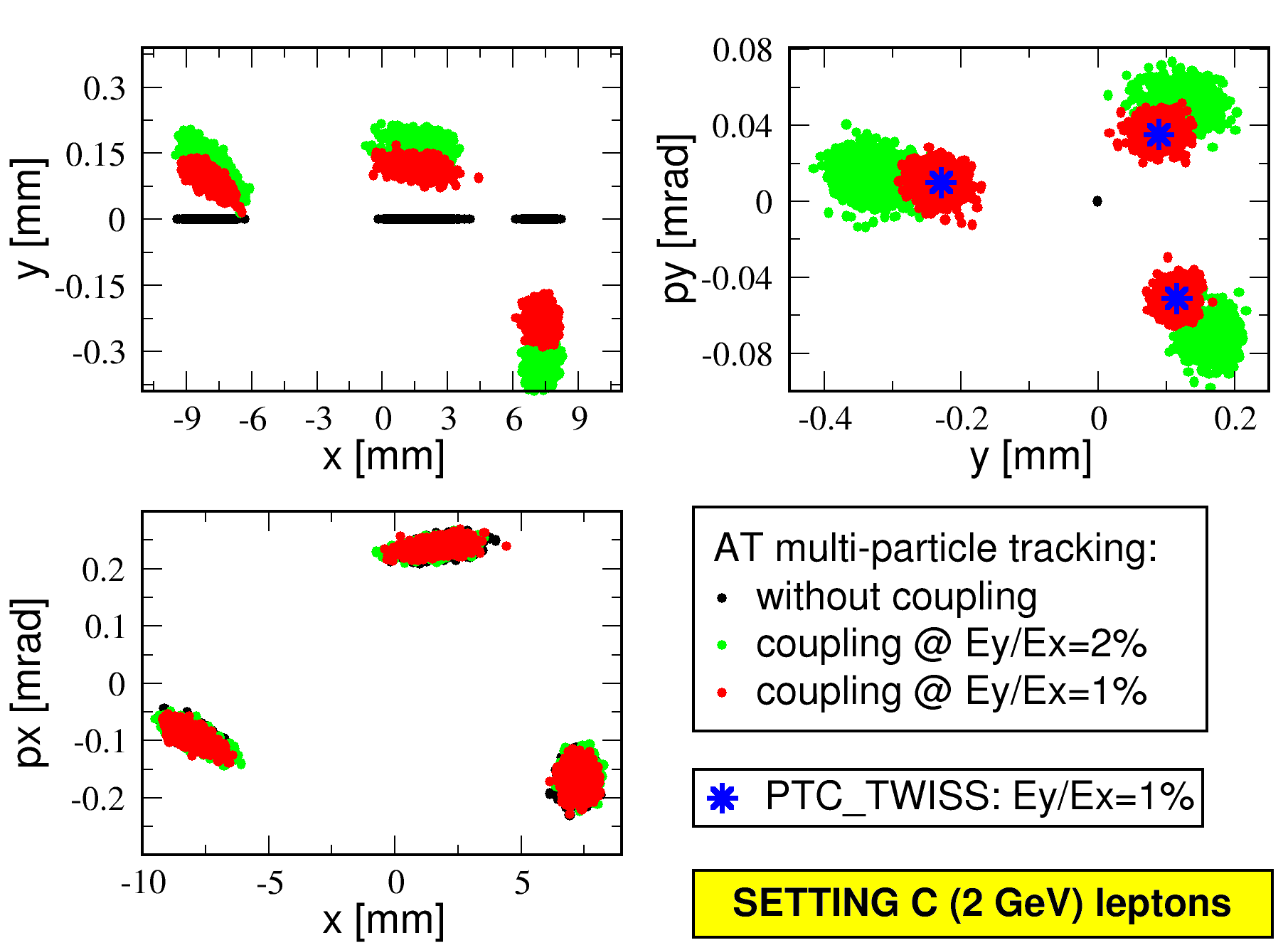}\ \ \
  \includegraphics[trim=2truemm 2truemm 2truemm 2truemm, width=0.49\linewidth,angle=0,clip=]{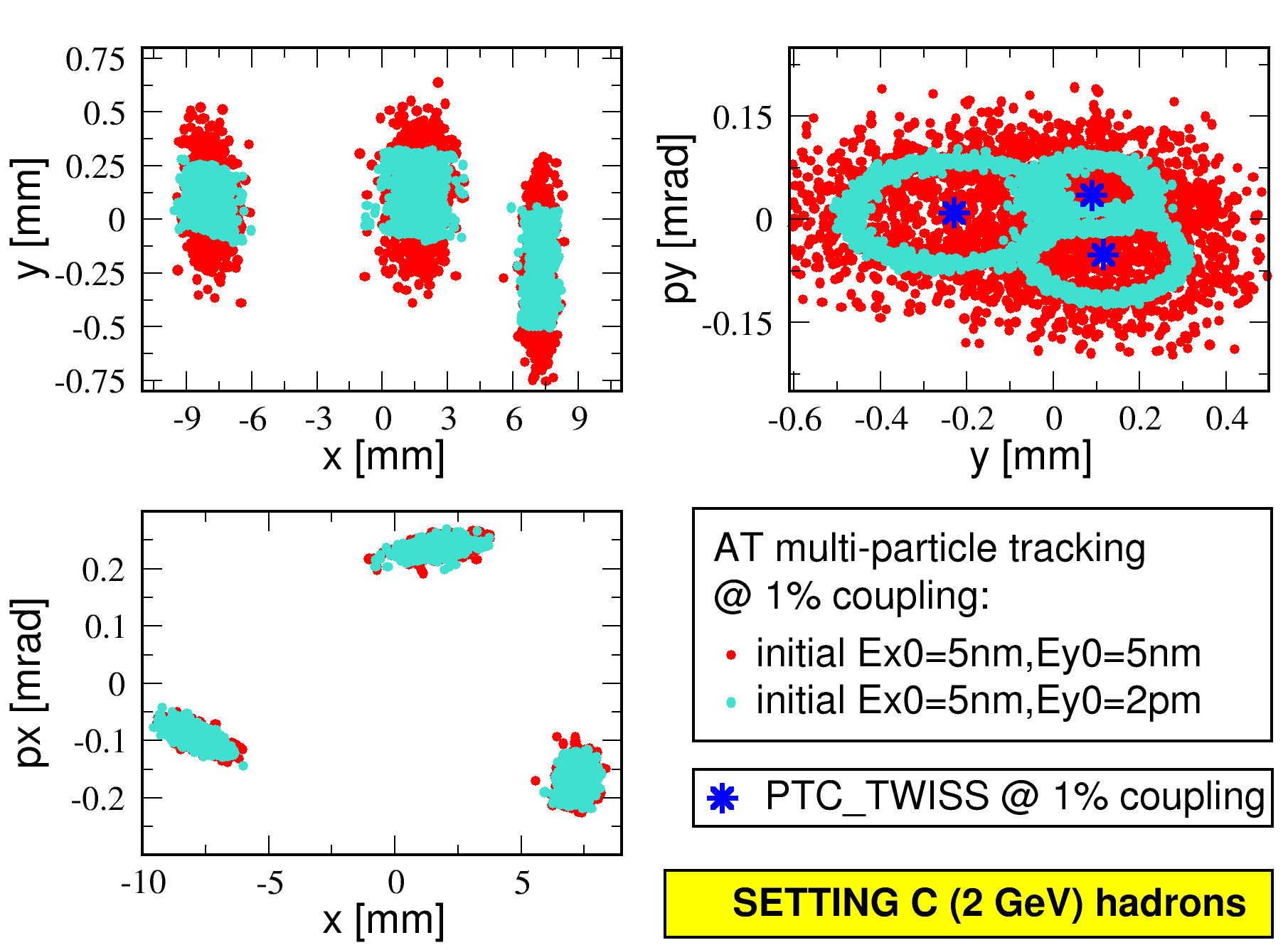}
  \caption{Left: Configuration, vertical, and horizontal phase-space projections of multi-particle distributions of electrons at equilibrium for different amount of linear  betatron coupling, defined as the ratio between the equilibrium vertical and horizontal emittances of a beam around the nominal closed orbit. Right: Same projections for a proton ensemble with the skew quadrupole having the same setting that generated $1$\% linear betatron coupling with the electrons and two different sets of initial emittances. The position of the vertical fixed points computed by MAD-X-PTC is denoted in both cases by the blue stars. }
  \label{Fig:IslandCoupling}
  \end{center}
\end{figure}

\subsection{Stable islands and linear betatron coupling} \label{sec:inter}

In Section~\ref{sec:scenario2}, it was mentioned that the orbit beamlets can be transferred to the vertical plane by means of either linear betatron coupling or a vertical AC dipole. Both MAD-X-PTC and AT multi-particle simulations indicate that the resonance islands in the horizontal phase space can be transferred to the vertical phase space by adding linear coupling. This result is by far not trivial, since Hamiltonian analytical models of fixed points and single-particle dynamics around them in a pure 2D system are well established, whereas, to the best of our knowledge, the theoretical extension to 4D structures has been rarely investigated in detail (see, e.g.\ Ref.~\cite{PhysRevE.50.R4298} for a result in 4D Hamiltonian systems), and the interplay with linear betatron coupling has never been addressed. Our study of this aspect has been merely numerical with no attempt to build a complete theoretical framework. For a given lattice setting, a skew quadrupole was added so to generate a certain amount of linear coupling, which for lepton machines is usually identified by the ratio between the equilibrium emittances in the two planes $E_y/E_x$: the smaller the ratio, the weaker the coupling. MAD-X-PTC was then used to compute the closed orbit in the vertical plane by giving as input the position of one fixed point in the horizontal phase space and including a skew quadrupole generating a coupling of $1$\%. In parallel, a multi-particle distribution with matched optical parameters and horizontal emittance at equilibrium was placed around the same horizontal fixed point and allowed to evolve over several hundreds of thousands turns in the very same lattice including the skew quadrupole. The final particle distributions in the $(x,y)$ configuration space and both transverse phase spaces are displayed in the left plots of Fig.~\ref{Fig:IslandCoupling} along with the fixed-point positions computed by MAD-X-PTC. The position of the fixed points in the vertical phase space are in excellent agreement. The case without linear coupling produces a central spot in the vertical phase space (black dots) and no vertical fixed points. With a coupling of $1$\% the fixed points appear both in the vertical phase space and in the configuration plane. A larger coupling of $2$\% further separates those fixed points and islands in the vertical plane. In neither case is the horizontal phase space very much affected, owing to the small strength of the linear coupling.

Similarly to the stable energy offset discussed above, the generation of vertical islands is a peculiarity of beams subjected to radiative effects. By simulating the lattice with the skew quadrupole having the same setting that generated $1$\% linear betatron coupling with the electrons, hadron beams of the same energy exhibit annular distributions in the vertical phase space as long as the initial vertical emittance remains small. For a round beam, i.e.\ with the same initial emittance in the horizontal and vertical planes, the vertical phase-space distribution is completely smeared out, as shown in the right portraits of Fig.~\ref{Fig:IslandCoupling}. This topic certainly deserves further consideration.

\subsection{How to populate the resonance islands?} \label{sec:Populate}

Having proven the possibility of having stable lepton beams inside resonance islands, it remains to be shown how to populate them, i.e.\ how to put electrons in the islands. In the previous simulations, the beam was artificially put in the island and allowed to evolve. In this section, possible schemes to fill the islands are presented, each with its own features, potential, and limitations. 
All simulations were performed with the same baseline linear optics and an electron beam energy of 2 GeV to ensure the survival of the beam within the islands.
\begin{figure}[htb]
  \begin{center}
  \includegraphics[trim=0truemm 0truemm 0truemm 0truemm, width=0.22\linewidth,angle=0,clip=]{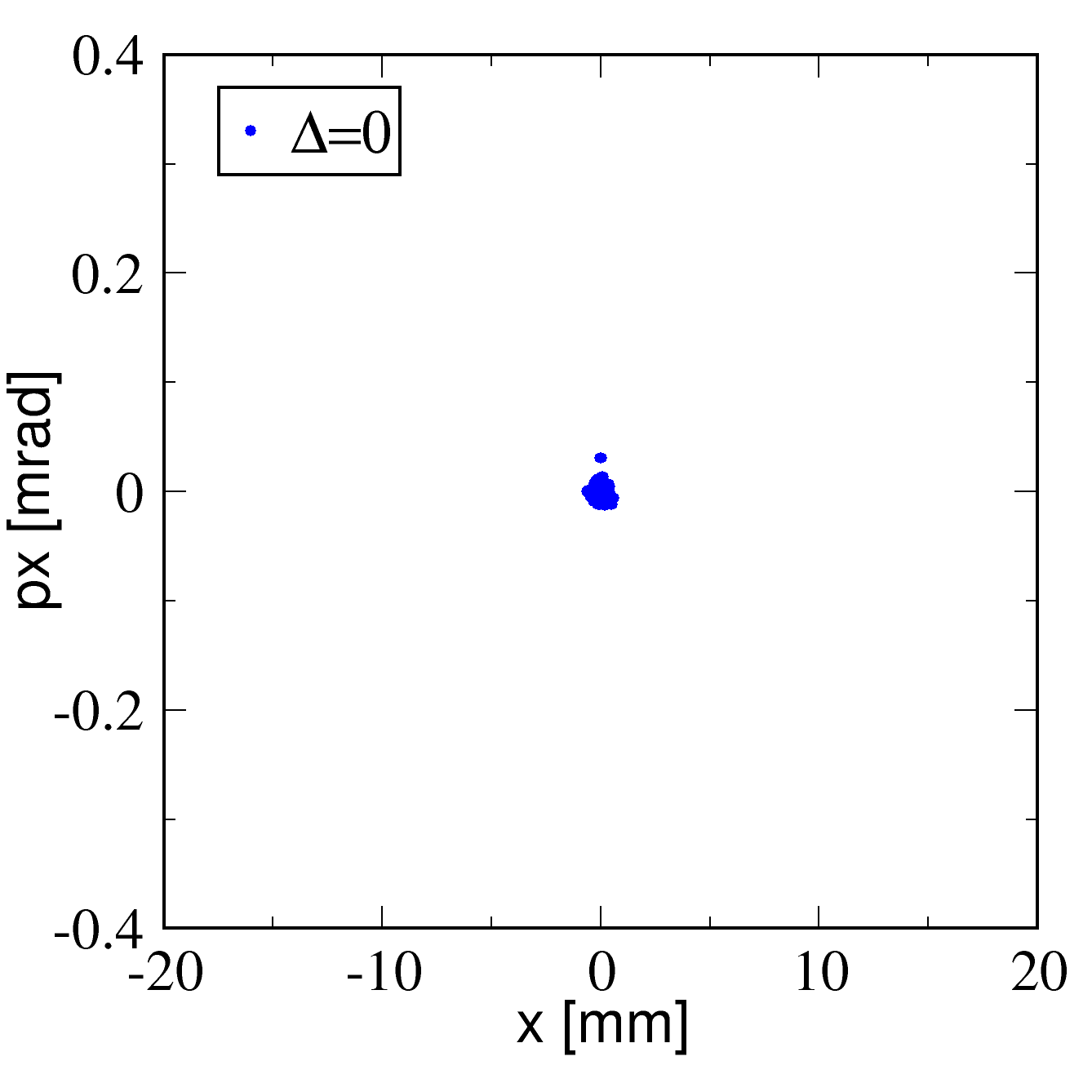}
  \includegraphics[trim=0truemm 0truemm 0truemm 0truemm, width=0.22\linewidth,angle=0,clip=]{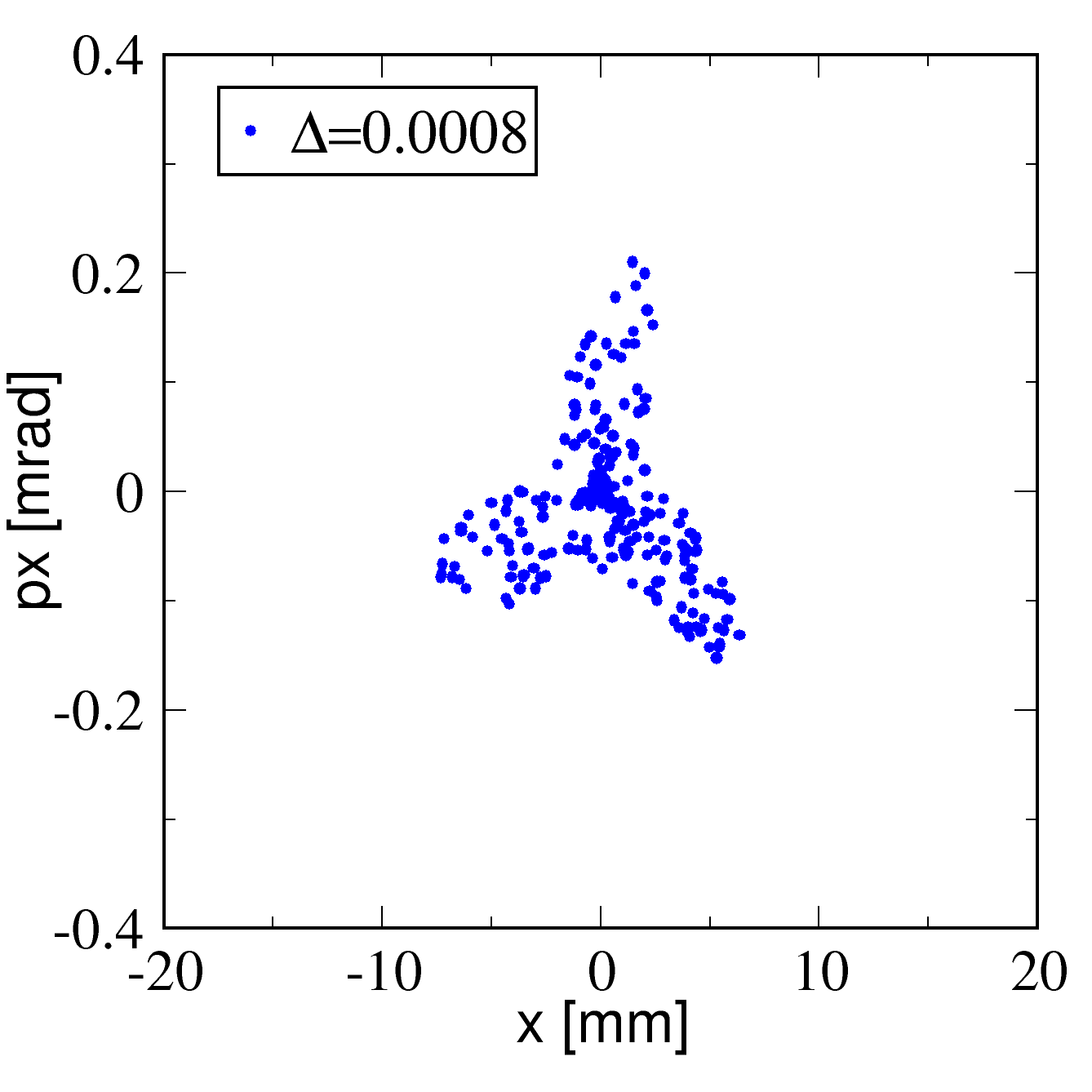}
  \includegraphics[trim=0truemm 0truemm 0truemm 0truemm, width=0.22\linewidth,angle=0,clip=]{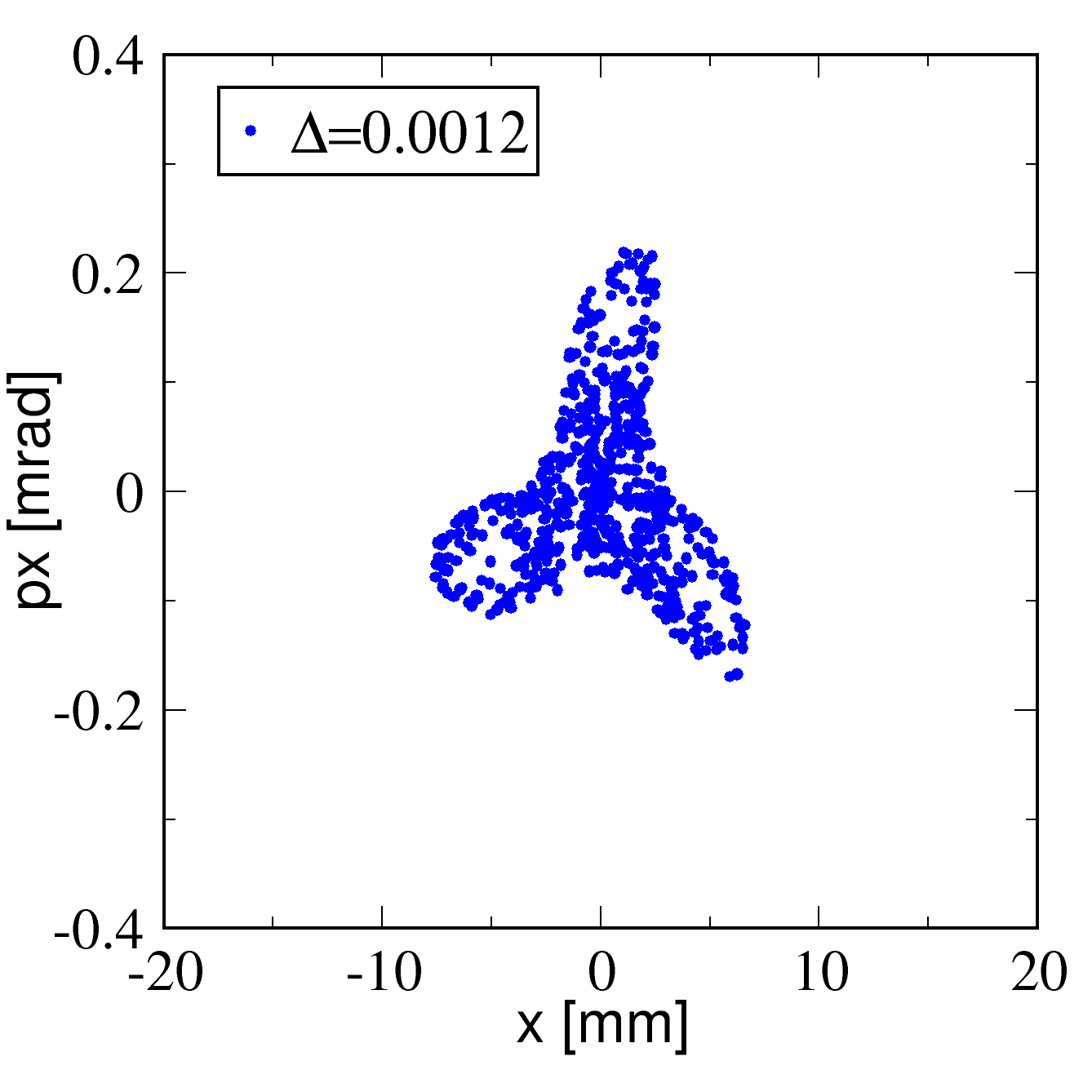}
  \includegraphics[trim=0truemm 0truemm 0truemm 0truemm, width=0.22\linewidth,angle=0,clip=]{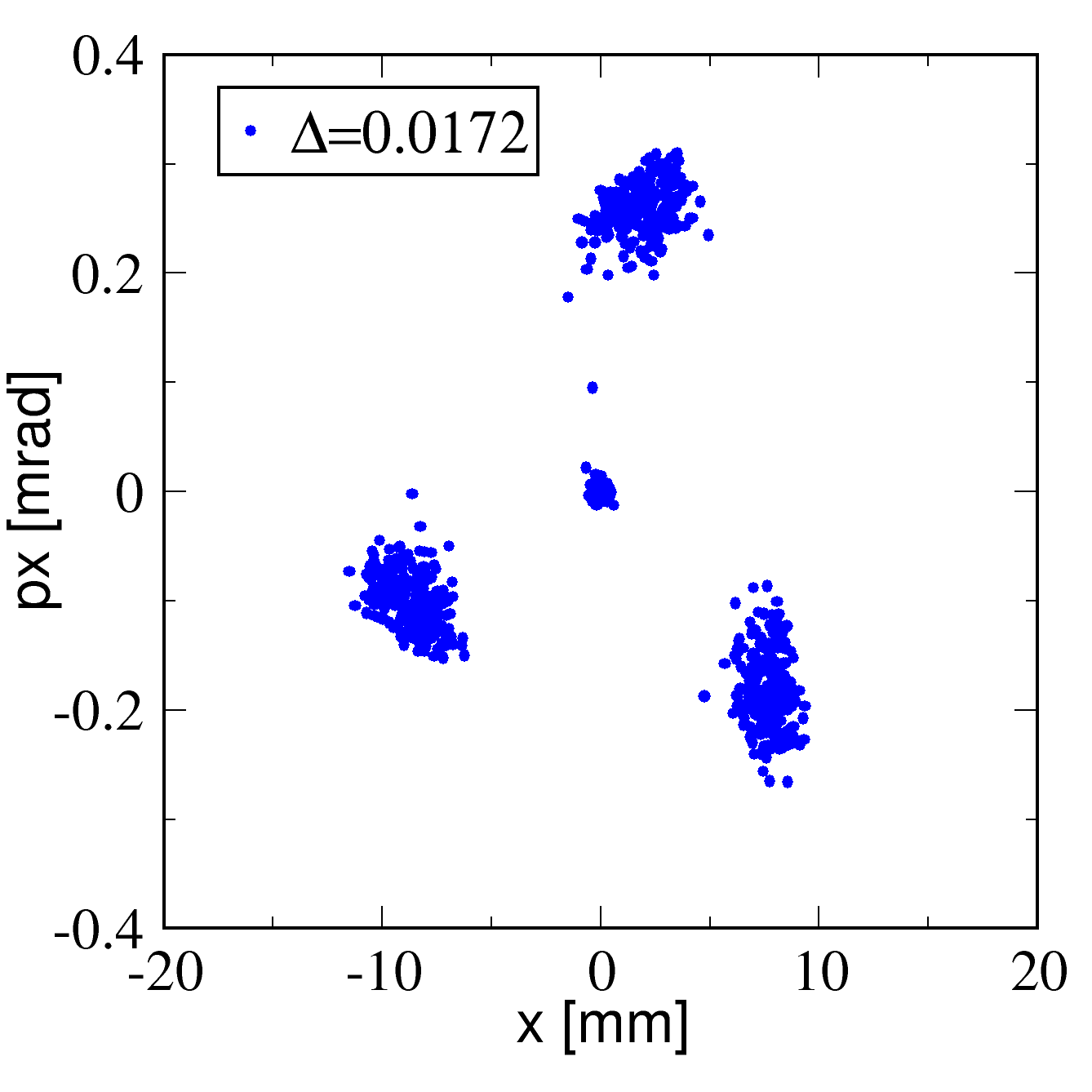} \\ \ \\
  \includegraphics[trim=0truemm 0truemm 0truemm 0truemm, width=0.22\linewidth,angle=0,clip=]{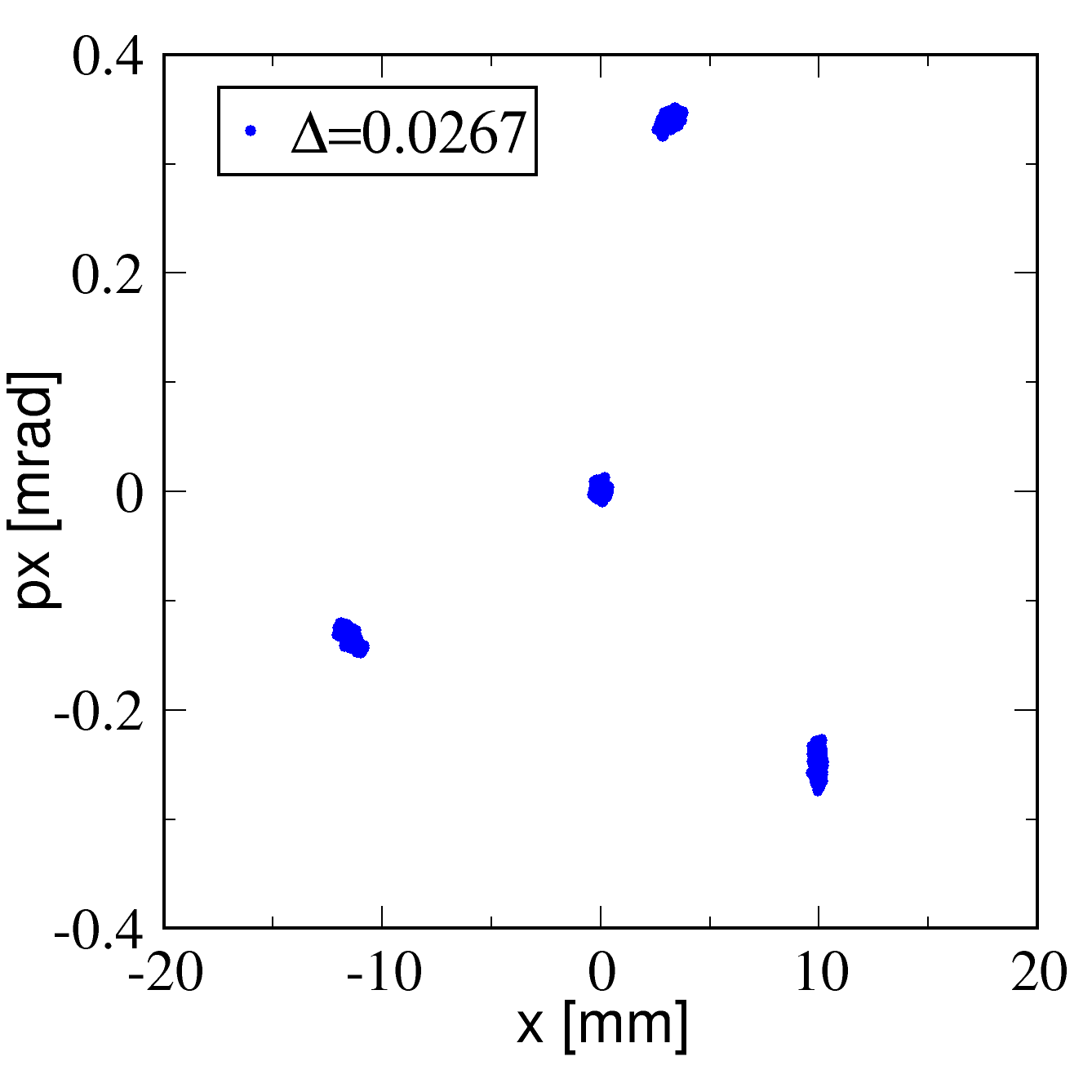}\ \ \ 
  \includegraphics[trim=0truemm 0truemm 0truemm 0truemm, width=0.35\linewidth,angle=0,clip=]{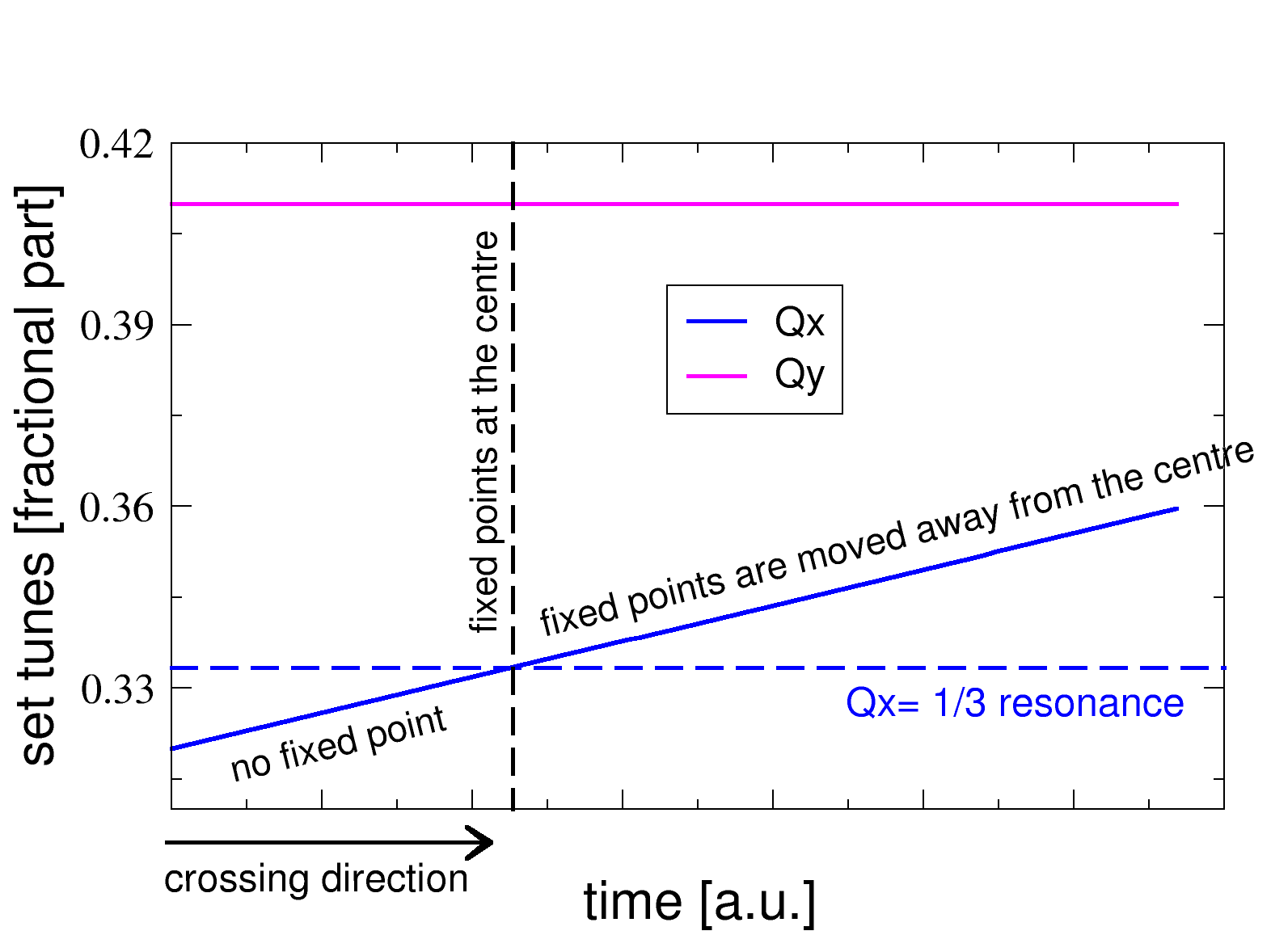}\ \ \ 
  \includegraphics[trim=0truemm 0truemm 0truemm 0truemm, width=0.35\linewidth,angle=0,clip=]{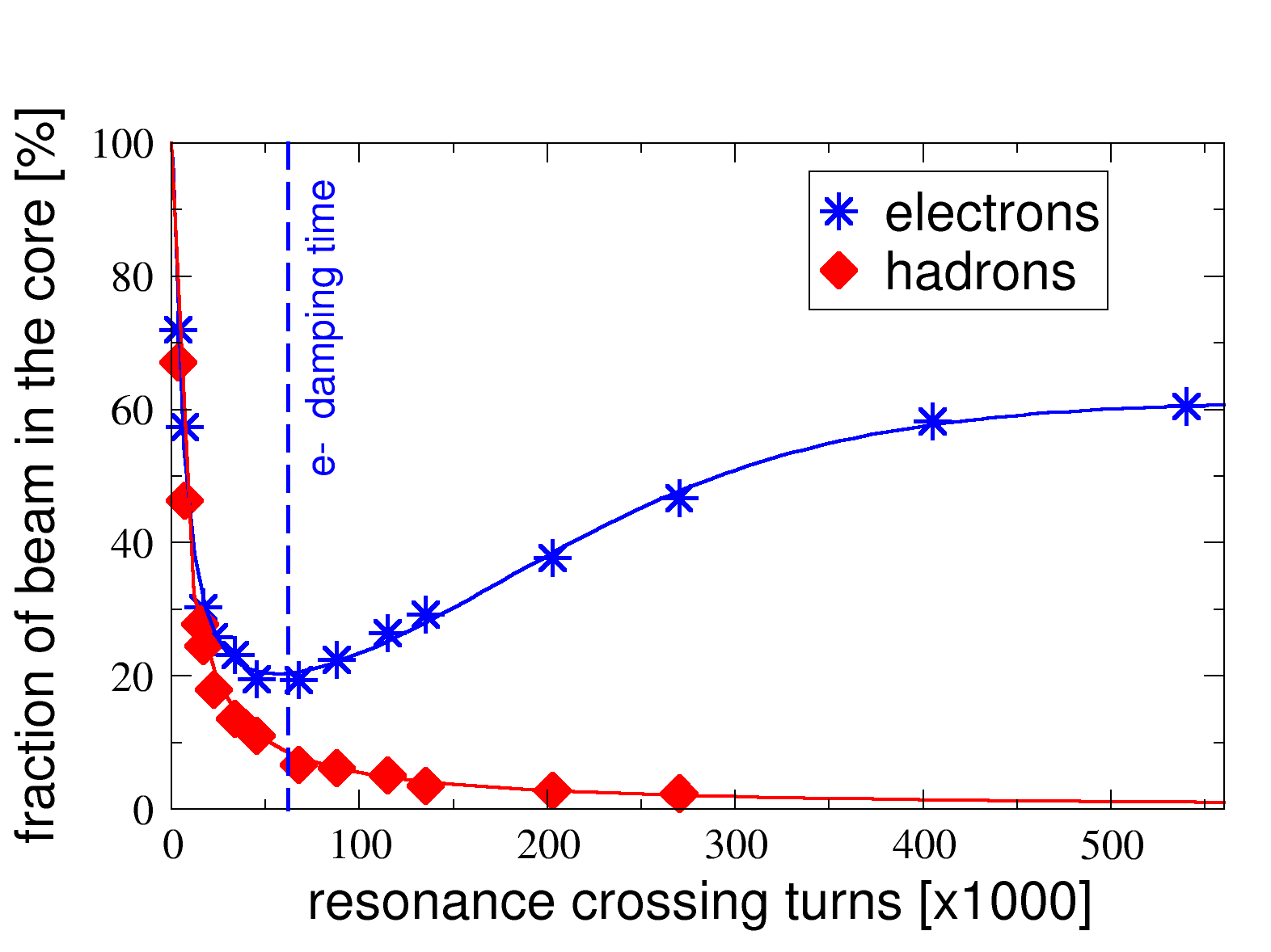}
  \caption{Simulation of a {\sl standard resonance crossing}. The particle distribution in the horizontal phase space is displayed at the beginning (all electrons inside the core, top left), at the end (beam split and separated in four beamlets, three islands and the core, bottom left) and at three intermediate moments of the crossing (top). The ramp of the horizontal tune is displayed in the bottom-centre plot. The bottom right plot shows the dependence of the fraction of the beam remaining inside the core upon the resonance crossing speed (represented by the number of turns set to generate the previous tune ramp). The red curve refers to protons with no radiative effects whereas the blue curve results from the same simulations with electrons including radiation damping and quantum diffusion.}
  \label{Fig:ResCross}
  \end{center}
\end{figure}
%

\subsubsection*{Scheme \#1: standard resonance crossing} 
A first approach to populate the islands consists of moving electrons from the standard central orbit by trapping them around the moving fixed points, as it is done for protons in the CERN PS~\cite{Borburgh:2137954,PhysRevAccelBeams.20.014001,PhysRevAccelBeams.22.104002,PhysRevLett.88.104801,PhysRevSTAB.7.024001}. The distance of the fixed points from the central orbit scales approximately with $\sim\Delta=Q_x-\bar{Q}_x$, see Eq.\eqref{eq:fixedpoint1}, and the fixed points are within the initial beam distribution if $\Delta\simeq0$. By varying $Q_x$ with dedicated quadrupoles and increasing $\Delta$ the fixed points are moved apart. If this is performed slowly, i.e.\ adiabatically compared to the betatron motion, electrons may be trapped and transported by following these fixed points while they drift at larger amplitudes. An example of this {\sl standard resonance crossing} scheme is shown in Fig.~\ref{Fig:ResCross}. The evolution of the horizontal phase space distribution at different crossing times (denoted by the time-varying $\Delta$) is shown along with the ramp of the horizontal tune. The final distribution comprises four split beamlets: three equally populated islands and a fraction of electrons that remains around the nominal closed orbit. It is well-known that hadron beams, not experiencing radiative effects, tend to evacuate the region around the nominal closed orbit when the adiabaticity of the resonance crossing is increased~\cite{PhysRevSTAB.7.024001} (i.e.\ when the crossing speed is reduced). This is shown by the red curve in the bottom-right plot of Fig.~\ref{Fig:ResCross} which displays the fraction of the beam remaining around the nominal closed orbit against the duration of the resonance crossing. The blue curve in the same plot refers to the same scan but performed with electrons, hence with radiative effects included. When the resonance is crossed over a small number of turns, the radiative effects are too slow to influence the dynamics and the capture is not so efficient, with most of the particles remaining inside the core. When the crossing time is comparable to the damping time, a minimum number of particles are left inside the core (at about $20$\% of the total). For slower resonance crossing, the fraction of the beam remaining inside the core increases to reach a plateau at about $60$\% of the total, a behaviour radically different from the hadron case. We speculate that this is due to the dominant effect of radiation damping and diffusion over the trapping inside the islands, with radiative effects making the electrons lose memory of their proximity to fixed points which are still too close to the central orbit. 
\begin{figure}
  \begin{center}
  \includegraphics[trim=0truemm 0truemm 0truemm 0truemm, width=0.23\linewidth,angle=0,clip=]{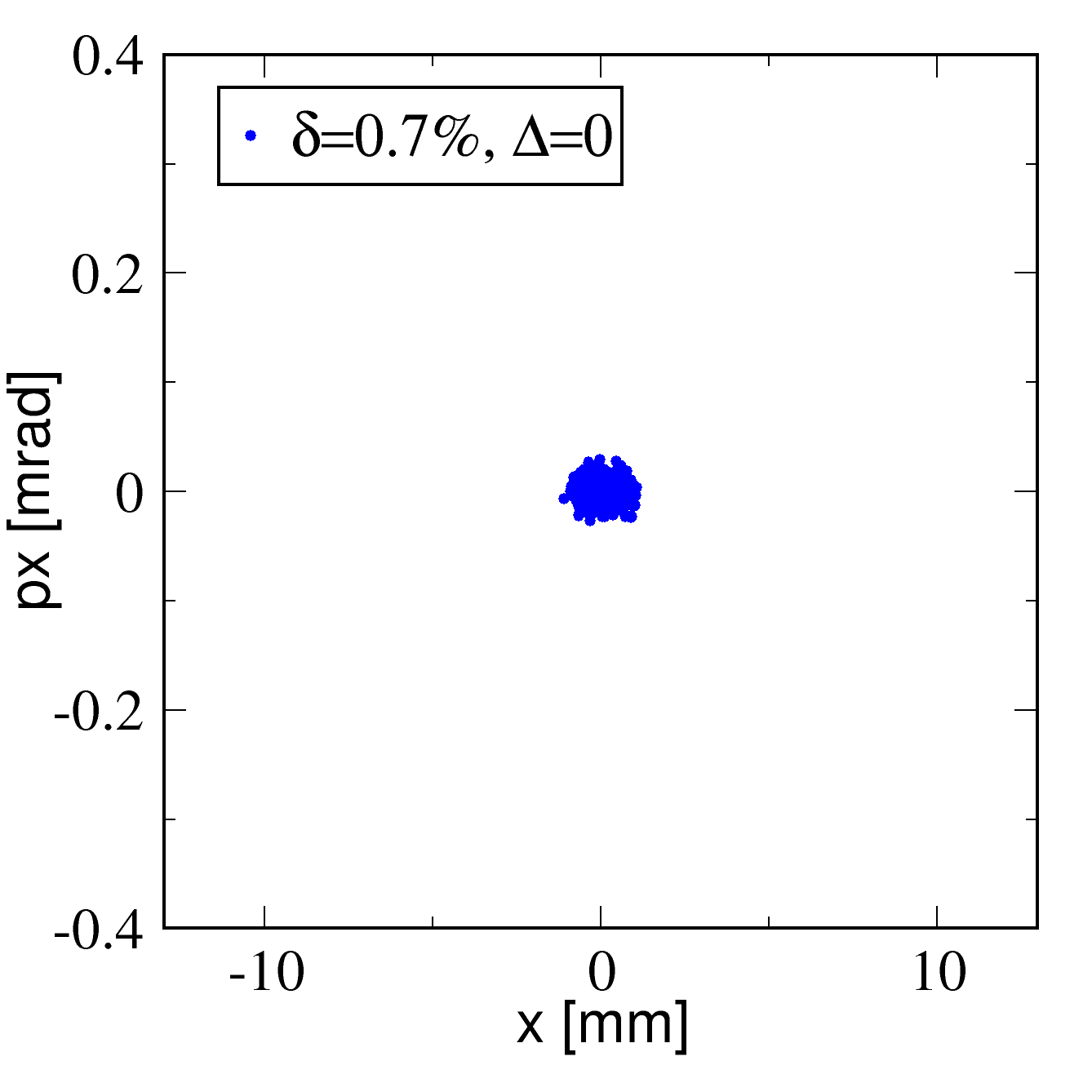}
  \includegraphics[trim=0truemm 0truemm 0truemm 0truemm, width=0.23\linewidth,angle=0,clip=]{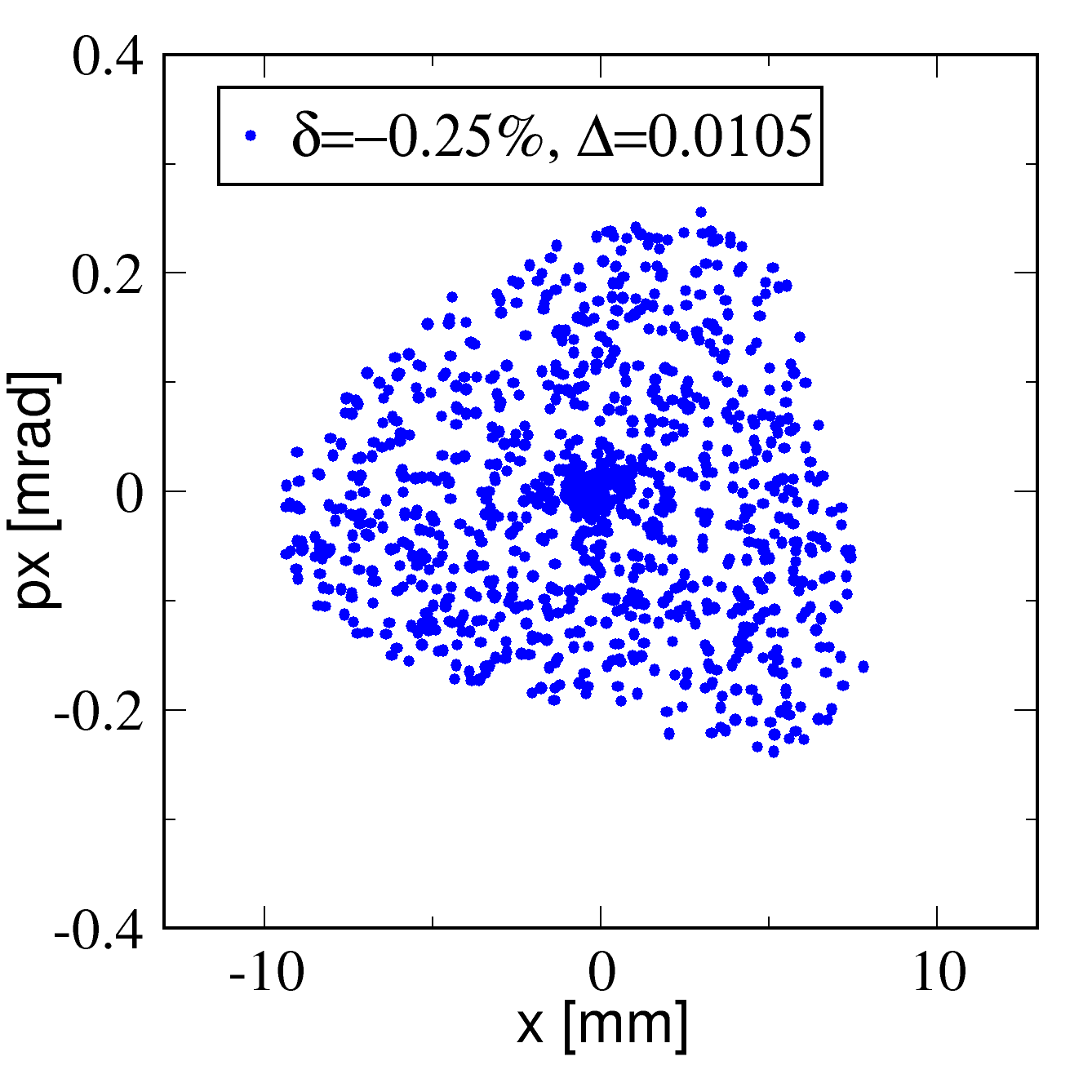}
  \includegraphics[trim=0truemm 0truemm 0truemm 0truemm, width=0.23\linewidth,angle=0,clip=]{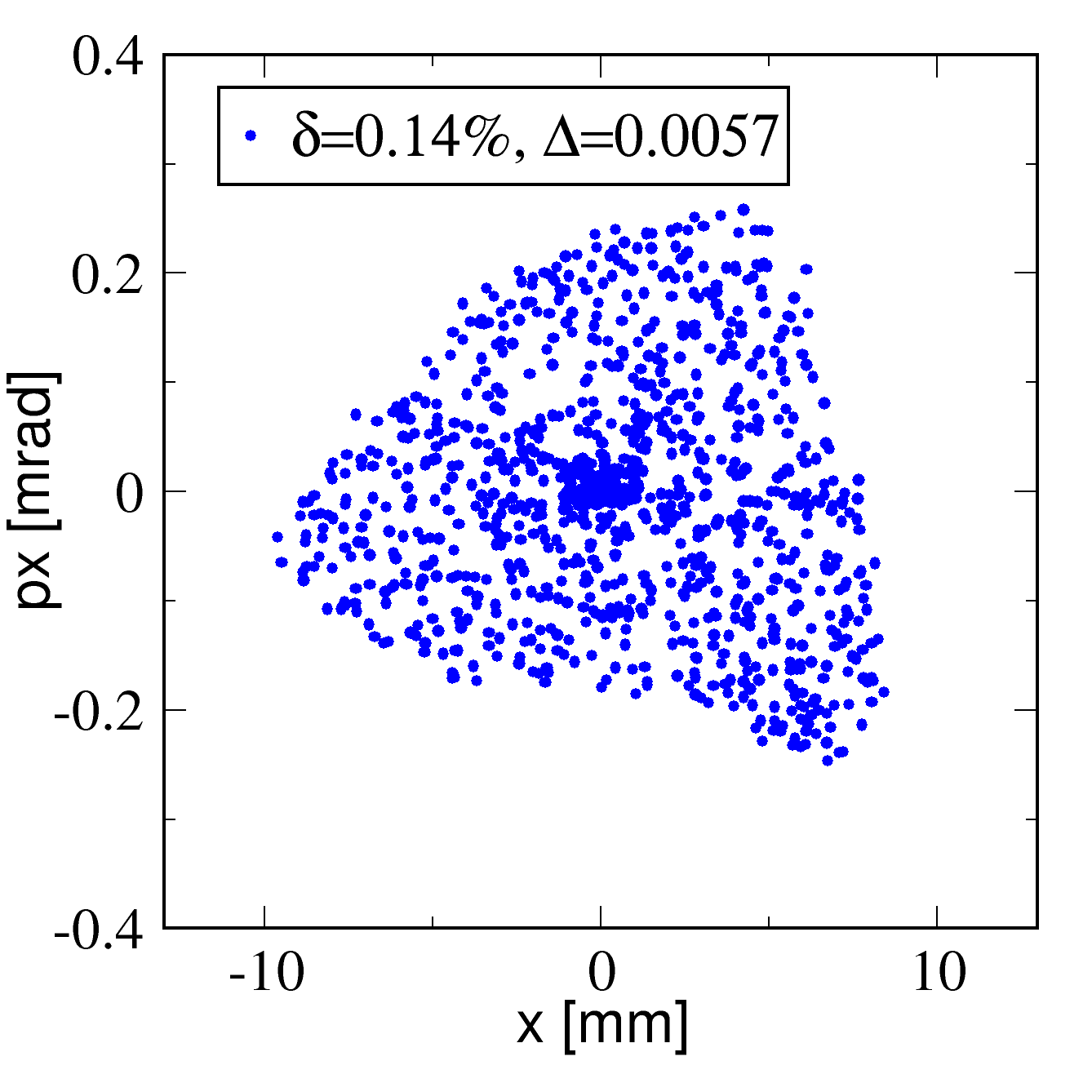}
  \includegraphics[trim=0truemm 0truemm 0truemm 0truemm, width=0.23\linewidth,angle=0,clip=]{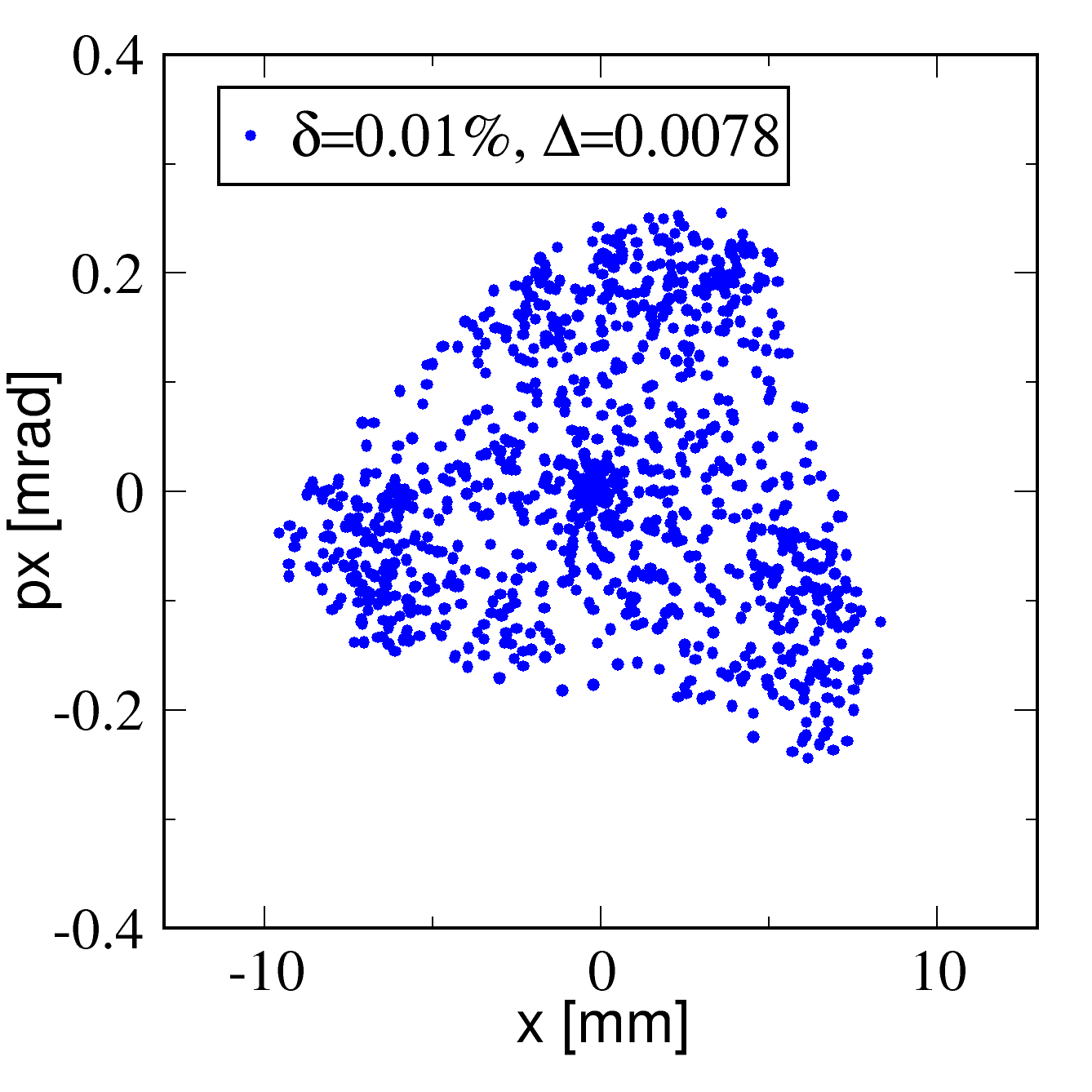} \\ \ \\
  \includegraphics[trim=0truemm 0truemm 0truemm 0truemm, width=0.24\linewidth,angle=0,clip=]{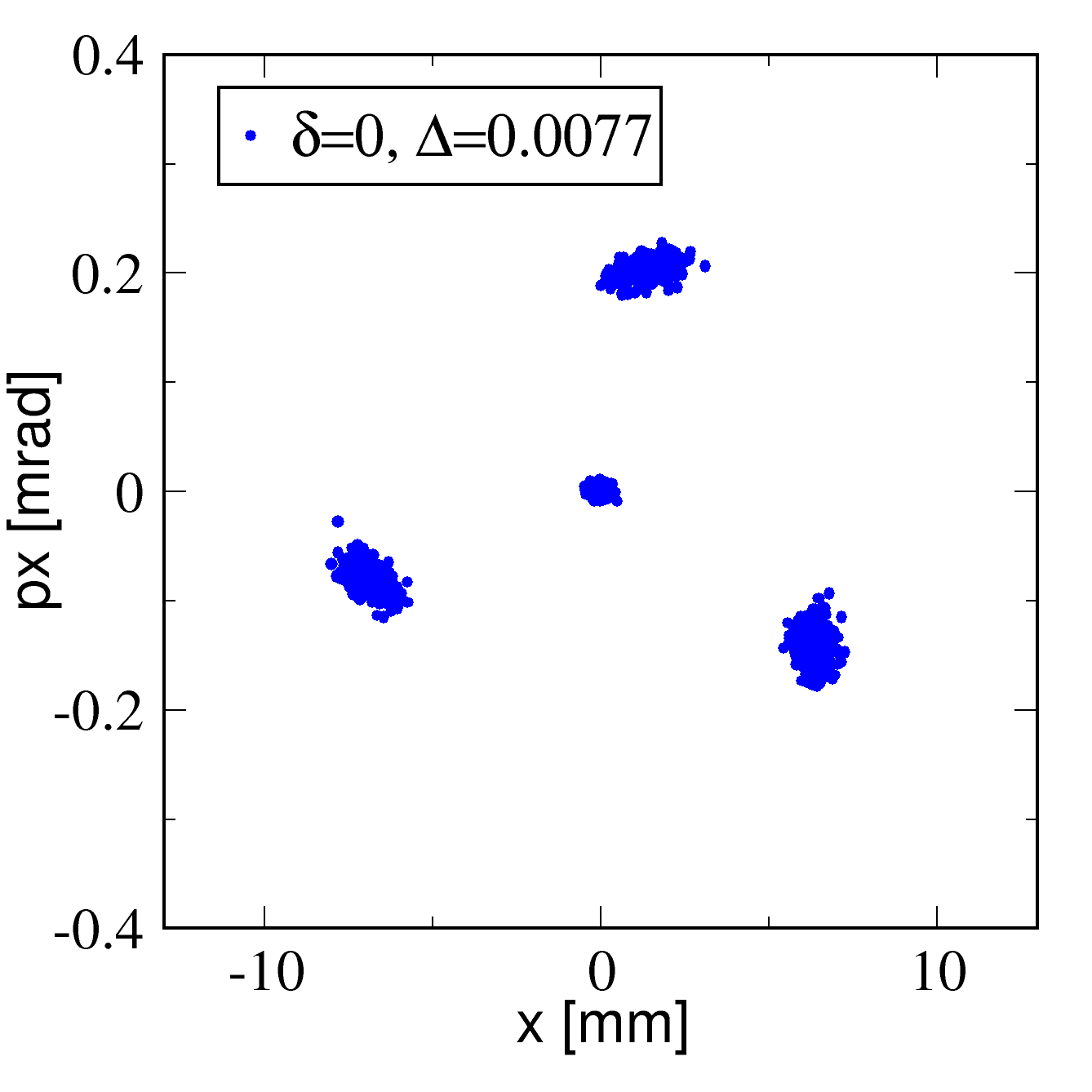}
  \includegraphics[trim=0truemm 0truemm 0truemm 0truemm, width=0.24\linewidth,angle=0,clip=]{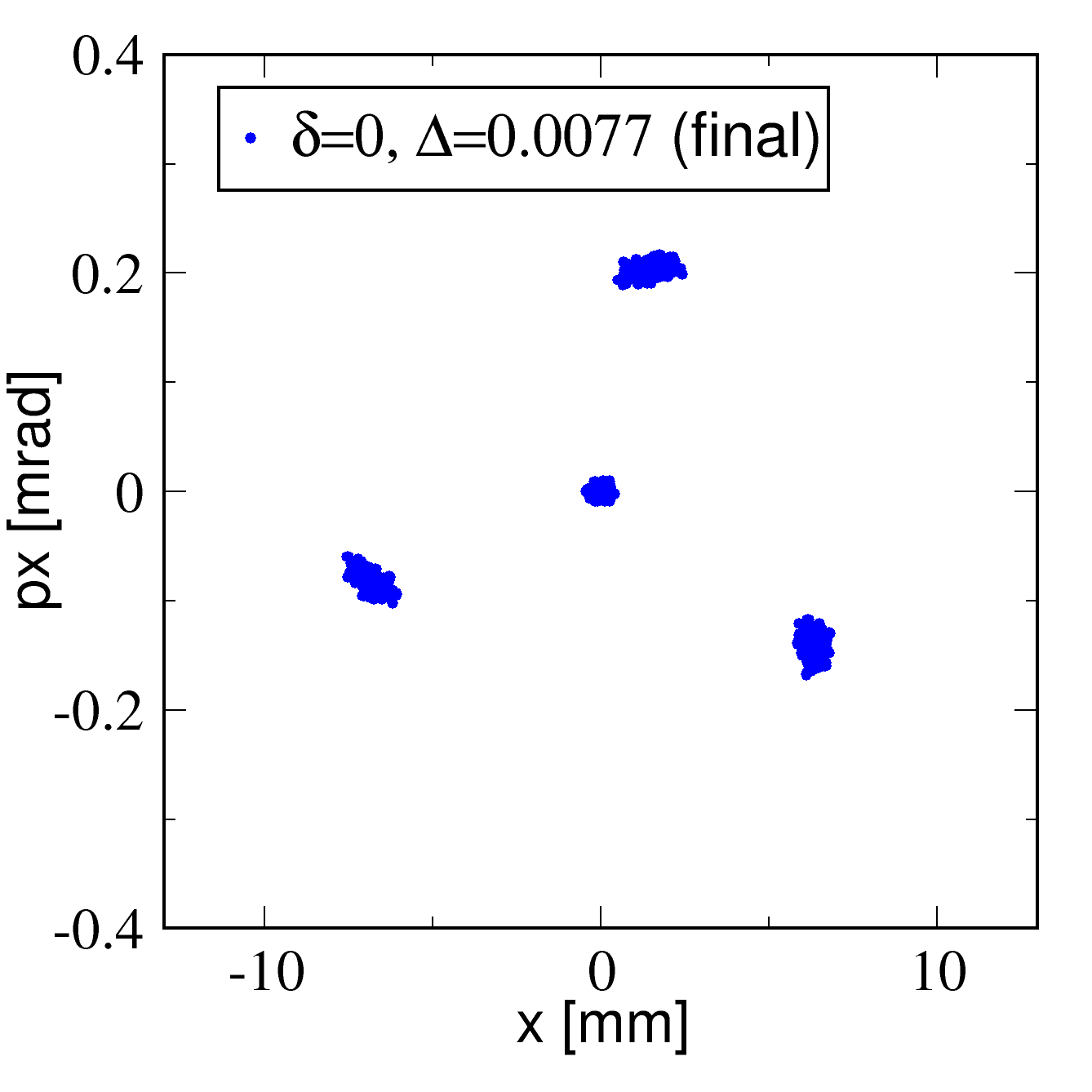}
  \includegraphics[trim=0truemm 0truemm 0truemm 0truemm, width=0.40\linewidth,angle=0,clip=]{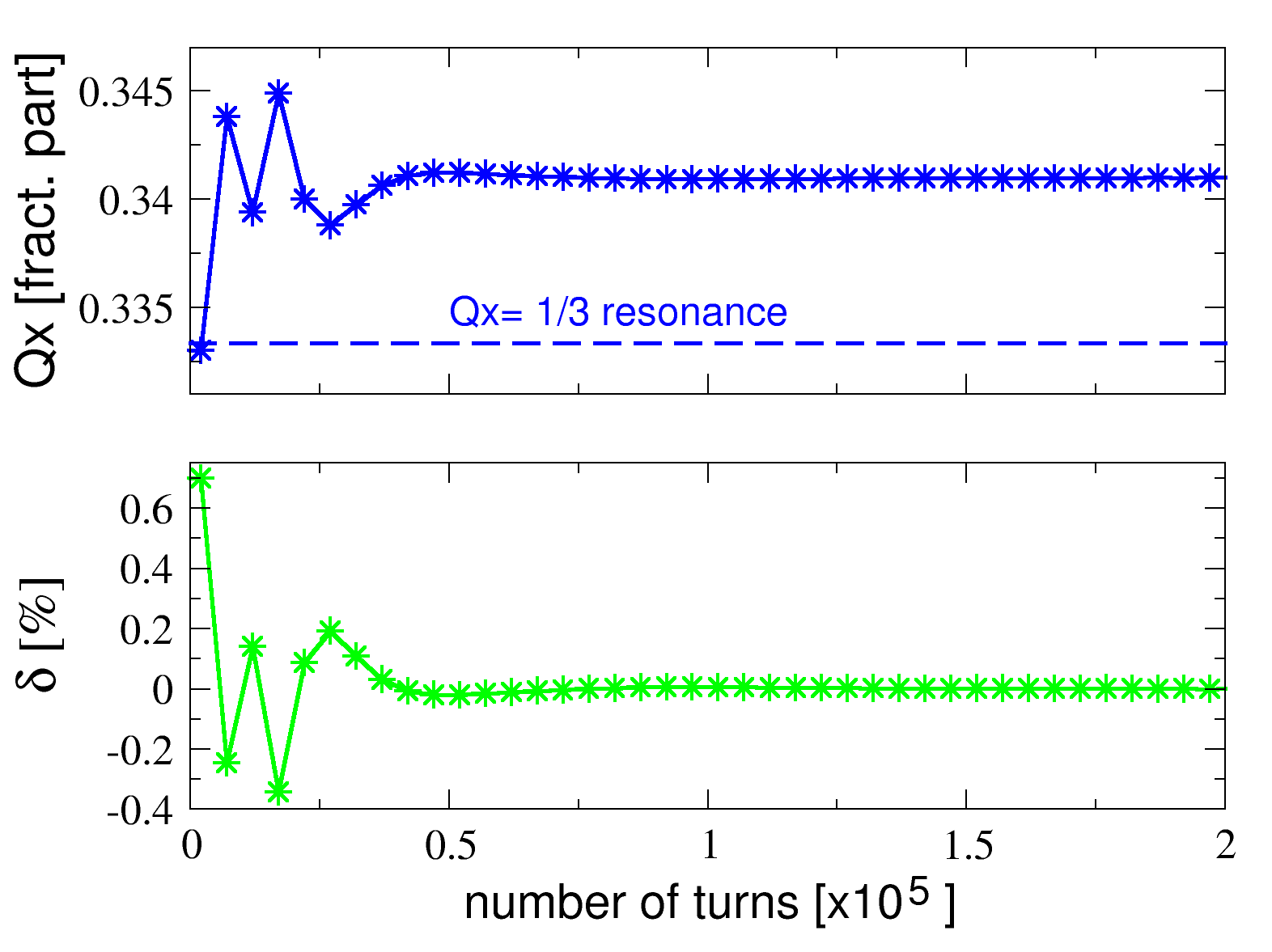}
\caption{Simulation of a {\sl chromatic resonance crossing}. The particle distribution in the horizontal phase space is displayed at the beginning (all electrons inside the core, top left), at the end (beam split and separated in four beamlets, three islands and the core, bottom centre) and at four intermediate moments of the crossing. The bottom-right plots show the evolution of the relative beam energy offset $\delta$ (green curve): radiation damping reduces the initial offset of $0.7$\% to zero after some damped synchrotron oscillations. Because of non-zero chromaticity, the variation of $\delta$ results in a change of the horizontal tune with a single crossing of the 1/3 resonance (blue curve).}
  \label{Fig:ChromCross}
  \end{center}
\end{figure}
%
 
\subsubsection*{Scheme \#2: chromatic resonance crossing} 
A second approach to splitting the central beam into separate beamlets makes use of the non-zero horizontal chromaticity, and thus of the dependence of the horizontal tune on the relative energy offset $\delta=\Delta p/p$, namely
\begin{equation}\label{eq:chrom1}
Q_x(\delta)=Q_x+Q_x'\delta +O(\delta^2)\quad \Rightarrow \quad \Delta(\delta)=Q_x(\delta)-\bar{Q}_x \, .
\end{equation} 

The idea behind this {\sl chromatic resonance crossing} is to inject a beam off-energy and let it reach the nominal equilibrium energy after some longitudinal damping times. By setting properly both the linear chromaticity $Q_x'$ and the initial energy offset $\delta_0$, during the synchrotron oscillations the resonance $\bar{Q}_x$ shall be crossed only once until the synchrotron coherent oscillations are damped by synchrotron radiation. For the sake of simplicity, we report results from simulations of on-axis, off-energy injections, but this should give indications also about the more general case of interest for light sources of off-axis injection for a typical top-up refill (see~\cite{Nakamura:EPAC90,Chen:EPAC96-MOP115G,Spring-8-06,PhysRevSTAB.10.123501,Ohkuma:EPAC08-MOZCG01,PEAKE2008143,PhysRevSTAB.13.020705,PhysRevSTAB.15.050705} and references therein). An example of on-axis chromatic crossing is shown in Fig.~\ref{Fig:ChromCross}. $10^3$ electrons are placed on the nominal closed orbit with an initial energy offset of $0.7$\%. Synchrotron motion and damping make the energy offset $\delta$ evolve as reported in the bottom-right plot of that figure (green curve). The resulting horizontal tune (blue curve in the same plot) crosses the resonance within the first few thousand turns before stabilising at the nominal value of $Q_x=0.342$. The evolution of the corresponding horizontal phase-space distribution is shown in the five portraits of Fig.~\ref{Fig:ChromCross}. No electron is lost during this process. As for the standard resonance crossing, the initial beam is split into four beamlets, three in the islands (with a share of about $24$\% each) and one around the nominal closed orbit containing the remaining $28$\% of the electrons. The sharing between islands and the core depends mainly on the synchrotron frequency, since it dominates the effective resonance crossing speed. This is a rather rigid parameter that is often difficult to vary in most circular accelerators. Horizontal chromaticity, however, can be rather easily varied, allowing the adjustment of the resonance crossing speed for a given synchrotron frequency (i.e.\ the slope of the first part of the tune variation, the blue curve of Fig.~\ref{Fig:ChromCross}). The lower the horizontal chromaticity, the slower the effective resonance crossing. The initial energy offset also plays a minor role. Once again, radiation damping is a necessary element for the formation of stable islands. Without the damping of the energy oscillations, particles would periodically cross the resonance, resulting in periodic trapping and detrapping. This is confirmed by numerical simulations of the equivalent hadronic beam distribution that show how the beamlets do not survive the periodic resonance crossing, which eventually generates a complete beam loss after several synchrotron periods.
 
It is worth stressing that a major operational difference between {\sl standard} and {\sl chromatic} resonance crossings is that the former affects all bunches, as the change of the quadrupole strength affects all buckets, whereas the latter can be performed on a bunch-by-bunch basis by injecting individual bunches off-energy into the storage ring. 


\subsubsection*{Scheme \#3: direct injection into the resonance island} 

A third {\sl straightforward} way to fill the stable islands would be to inject on-energy electrons directly into the islands. This can be done by varying the beam trajectory in the transfer line and the strength of the injection septum so to ensure that the horizontal position and angle of the incoming beam matches those of one of the fixed points at the exit of the injection septum. As for the chromatic crossing, this manipulation can be performed in a bunch-by-bunch fashion with the additional flexibility of being capable of filling only one or some of the available islands, and of leaving the nominal closed orbit empty. In the case of the original ESRF storage ring, the standard horizontal position and angle of the injected beam are $x=-8$~mm and $p_x=0$~mrad, respectively. The final phase-space portrait of Fig.~\ref{Fig:ChromCross} shows a fixed point at $x_{\text{FP}}=-7$~mm and $p_{x, \text{FP}}=-0.05$~mrad that could be used for direct injection by suitably adjusting the settings of the injection elements. This option, however, suffers from a major drawback. In general, the horizontal emittance and the bunch length of the incoming beam are usually much larger than the corresponding equilibrium values of that of the receiving storage ring. Hence, the chance that electrons survive around the fixed point are not high, unless the optical conditions inside the stable islands fulfil some specific constraints. Of course, light sources with shorter incoming bunches and with horizontal emittance compatible with the surface of the islands may be able to consider this option. 
  
\subsubsection*{Scheme \#4: using an AC dipole} 

A fourth, more exotic, option to populate stable islands foresees a weak excitation of the beam with a horizontal AC dipole at the same frequency as, or close to, the resonant tune $\bar{Q}_x=1/N$. There are three main differences between this approach and the excitation of a resonant closed orbit discussed in Section~\ref{sec:scenario2}. Firstly, the physics behind this approach is based on a double-resonance condition, with one of the stable resonance islands generated by a suitable nonlinear amplitude-dependent detuning and the other being the resonant closed orbit. The scheme discussed in Section~\ref{sec:scenario2} is based on a single-resonance effect. While the mathematical description of a 2D system under the influence of a single resonance is rather well established, to the best of our knowledge, the same cannot be said of a 2D system with a double-resonance condition. This means that suitable settings (tune, strength of the nonlinear magnets, strength of the AC dipole, and frequency) are determined by a trial-and-error approach. The second main difference is that as a consequence of this double-resonance nature, the peak excitation amplitude is by far weaker (of the order of a few $\mu$rad for the simulations shown) compared to that needed to create a periodic oscillating closed orbit ($\sim$ hundreds of $\mu$rad). The third main difference is that if the final equilibrium emittance is sufficiently small compared to the surface of the stable island and the electrons do not spill out from them, then $N$ beamlets are created that would survive even after switching off the AC dipole excitation. 

\begin{figure}[!t]
  \begin{center}
  \includegraphics[trim=6truemm 178truemm 15truemm 18truemm, width=0.45\linewidth,angle=0,clip=]{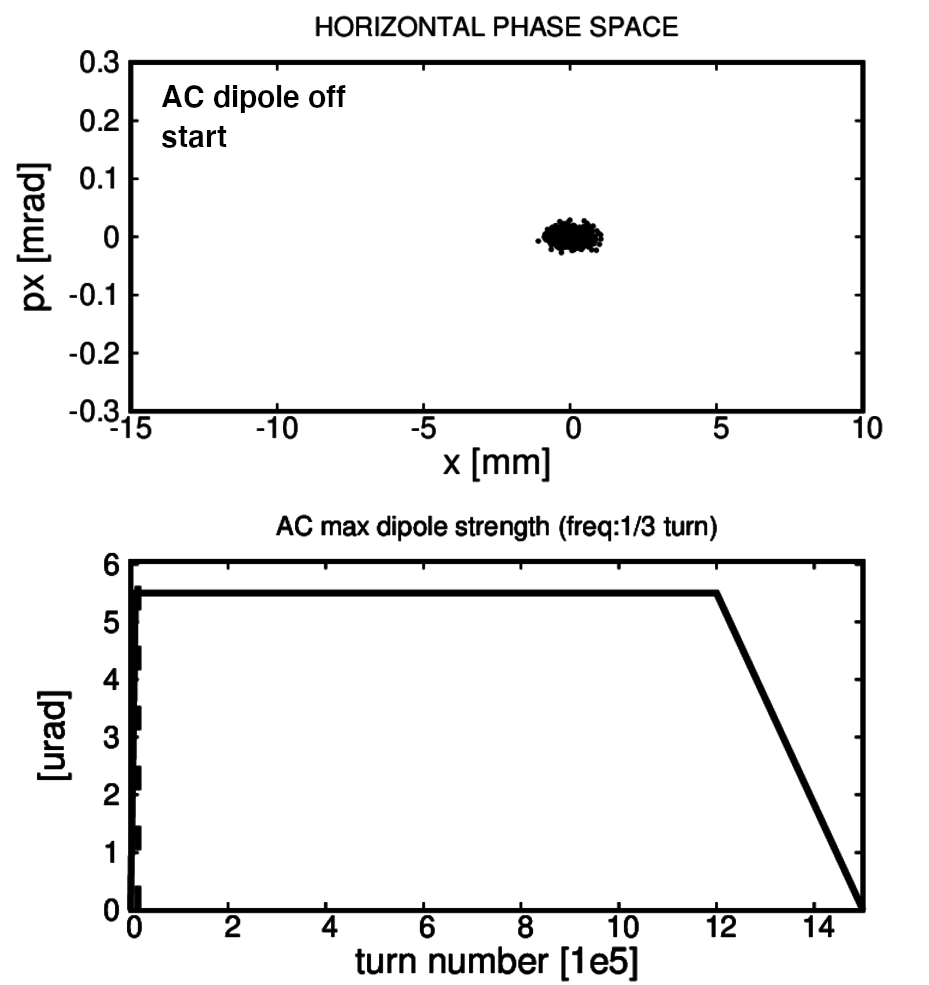}
  \includegraphics[trim=6truemm 178truemm 15truemm 18truemm, width=0.45\linewidth,angle=0,clip=]{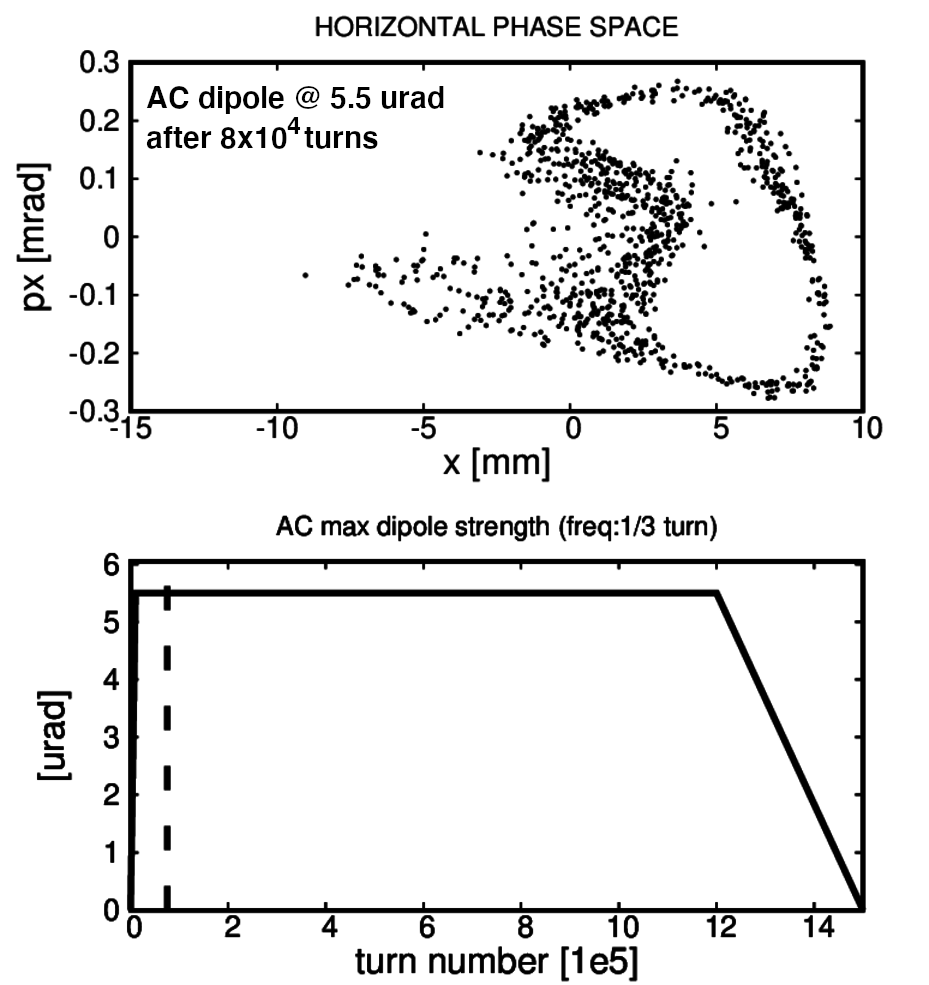}\\ \ \\
  \includegraphics[trim=6truemm 178truemm 15truemm 18truemm, width=0.45\linewidth,angle=0,clip=]{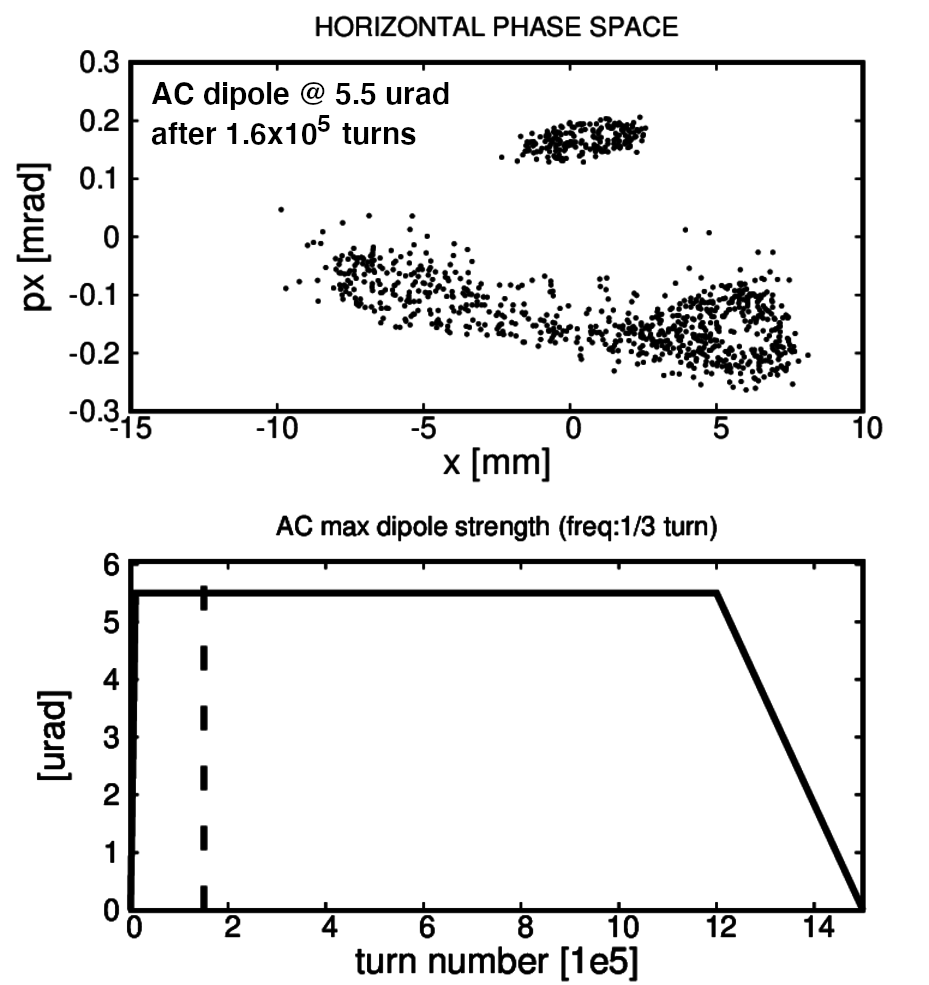}
  \includegraphics[trim=6truemm 178truemm 15truemm 18truemm, width=0.45\linewidth,angle=0,clip=]{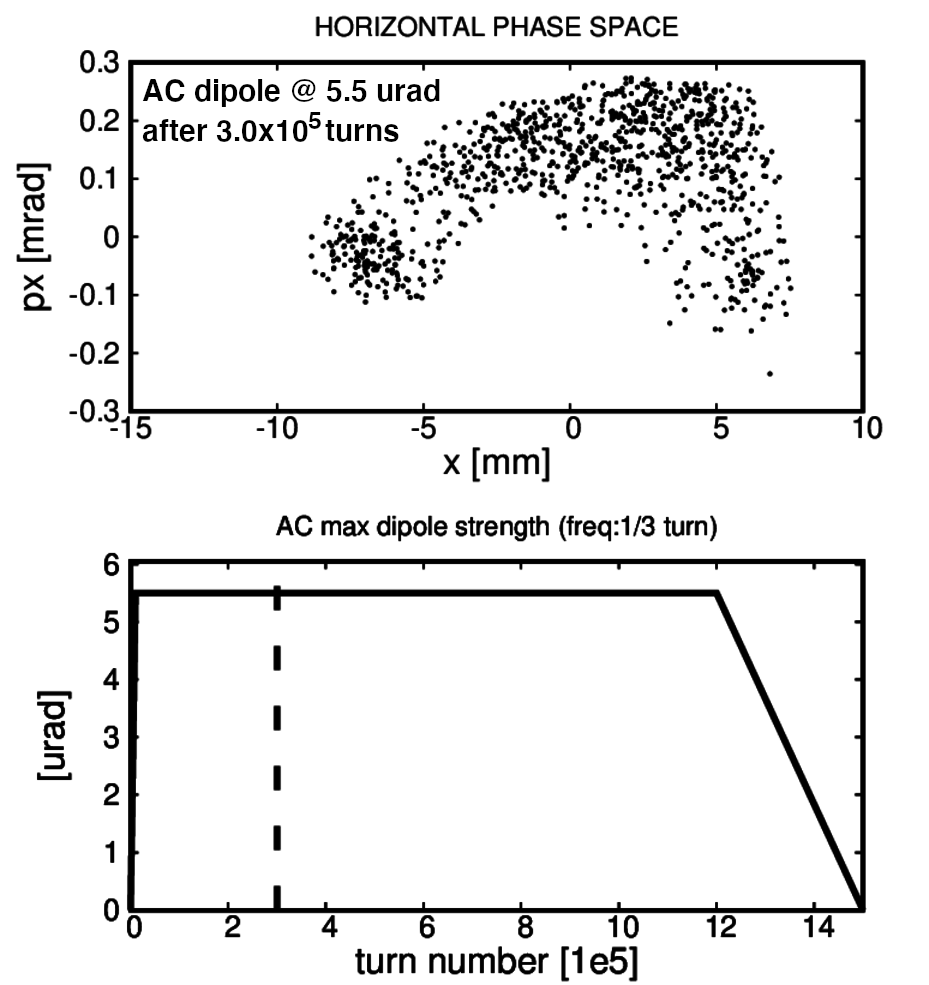}\\ \ \\
  \includegraphics[trim=6truemm 178truemm 15truemm 18truemm, width=0.45\linewidth,angle=0,clip=]{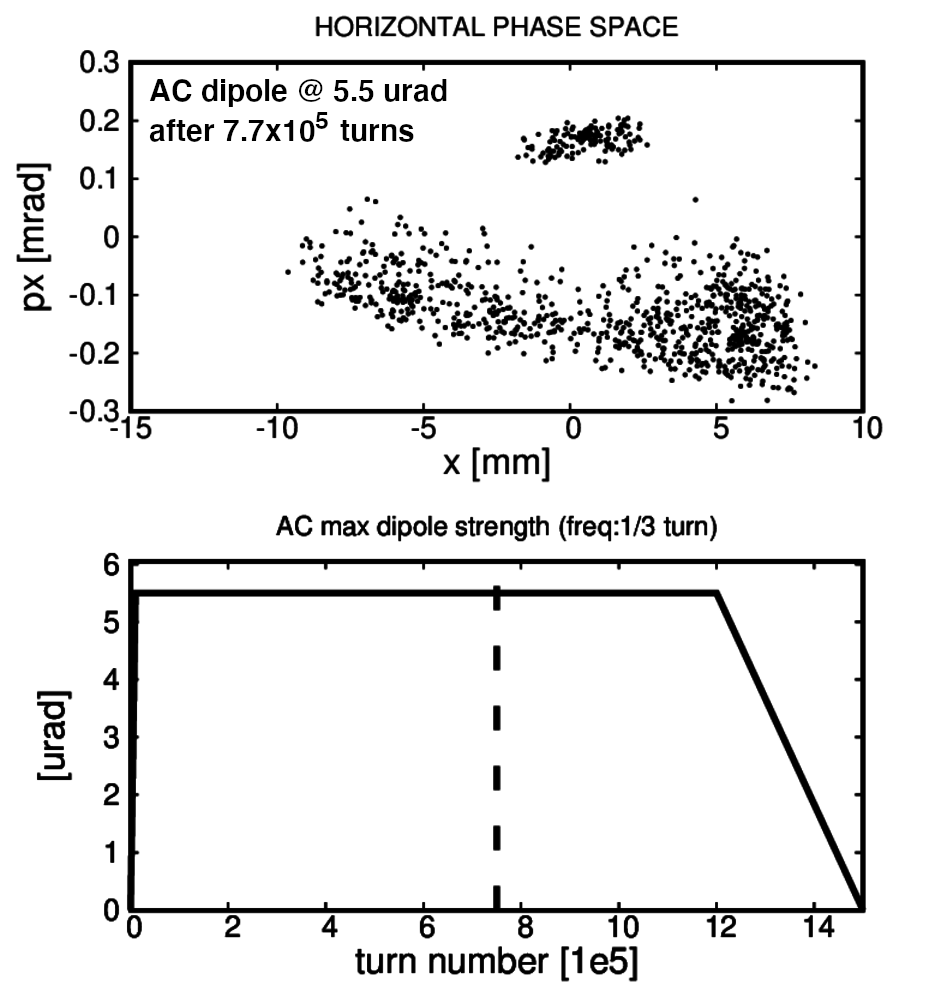}
  \includegraphics[trim=6truemm 178truemm 15truemm 18truemm, width=0.45\linewidth,angle=0,clip=]{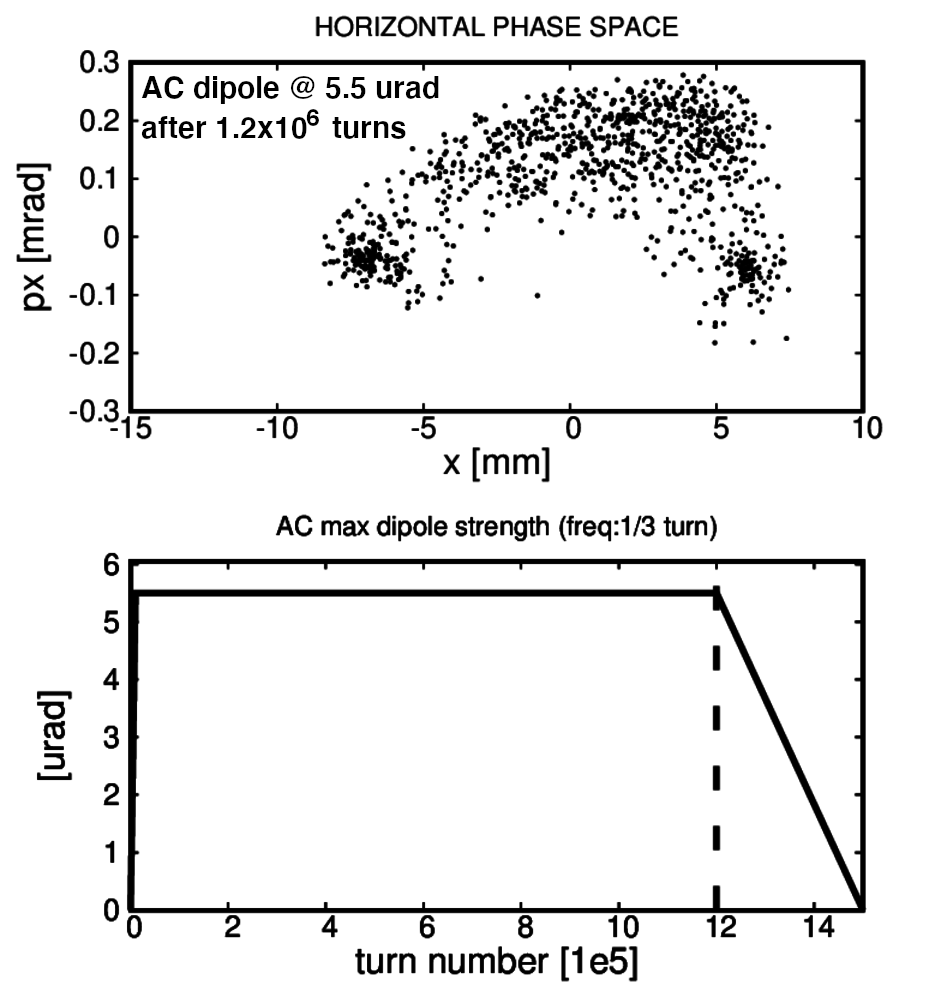}\\ \ \\
  \includegraphics[trim=6truemm 178truemm 15truemm 18truemm, width=0.45\linewidth,angle=0,clip=]{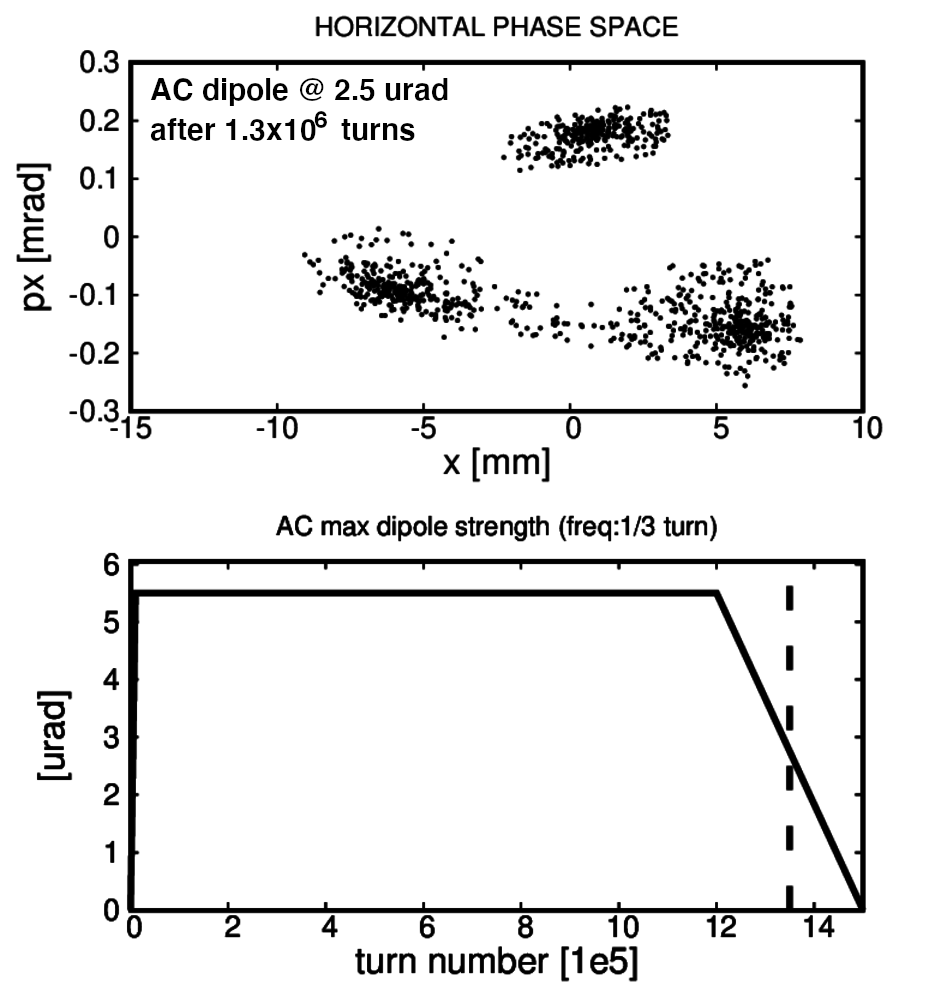}
  \includegraphics[trim=6truemm 178truemm 15truemm 18truemm, width=0.45\linewidth,angle=0,clip=]{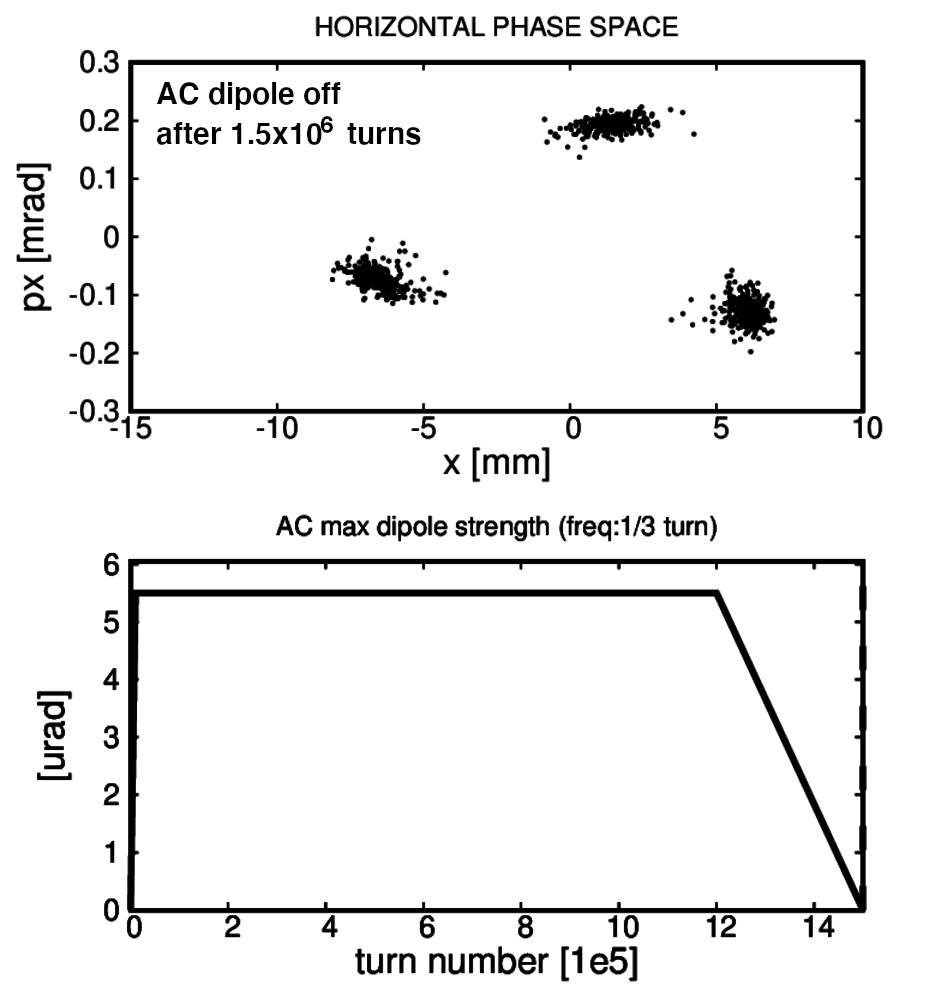}
  \vspace{-3mm}
  \caption{Simulation of electron beam splitting by means of an AC dipole excitation at a frequency of $f_\mathrm{AC}=1/3 + 5\times10^{-5}$ in tune unit and maximum strength of $5.5\ \mu$rad. In each plot, the horizontal phase-space distribution is shown at eight moments of the AC dipole excitation cycle. The first and last plots correspond to the beginning and the end of the process (after $1.5\times10^6$ turns). When the AC dipole is eventually turned off, three beamlets are formed of about the same intensity, with no beam around the nominal closed orbit (bottom right plot).
  \label{Fig:ACDIP-sim1}}
  \end{center}
\end{figure}

\begin{figure}[!ht]
  \begin{center}
  \includegraphics[trim=5truemm 4truemm 10truemm 17truemm, width=0.325\linewidth, height=0.18\linewidth, angle=0,clip=]{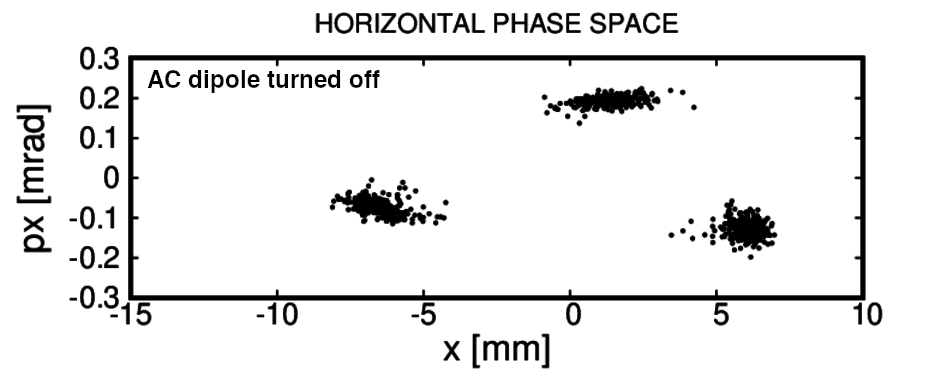}
  \includegraphics[trim=5truemm 4truemm 10truemm 17truemm, width=0.325\linewidth, height=0.18\linewidth, angle=0,clip=]{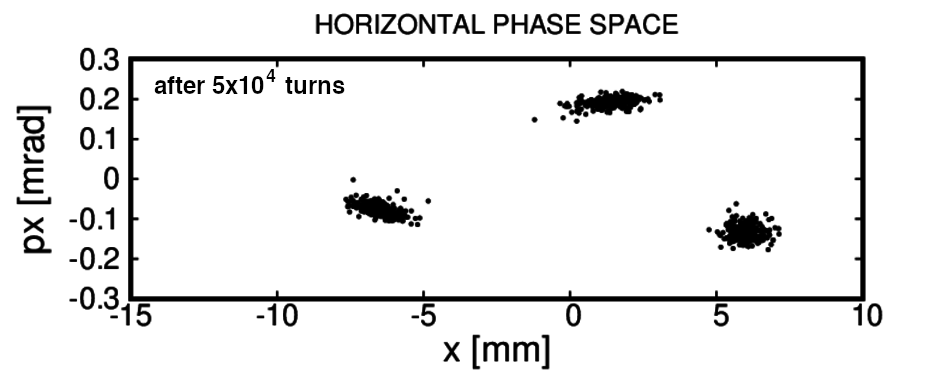}
  \includegraphics[trim=5truemm 4truemm 10truemm 17truemm, width=0.325\linewidth, height=0.18\linewidth, angle=0,clip=]{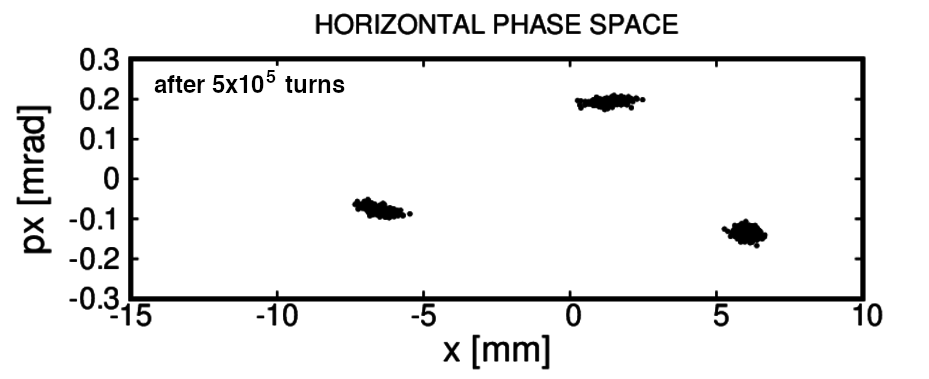}
  \vspace{-3mm}
  \caption{Simulated horizontal phase-space distribution after turning off the AC dipole. The final distribution of particles of Fig.~\ref{Fig:ACDIP-sim1} is eventually tracked without excitation to assess the final equilibrium. The three beamlets survive and a lower equilibrium emittance is eventually obtained.   \vspace{3mm}
  \label{Fig:ACDIP-sim1B}}
  \end{center}
\end{figure}

Fig.~\ref{Fig:ACDIP-sim1} presents the results of multi-particle simulations generating three stable islands by means of an AC dipole excitation. The horizontal phase space and the peak strength of the AC dipole are plotted at eight different moments during the excitation cycle and after reducing the AC dipole strength to zero. The beam is initially on the nominal closed orbit and the lattice configuration corresponds to Setting~A of Fig.~\ref{Fig:Ex-SET6-9}, thus featuring three stable islands not yet populated. The peak AC dipole strength is set to quickly increase from zero to $5.5\ \mu$rad in $10^3$ turns. This strength alone, i.e.\ without the stable islands generated by the nonlinear magnets, would neither be enough to generate separated orbit beamlets similar to those of Fig.~\ref{ResOrb01} nor to place the electrons directly into the resonance islands, the stable fixed points being at more than $100\ \mu$rad in absolute terms, as shown in the bottom-left phase-space portrait of Fig.~\ref{Fig:Ex-SET6-9}. The excitation frequency is set close to the 1/3 resonance, i.e.\ $f_\mathrm{AC}=1/3 + 5\times10^{-5}$. When the AC dipole is eventually turned off in $3\times10^5$ turns, the nominal closed orbit is empty and all electrons are locked inside the three stable islands generated by the nonlinear magnets, with almost perfect intensity sharing (within $1$\%). To validate the survival of the three beamlets and their equilibrium state, the final particle distribution of Fig~\ref{Fig:ACDIP-sim1} was tracked for an additional $5\times10^5$ turns, corresponding to several damping times. As shown by the three phase-space distributions of Fig.~\ref{Fig:ACDIP-sim1B}, the three beamlets do indeed survive and reach equilibrium.

%
\begin{figure}[!ht]
  \begin{center}
  \includegraphics[trim=6truemm 178truemm 15truemm 18truemm, width=0.40\linewidth,angle=0,clip=]{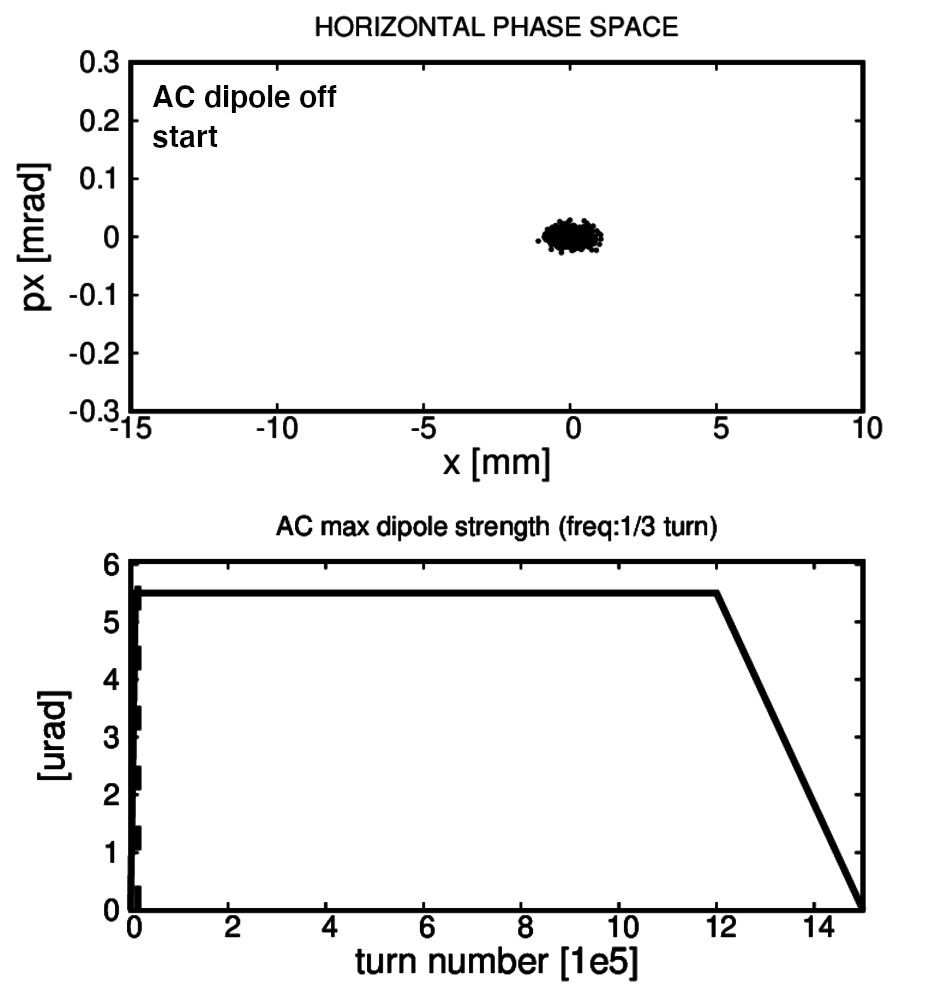}
  \includegraphics[trim=6truemm 178truemm 15truemm 18truemm, width=0.40\linewidth,angle=0,clip=]{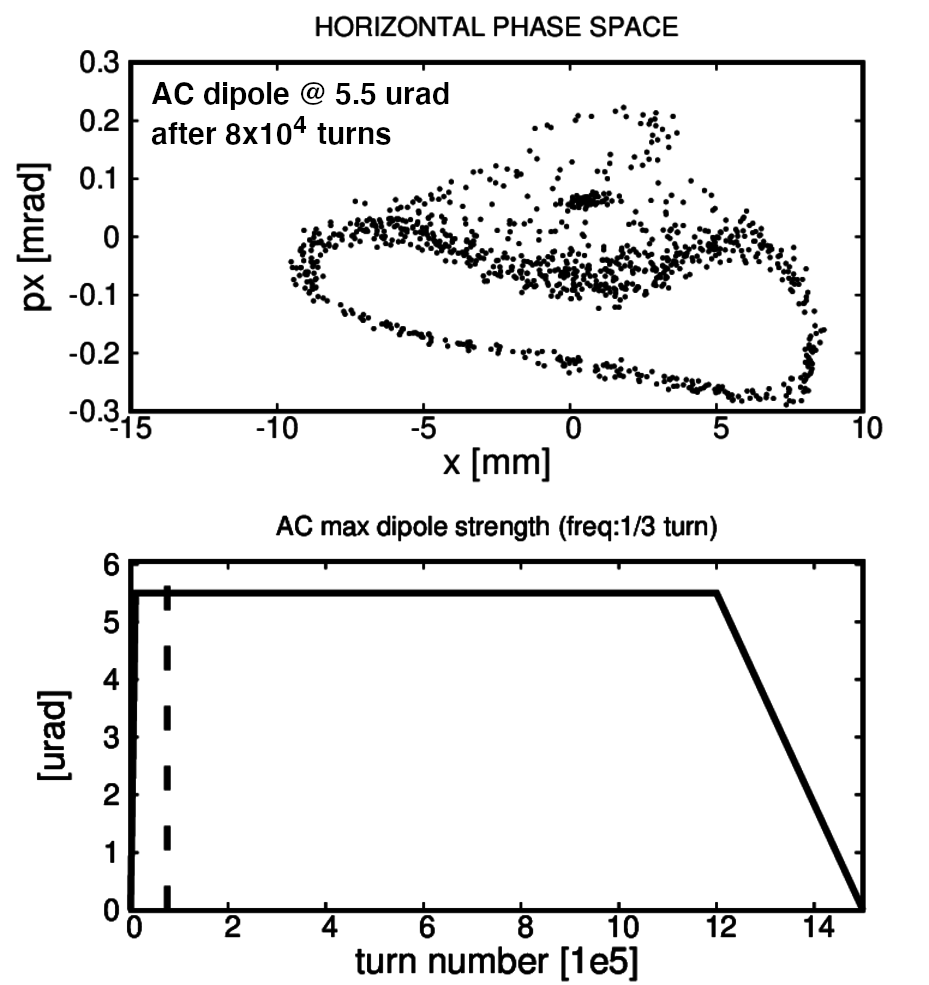}\\ \ \\
  \includegraphics[trim=6truemm 178truemm 15truemm 18truemm, width=0.40\linewidth,angle=0,clip=]{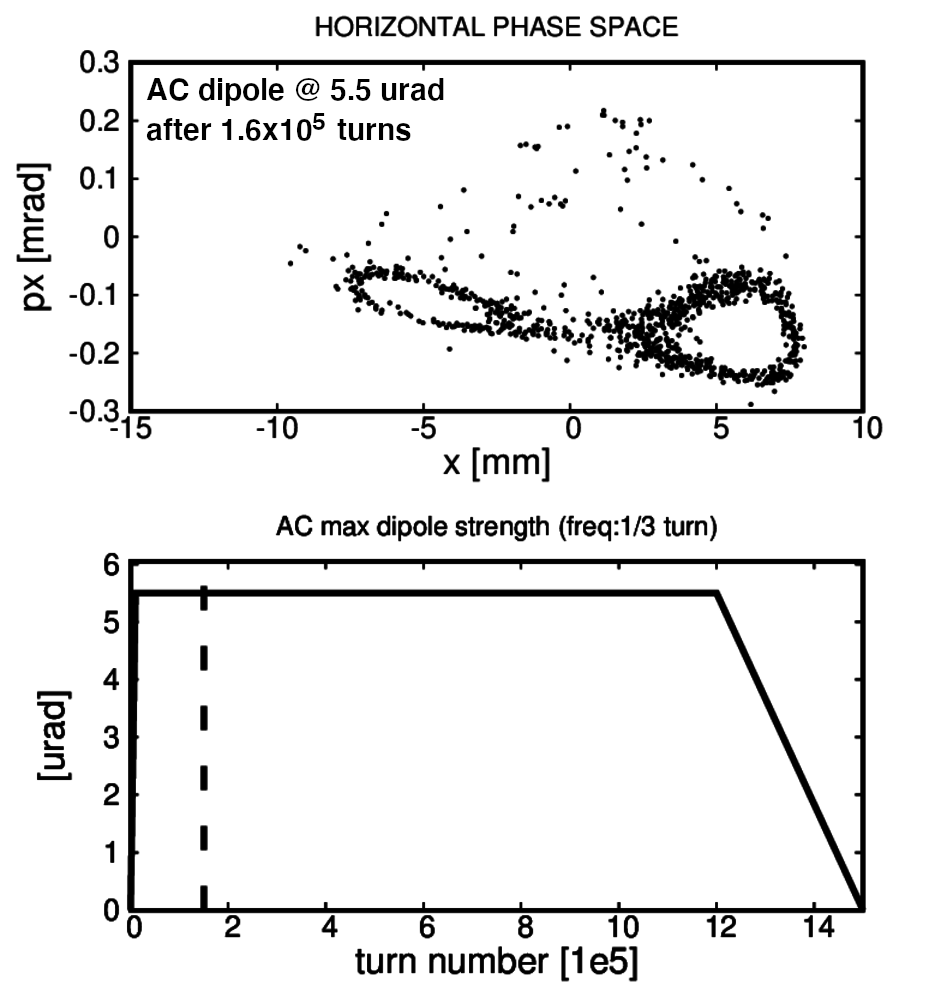}
  \includegraphics[trim=6truemm 178truemm 15truemm 18truemm, width=0.40\linewidth,angle=0,clip=]{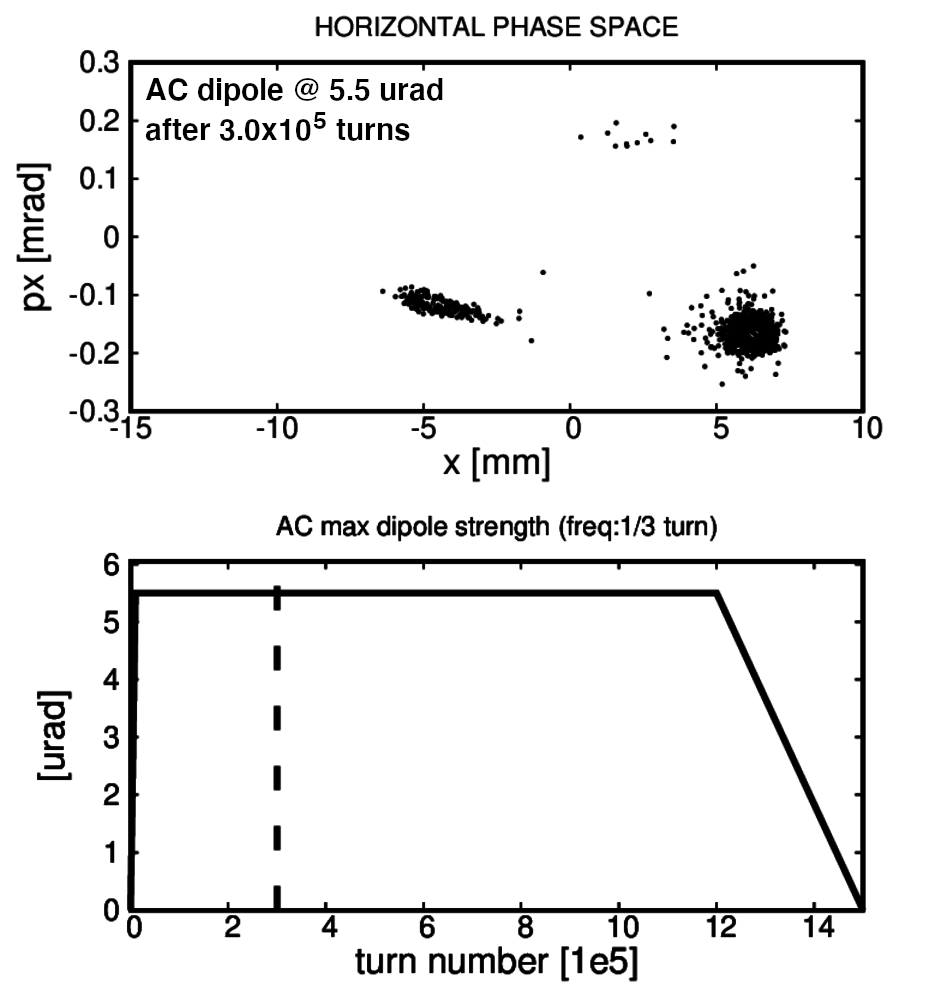}\\ \ \\
  \includegraphics[trim=6truemm 178truemm 15truemm 18truemm, width=0.40\linewidth,angle=0,clip=]{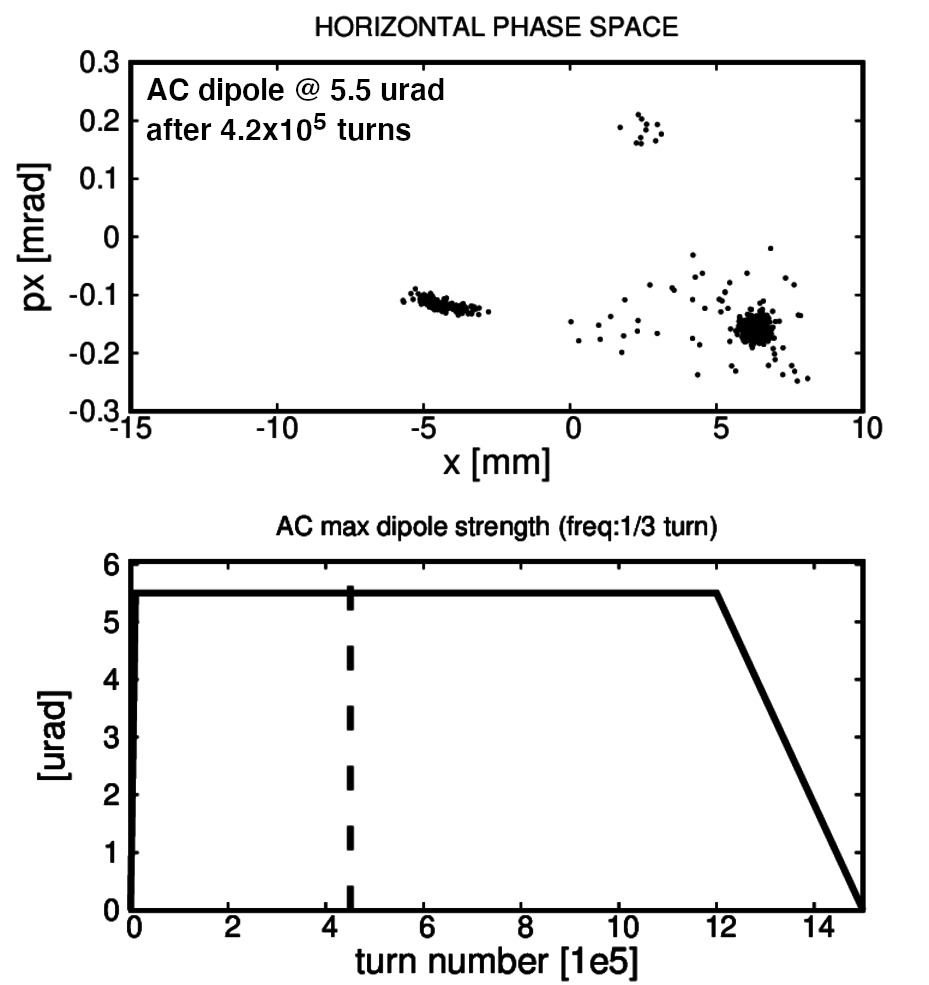}
  \includegraphics[trim=6truemm 178truemm 15truemm 18truemm, width=0.40\linewidth,angle=0,clip=]{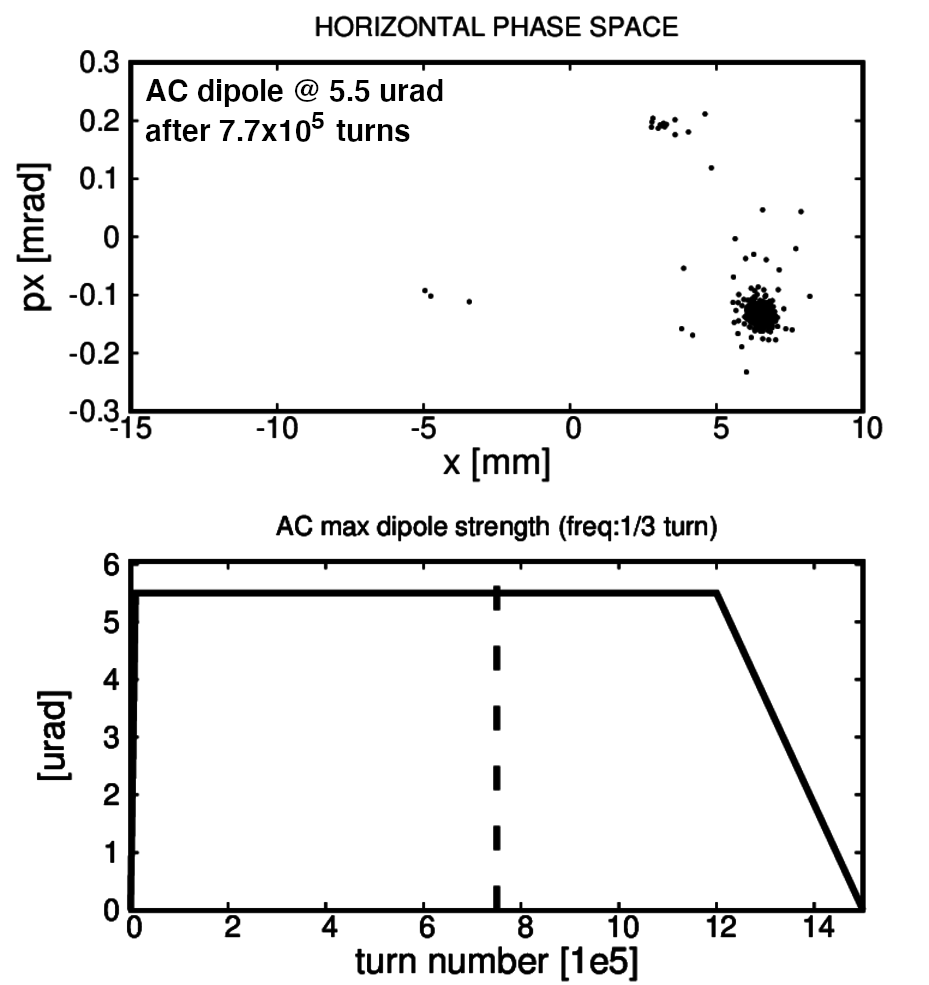}\\ \ \\
  \includegraphics[trim=6truemm 178truemm 15truemm 18truemm, width=0.40\linewidth,angle=0,clip=]{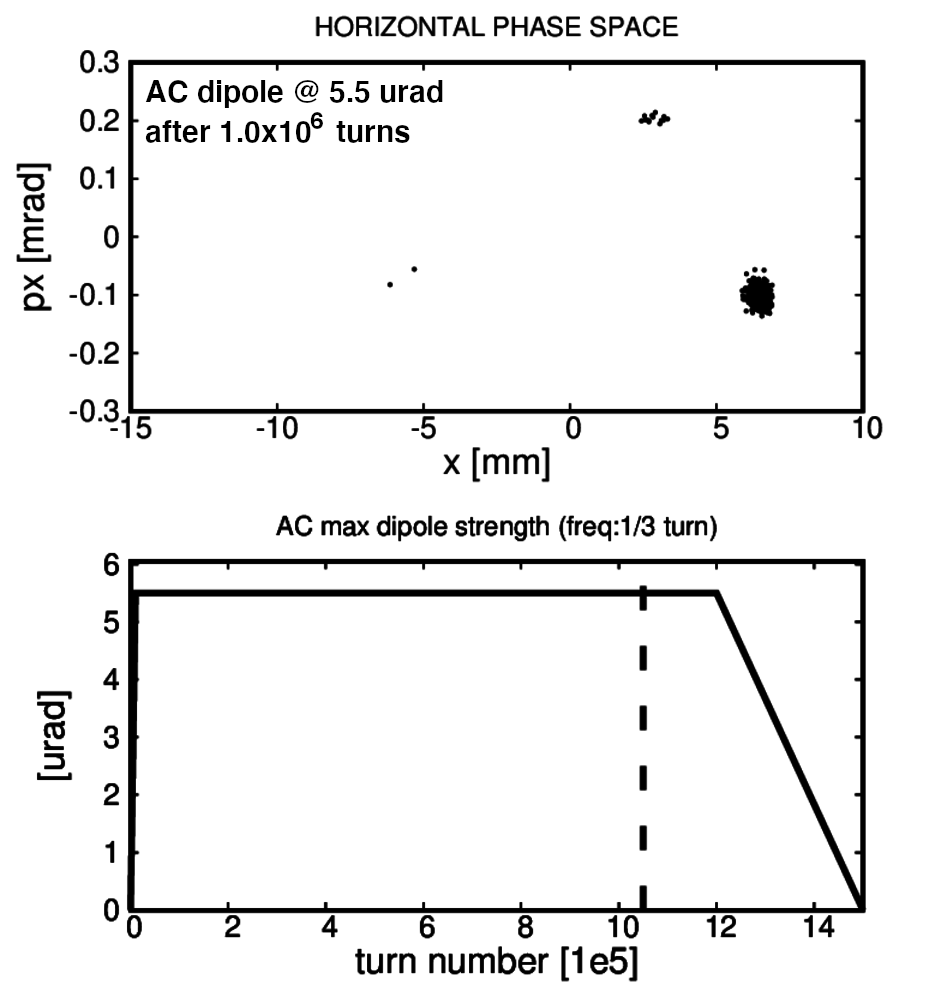}
  \includegraphics[trim=6truemm 178truemm 15truemm 18truemm, width=0.40\linewidth,angle=0,clip=]{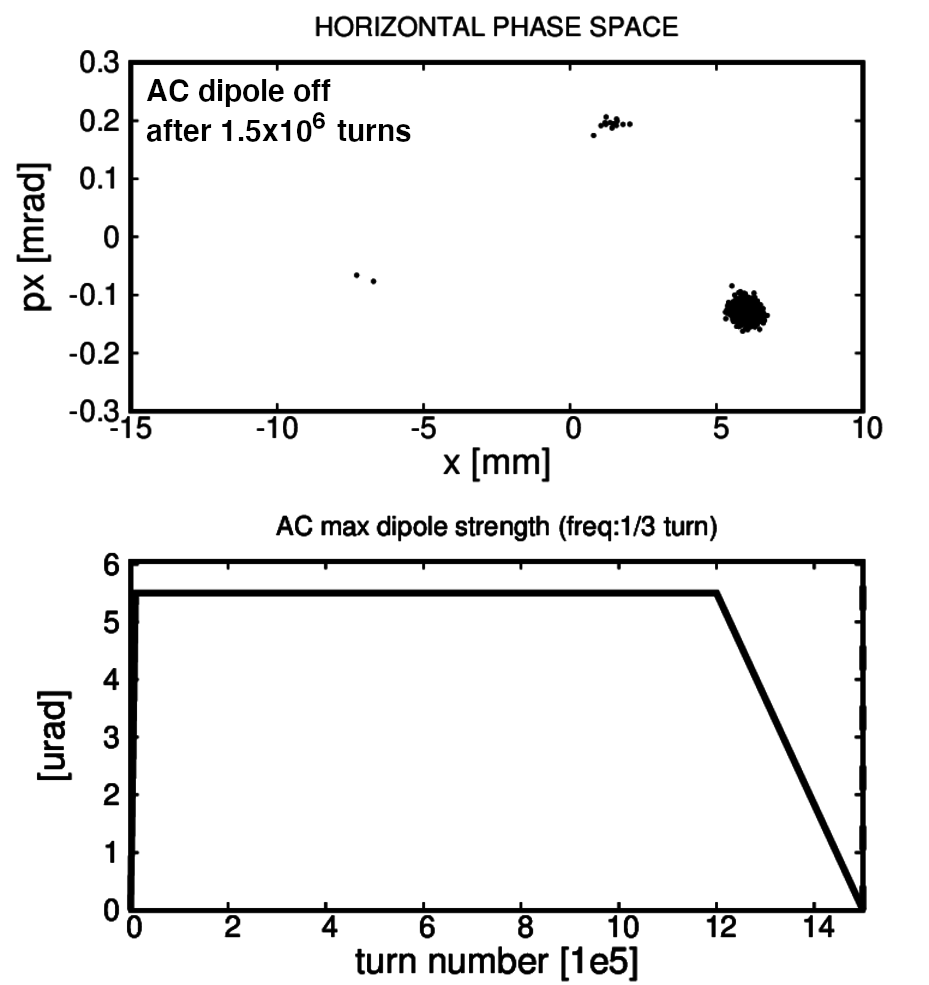}
  \caption{Same simulation configuration as that of Fig.~\ref{Fig:ACDIP-sim1}, but this time with a different AC-dipole frequency of $f_\mathrm{AC}=1/3- 3\times10^{-6}$. Instead of the almost perfect sharing between the three islands shown in Fig.~\ref{Fig:ACDIP-sim1}, more than $98$\% of the electrons are trapped in a single island.}
  \label{Fig:ACDIP-sim2}
  \end{center}
\end{figure}

It is worth pointing out that the sharing between the islands is highly dependent on the AC dipole parameters, namely its location, peak strength, frequency, and ramp profile. This feature offers the possibility of creating different patterns of island filling. The settings used in the numerical simulations of Fig.~\ref{Fig:ACDIP-sim1} resulted in an almost perfect sharing between the three beamlets, each containing $1/3$ of the initial beam intensity. Fig.~\ref{Fig:ACDIP-sim2} shows a different result with more than $98$\% of the electrons populating a single island, a bit more than $1$\% in another, and an almost missing third beamlet. This interesting result was obtained by just changing the AC dipole frequency to $f_\mathrm{AC}=1/3 - 3\times10^{-6}$. Further trimming of other parameters, such as the strength or ramp, could produce two completely empty islands with $100$\% of the electrons in a single one, thus resulting in users seeing x-rays only every 3 turns. This result would be equivalent to what obtained with the approach of Section~\ref{sec:scenario2}, although with two notable differences: the AC dipole strength would be more than one order of magnitude lower, and it would need to be active only for a short time and not continuously. 
 
\section{Comparison of the two approaches} \label{sec:comparing}
Table~\ref{tab:comparison} gathers together the main similarities and differences between the two approaches discussed in Section~\ref{sec:scenario2} and Section~\ref{sec:scenario1} for comparison.
\begin{table}[!ht]
\centering
\begin{tabularx}{\textwidth}{X X}
\hline\hline\\ 
\vspace{-0.45 cm}  {\bf \hspace{2.5cm} Resonant orbit}  &\vspace{-0.4 cm} {\bf \hspace{2.5cm}  Resonance islands}   \vspace{0.15 cm} \\
\hline\hline\vspace{-0.18 cm} \\ \vspace{0.05 cm}
  \vspace{-0.45 cm} Even if $N$ beamlets appear on a screen, the electron beam is unsplit. 
&\vspace{-0.4 cm} 
If $N$ beamlets appear on a screen, the electron beam is split in $N$ separated beamlets.  \vspace{0.15 cm}\\
\hline\vspace{-0.18 cm} \\ \vspace{0.05 cm}
  \vspace{-0.45 cm} A beamline taking synchrotron radiation from one orbit beamlet sees the photons every $N$ machine revolutions, i.e.\ $f_{\gamma}=f_\mathrm{rev}/N$.
& \vspace{-0.4 cm} 
A beamline taking synchrotron radiation from one resonance island sees the photons every machine revolution, i.e.\  $f_{\gamma}= f_\mathrm{rev}$. This statement may not be true if an AC dipole excitation is used to populate one or some of the available resonance islands, as discussed in Section~\ref{sec:scenario1}. \vspace{0.15 cm}\\
\hline\vspace{-0.18 cm} \\ \vspace{0.05 cm}
  \vspace{-0.45 cm} A beamline taking synchrotron radiation from a bunch on one orbit beamlet sees a photon intensity per turn equal to that of the nominal bunch, i.e.\ $I_{\gamma, \text{1 turn}}=I_{\text{bunch}, \text{1 turn}}$.
&\vspace{-0.4 cm} A beamline taking synchrotron radiation from one bunch split in $N$ resonance islands sees a fraction of the photon intensity per turn, i.e.\ $I_{\gamma, \text{1 turn}}=I_{\text{bunch}, \text{1 turn}}/N$. This statement may not be true if an AC dipole excitation is used to populate one or some of the available resonance islands, as discussed in Section~\ref{sec:scenario1}. \vspace{0.15 cm}\\
 \hline\vspace{-0.18 cm} \\
 \end{tabularx}
 
\newcolumntype{C}{>{\centering\arraybackslash}X}%
\begin{tabularx}{\textwidth}{C}
  \vspace{-0.45 cm} The photon intensity integrated over $N$ turns from a bunch  is equal to the nominal photon intensity per bunch over a single turn, i.e.\ $I_{\gamma, \text{N turns}}=I_{\text{bunch}, \text{1 turn}}$. \vspace{0.15 cm}\\
 \hline\vspace{-0.18 cm} \\ \vspace{0.05 cm}
  \vspace{-0.45 cm} The synchrotron radiation is emitted off axis and/or with a sizeable angle (canted-like beamline needed). \vspace{0.15 cm}\\
 \hline\vspace{-0.18 cm} \\ 
 \end{tabularx}

 \vspace{-0.05 cm}
\begin{tabularx}{\textwidth}{X X}
  \vspace{-0.40 cm} Very weak dependence on nonlinear optics.
& \vspace{-0.40 cm} Very strong dependence on nonlinear optics.
 \vspace{0.15 cm}\\
 \hline\vspace{-0.18 cm} \\ \vspace{0.05 cm}
  \vspace{-0.45 cm} Needs continuous AC dipole excitation with peak strength $\propto 100\ \mu$rad.
& \vspace{-0.4 cm} Does not need AC dipole if either resonance crossing, or chromatic crossing or direct injection into the island is used. Alternatively, a pulsed AC dipole excitation lasting only $\sim 1$ s with peak strength $\sim 5\ \mu$rad may be sufficient.
 \vspace{0.15 cm}\\
 \hline\vspace{-0.18 cm} \\ \vspace{0.05 cm}
  \vspace{-0.45 cm} Individual bunches may be moved on orbit beamlets (provided the AC dipole can be operated on selected RF buckets).
& \vspace{-0.4 cm}  If either resonance crossing or chromatic crossing is used, all bunches are split. If either direct injection or AC dipole excitation is used, islands can be populated on selected bunches.
 \vspace{0.15 cm}\\
\hline
\end{tabularx}
\caption{Summary of the main differences and similarities of electron beams and x-rays obtained from the approaches discussed in detail in Section~\ref{sec:scenario2} (resonant orbit) and Section~\ref{sec:scenario1} (resonance islands). The values of the strength of the AC dipole depend on the magnetic lattice: they refer here to the original ESRF storage ring.}
\label{tab:comparison}
\end{table}

\section{Possible applications of resonance islands in synchrotron light sources} \label{sec:applications}

From the operational point of view, BESSY~II has been a pioneer machine in providing users with off-axis bunches inside resonance islands~\cite{TRIBs-BESSY1:ipac19}. For a list of current applications and benefits to synchrotron light source beamlines see Refs.~\cite{TRIBs-BESSY1:ipac19,VSR-tds,Nature-BESSY-II}. In this section, a few novel proposals are discussed. 

The first proposal is based on the use of an additional closed orbit linked to stable islands in static mode, i.e.\ without any mechanism to cross a resonance. Under these conditions, a reversed chromatic crossing similar to what presented in Fig.~\ref{Fig:ChromCross} could be implemented for transparent top-up injection with no need of kicker magnets to generate a closed-orbit bump. Provided that the bunch length and the horizontal emittance of the injected particles are sufficiently small to survive some synchrotron periods, the incoming beam could be injected directly inside a stable island, such as the one in the bottom-left phase space of Fig.~\ref{Fig:ChromCross}, but with an energy offset. To avoid hitting the septum magnet at the end of the injection process, one could introduce a pair of pulsed magnets (e.g.\ quadrupoles or sextupoles) as proposed in~\cite{PhysRevSTAB.18.074001}. This pair of pulsed magnets would be used during the injection process to create a closed bump in the beamlet positions to reach the appropriate injection conditions. Upon completion of the injection process, these pulsed magnets would be switched off, thus moving the beamlet positions inside the septum magnet, after which the beamlets would merge spontaneously into the main beam around the nominal closed orbit, thus topping up the circulating (and unperturbed) bunch, thanks to the natural longitudinal damping. The benefit of such a scheme is that the only possible perturbation to the circulating beam is generated by the pair of pulsed magnets. The choice of the type of magnetic field can mitigate the possible perturbation, as in the case of pulsed quadrupoles, the circulating beam might suffer from dipolar feed-down, whereas for sextupoles, a quadrupolar feed-down might be generated. This is certainly much less harmful to the users than a dipolar kick. Pulsed octupolar magnets, which would generate only sextupolar feed-down, would be even more transparent for users.

The second proposal is based on the difference in linear optics between the dynamics around the nominal closed orbit and that inside the islands. This can be exploited to generate alternate sequences of short and long bunches without the need of special RF systems (for instance, harmonic cavities that would stretch the entire train, such as the two higher harmonic cavity systems of Ref.~\cite{VSR-tds}). In the proposed scheme, simultaneous operation with short and long bunches can be achieved by means of resonance islands of appropriate properties generated only by standard magnetic elements, with a clear advantage in terms of simplicity and operational efficiency. The nominal lattice is already optimised to generate the lowest possible horizontal equilibrium emittance for the beam around the nominal closed orbit. Hence, the equilibrium emittance of the beam in the islands will be very likely larger, as the dipolar and quadrupolar feed-down fields stemming from quadrupoles and sextupoles might be hard to set as to fulfil the conditions to minimise the transverse equilibrium emittance. On the other hand, different sextupole settings may be found not only to generate a suitable phase-space topology, but also to alter the path length and dispersion of the islands, and thus their momentum compaction, which in turn changes the bunch length of the beamlets inside the stable islands. It is worth mentioning that this approach has recently been proposed as the basis for a novel, non-adiabatic, transition crossing gymnastics~\cite{GiovannozziEPJPlusTransition}. 

The third proposal is based on the use of split beams in combination with canted beamlines. A single undulator, acting on multiple beamlets, would generate x-rays oriented like the source beamlets. In this way, several canted beamlines could be operated simultaneously without the need of the orbit bump and the pairs of undulators that are present in the standard scheme. In the proposed scheme, up to $N$ canted beamlines could take beam at the same time (where $N$ is the order of the resonance used to split the beam) with a single undulator and no orbit bump, which is a considerable operational improvement. To note that the intensity for each beamline is divided by $N$ with respect to the standard operational mode.

Finally, it is possible to generalise what was done in~\cite{Nature-BESSY-II} and to propose an application that provides fast switching (on a turn-by-turn basis and even faster) of general x-ray properties, not only helicity as reported in~\cite{Nature-BESSY-II}. Two configurations can be envisaged, namely, electrons inside a single island of a resonance of order $r$, or electrons with an AC dipole that excites a periodic orbit of period $r$. In both cases, at a given location of the ring, the electrons pass at different positions each turn, and each position repeats regularly with a periodicity of $r$ turns. If at that section an insertion device is installed, featuring properties that depend on the transverse position, the light emitted will switch its properties on a turn-by-turn basis. More than that, if two bunches in different buckets are located in different islands (or the AC dipole has different phases for different buckets), the switching of the light properties would occur on a sub-turn time scale, thus opening new options for extremely fast switching of light properties in a storage-ring-based synchrotron light source. 

\section{Conclusions} \label{sec:conc}

In the wake of the beam manipulations pioneered at the MLS and BESSY~II facilities, recently tested also at the MAX~IV light source, we have explored in a numerical manner different scenarios capable of generating stable electron beams on periodic off-axis closed orbits, with the aim of providing a new means for the efficient use of high-quality electron beams in synchrotron storage rings.

A resonant orbit excitation can be used to generate an orbit distortion that closes after $N$ turns. This would result in $N$ so-called orbit beamlets appearing on standard beam imaging systems, hence giving the impression of having $N$ beamlets circulating simultaneously. We have shown that by using this technique, the beam occupies the $N$ positions on different turns, but is not split into $N$ beamlets, and that the resulting orbit beamlets are unrelated to the stable resonance islands of phase space generated by nonlinear magnets and non-zero amplitude-dependent detuning. 

Resonance islands around stable fixed points of order $N$ are created in phase space by suitable sextupole settings. Several ways to populate them with electron beams have been considered and studied in detail. The dynamics of electrons inside the islands has been studied numerically, revealing a conditional stability whose physics detail is yet to be fully explored, due to the combined effect of strong nonlinearities, damping, and diffusion, which is far from the typical regimes of electron beams in synchrotron light sources. It is clear that the presence of radiation damping and quantum diffusion alters the Hamiltonian description of the electron dynamics in the resonance islands, with respect to the case of proton beams. It is also clear that, except in the case of extreme conditions of damping and quantum diffusion, the fixed points are almost unaffected and behave very similarly in both the electron and proton case. Numerical comparisons of the dynamics of protons and electrons revealed intriguing features of the latter, which may be exploited to improve the performance and flexibility of existing and future storage-ring-based synchrotron light sources. 

section{Acknowledgements}

We would like to thank Simone Liuzzo and Nicola Carmignani for their support in modifying the AT code to perform optics calculations on fixed points and in setting up multi-particle simulations. We are also grateful to Andrea Latina, Piotr Skowronski, and Riccardo De Maria for their support for MAD-X-PTC. 
\section*{Data availability}
Data sharing not applicable to this article as no data sets were generated or analysed during the current study.

\bibliographystyle{unsrt}
\bibliography{mybibliography}

\providecommand{\noopsort}[1]{}\providecommand{\singleletter}[1]{#1}%
\begin{thebibliography}{10}
  
\bibitem{Garoby:1998kea}
R.~Garoby.
\newblock {Bunch Merging and Splitting Techniques in the Injectors for High
  Energy Hadron Colliders}.
\newblock In I.~N. Meshkov, editor, {\em {17th International Conference on
  High-Energy Accelerators}}, pages 172--174, Dubna, 1998. Joint Inst. Nucl.
  Res.

\bibitem{Garoby:EPAC00-WEOAF102}
R.~Garoby, S.~Hancock, and J.~L. Vallet.
\newblock Demonstration of bunch triple splitting in the cern ps.
\newblock In {\em Proc. 7th European Particle Accelerator Conf. (EPAC'00)},
  pages 304--306. JACoW Publishing, Jun. 2000.

\bibitem{Burnet:1359959}
J.-P. Burnet, C.~Carli, M.~Chanel, R.~Garoby, S.~Gilardoni, M.~Giovannozzi,
  S.~Hancock, H.~Haseroth, K.~Hübner, D.~Küchler, J.~Lewis, A.~Lombardi,
  D.~Manglunki, M.~Martini, S.~Maury, E.~Métral, D.~Möhl, G.~Plass,
  L.~Rinolfi, R.~Scrivens, R.~Steerenberg, C.~Steinbach, M.~Vretenar, and
  T.~Zickler.
\newblock {\em {Fifty years of the CERN Proton Synchrotron: Volume 1}}.
\newblock CERN Yellow Reports: Monographs. CERN, Geneva, 2011.

\bibitem{Laxdal:EPAC92}
R.~E. Laxdal and W.~Joho.
\newblock Longitudinal splitting of bunches in a cyclotron by superposition of
  different rf harmonics.
\newblock In {\em Proc. 3rd European Particle Accelerator Conf. (EPAC'92)},
  pages 590--593. JACoW Publishing, Mar. 1992.

\bibitem{PhysRevE.49.2484}
J.~A. Ellison, H.-J. Shih, and M.~Kummer.
\newblock Theoretical study of longitudinal beam splitting and related
  phenomena.
\newblock {\em Phys. Rev. E}, 49:2484--2487, Mar 1994.

\bibitem{Borburgh:2137954}
J.~Borburgh, S.~Damjanovic, S.~Gilardoni, M.~Giovannozzi, C.~Hernalsteens,
  M.~Hourican, A.~Huschauer, K.~Kahle, G.~Le~Godec, O.~Michels, and
  G.~Sterbini.
\newblock {First implementation of transversely split proton beams in the CERN
  Proton Synchrotron for the fixed-target physics programme}.
\newblock {\em EPL}, 113(3):34001. 6 p, 2016.

\bibitem{PhysRevAccelBeams.20.014001}
S.~Abernethy, A.~Akroh, H.~Bartosik, A.~Blas, T.~Bohl, S.~Cettour-Cave,
  K.~Cornelis, H.~Damerau, S.~Gilardoni, M.~Giovannozzi, C.~Hernalsteens,
  A.~Huschauer, V.~Kain, D.~Manglunki, G.~M\'etral, B.~Mikulec, B.~Salvant,
  J.-L. Sanchez~Alvarez, R.~Steerenberg, G.~Sterbini, and Y.~Wu.
\newblock Operational performance of the cern injector complex with
  transversely split beams.
\newblock {\em Phys. Rev. Accel. Beams}, 20:014001, 2017.

\bibitem{PhysRevAccelBeams.22.104002}
A.~Huschauer, H.~Bartosik, S.~Cettour Cave, M.~Coly, D.~Cotte, H.~Damerau,
  G.~P. Di~Giovanni, S.~Gilardoni, M.~Giovannozzi, V.~Kain,
  E.~Koukovini-Platia, B.~Mikulec, G.~Sterbini, and F.~Tecker.
\newblock {Advancing the CERN proton synchrotron multiturn extraction towards
  the high-intensity proton beams frontier}.
\newblock {\em Phys. Rev. Accel. Beams}, 22:104002, Oct 2019.

\bibitem{Giovannozzi:987493}
M.~Giovannozzi, M.J. Barnes, O.E. Berrig, A.~Beuret, J.~Borburgh, P.~Bourquin,
  R.~Brown, J.-P. Burnet, F.~Caspers, J.-M. Cravero, T.~Dobers, T.~Fowler, S.S.
  Gilardoni, M.~Hourican, W.~Kalbreier, T.~Kroyer, F.~Di~Maio, M.~Martini,
  V.~Mertens, E.~Métral, K.D. Metzmacher, C.~Rossi, J.-P. Royer, L.~Sermeus,
  R.~Steerenberg, G.~Villiger, and T.~Zickler.
\newblock {\em {The CERN PS multi-turn extraction based on beam splittting in
  stable islands of transverse phase space: Design Report}}.
\newblock CERN Yellow Reports: Monographs. CERN, Geneva, 2006.

\bibitem{PhysRevLett.88.104801}
R.~Cappi and M.~Giovannozzi.
\newblock Novel method for multiturn extraction: Trapping charged particles in
  islands of phase space.
\newblock {\em Phys. Rev. Lett.}, 88:104801, 2002.

\bibitem{PhysRevSTAB.7.024001}
R.~Cappi and M.~Giovannozzi.
\newblock Multiturn extraction and injection by means of adiabatic capture in
  stable islands of phase space.
\newblock {\em Phys. Rev. ST Accel. Beams}, 7:024001, 2004.

\bibitem{PhysRevSTAB.18.074001}
A.~Franchi and M.~Giovannozzi.
\newblock Novel technique for injecting and extracting beams in a circular
  hadron accelerator without using septum magnets.
\newblock {\em Phys. Rev. ST Accel. Beams}, 18:074001, Jul 2015.

\bibitem{our_paper3}
M.~Giovannozzi, L.~Huang, A.~Huschauer, and A.~Franchi.
\newblock A novel non-adiabatic approach to transition crossing in a circular
  hadron accelerator.
\newblock In preparation.

\bibitem{MarkusRies:phd}
M.~Ries.
\newblock {Nonlinear Momentum Compaction and Coherent Synchrotron Radiation at the Metrology Light Source}.
\newblock {PhD thesis, Humboldt University of Berlin, 2014}.

\bibitem{MLS-BESSY1:ipac15}
M.~Ries, J.~Feikes, T.~Goetsch, P.~Goslawski, J.~Li, M.~Ruprecht, A.~Schälicke, G.~Wüstefeld
\newblock {Transverse resonance island buckets at the MLS and BESSY II}.
\newblock In {\em Proc. 6th International Particle Accelerator Conf. (IPAC2015)},
  pages 138--140. JACoW Publishing, 2015.

\bibitem{TRIBs-BESSY1:ipac19}
P.~Goslawski , F.~Andreas , F.~Armborst , T.~Atkinson, J.~Feikes, A.~Jankowiak , J.~Li, T.~Mertens, M.~Ries, A.~Schälicke, G.~Schiwietz, G.~Wüstefeld.
\newblock {Two orbit operation at BESSY II during a user test week}.
\newblock In {\em Proc. 10th International Particle Accelerator Conf. (IPAC2019)},
  pages 3419--3421. JACoW Publishing, 2019.

\bibitem{TRIBs-MAXIV:ipac19}
P. F.~Tavares, E.~Al-Dmour,Å.~Andersson, J.~Breunlin, F.~Cullinan, E.~Mansten, S.~Molloy, D.~Olsson, D.K.~Olsson, M.~Sjöström, S.~Thorin.
\newblock {Status of the MAX IV accelerators}.
\newblock In {\em Proc. 10th International Particle Accelerator Conf. (IPAC2019)},
  pages 1185--1190. JACoW Publishing, 2019.

\bibitem{NOCE17}
M.~Giovannozzi, N.~Carmignani and A.~Franchi
\newblock {Could synchrotron light sources benefit from the experience at CERN with beams split in horizontal phase space?}
\newblock In {\em Proc. Nonlinear Dynamics and Collective Effects NOCE 2017 in Particle Beam Physics, Arcidosso, Italy}.  
   World Scientific 2017, ISBN: 978-981-3279-61-2. 

\bibitem{Helmut-diffusion} 
H.~Burkhardt
\newblock {Monte Carlo generation of the energy spectrum of synchrotron radiation}.
\newblock {CLIC-Note-709, EUROTEV-Report-2007-018, Geneva 8 June 2007}

\bibitem{Andrea-CouplingESRF}
A.~Franchi, L.~Farvacque, J.~Chavanne, F.~Ewald, B.~Nash, K.~Scheidt, and R.~Tomás.
\newblock {Vertical emittance reduction and preservation in electron storage rings via resonance driving terms correction}
\newblock {\em Phys. Rev. ST Accel. Beams}, 14:034002, March 2011.

\bibitem{VSR-tds}
{\em VV. ~AA. }
\newblock {Technical Design Study BESSY VSR}. 
\newblock {Helmholtz-Zentrum Berlin für Materialien und Energie GmbH}, June 2015.

\end{thebibliography}



\providecommand{\noopsort}[1]{}\providecommand{\singleletter}[1]{#1}%
\begin{thebibliography}{10}

\bibitem{Garoby:1998kea}
R.~Garoby.
\newblock {Bunch Merging and Splitting Techniques in the Injectors for High
  Energy Hadron Colliders}.
\newblock In I.~N. Meshkov, editor, {\em {Proc. 17th International Conference
  on High-Energy Accelerators}}, pages 172--174, Dubna, 1998. Joint Inst. Nucl.
  Res.

\bibitem{Garoby:EPAC00-WEOAF102}
R.~Garoby, S.~Hancock, and J.~L. Vallet.
\newblock Demonstration of bunch triple splitting in the {CERN} {PS}.
\newblock In {\em Proc. 7th European Particle Accelerator Conf. (EPAC'00)},
  pages 304--306. JACoW Publishing, Jun. 2000.

\bibitem{Burnet:1359959}
J.-P. Burnet, C.~Carli, M.~Chanel, R.~Garoby, S.~Gilardoni, M.~Giovannozzi,
  S.~Hancock, H.~Haseroth, K.~Hübner, D.~Küchler, J.~Lewis, A.~Lombardi,
  D.~Manglunki, M.~Martini, S.~Maury, E.~Métral, D.~Möhl, G.~Plass,
  L.~Rinolfi, R.~Scrivens, R.~Steerenberg, C.~Steinbach, M.~Vretenar, and
  T.~Zickler.
\newblock {\em {Fifty years of the {CERN} Proton Synchrotron: Volume 1}}.
\newblock {CERN} Yellow Reports: Monographs. {CERN}, Geneva, 2011.

\bibitem{Laxdal:EPAC92}
R.~E. Laxdal and W.~Joho.
\newblock Longitudinal splitting of bunches in a cyclotron by superposition of
  different rf harmonics.
\newblock In {\em Proc. 3rd European Particle Accelerator Conf. (EPAC'92)},
  pages 590--593. JACoW Publishing, Mar. 1992.

\bibitem{PhysRevE.49.2484}
J.~A. Ellison, H.-J. Shih, and M.~Kummer.
\newblock Theoretical study of longitudinal beam splitting and related
  phenomena.
\newblock {\em Phys. Rev. E}, 49:2484--2487, Mar 1994.

\bibitem{Borburgh:2137954}
J.~Borburgh, S.~Damjanovic, S.~Gilardoni, M.~Giovannozzi, C.~Hernalsteens,
  M.~Hourican, A.~Huschauer, K.~Kahle, G.~Le~Godec, O.~Michels, and
  G.~Sterbini.
\newblock {First implementation of transversely split proton beams in the
  {CERN} Proton Synchrotron for the fixed-target physics programme}.
\newblock {\em EPL}, 113(3):34001. 6 p, 2016.

\bibitem{PhysRevAccelBeams.20.014001}
S.~Abernethy, A.~Akroh, H.~Bartosik, A.~Blas, T.~Bohl, S.~Cettour-Cave,
  K.~Cornelis, H.~Damerau, S.~Gilardoni, M.~Giovannozzi, C.~Hernalsteens,
  A.~Huschauer, V.~Kain, D.~Manglunki, G.~M\'etral, B.~Mikulec, B.~Salvant,
  J.-L. Sanchez~Alvarez, R.~Steerenberg, G.~Sterbini, and Y.~Wu.
\newblock Operational performance of the {CERN} injector complex with
  transversely split beams.
\newblock {\em Phys. Rev. Accel. Beams}, 20:014001, 2017.

\bibitem{PhysRevAccelBeams.22.104002}
A.~Huschauer, H.~Bartosik, S.~Cettour Cave, M.~Coly, D.~Cotte, H.~Damerau,
  G.~P. Di~Giovanni, S.~Gilardoni, M.~Giovannozzi, V.~Kain,
  E.~Koukovini-Platia, B.~Mikulec, G.~Sterbini, and F.~Tecker.
\newblock {Advancing the {CERN} proton synchrotron multiturn extraction towards
  the high-intensity proton beams frontier}.
\newblock {\em Phys. Rev. Accel. Beams}, 22:104002, Oct 2019.

\bibitem{Giovannozzi:987493}
M.~Giovannozzi, M.J. Barnes, O.E. Berrig, A.~Beuret, J.~Borburgh, P.~Bourquin,
  R.~Brown, J.-P. Burnet, F.~Caspers, J.-M. Cravero, T.~Dobers, T.~Fowler, S.S.
  Gilardoni, M.~Hourican, W.~Kalbreier, T.~Kroyer, F.~Di~Maio, M.~Martini,
  V.~Mertens, E.~Métral, K.D. Metzmacher, C.~Rossi, J.-P. Royer, L.~Sermeus,
  R.~Steerenberg, G.~Villiger, and T.~Zickler.
\newblock {\em {The {CERN} {PS} multi-turn extraction based on beam splittting
  in stable islands of transverse phase space: Design Report}}.
\newblock {CERN} Yellow Reports: Monographs. {CERN}, Geneva, 2006.

\bibitem{PhysRevLett.88.104801}
R.~Cappi and M.~Giovannozzi.
\newblock Novel method for multiturn extraction: Trapping charged particles in
  islands of phase space.
\newblock {\em Phys. Rev. Lett.}, 88:104801, 2002.

\bibitem{PhysRevSTAB.7.024001}
R.~Cappi and M.~Giovannozzi.
\newblock Multiturn extraction and injection by means of adiabatic capture in
  stable islands of phase space.
\newblock {\em Phys. Rev. ST Accel. Beams}, 7:024001, 2004.

\bibitem{PhysRevSTAB.18.074001}
A.~Franchi and M.~Giovannozzi.
\newblock Novel technique for injecting and extracting beams in a circular
  hadron accelerator without using septum magnets.
\newblock {\em Phys. Rev. ST Accel. Beams}, 18:074001, Jul 2015.

\bibitem{GiovannozziEPJPlusTransition}
M.~Giovannozzi, L.~Huang, A.~Huschauer, and A.~Franchi.
\newblock A novel non-adiabatic approach to transition crossing in a circular
  hadron accelerator.
\newblock {\em The European Physical Journal Plus}, 136(11):1189, 2021.

\bibitem{MarkusRies:phd}
Markus Ries.
\newblock {\em Nonlinear Momentum Compaction and Coherent Synchrotron Radiation
  at the Metrology Light Source}.
\newblock PhD thesis, Humboldt University of Berlin, 2014.

\bibitem{MLS-BESSY1:ipac15}
M.~Ries, J.~Feikes, T.~Goetsch, P.~Goslawski, J.~Li, M.~Ruprecht,
  A.~Schälicke, and G.~Wüstefeld.
\newblock {Transverse resonance island buckets at the MLS and BESSY II}.
\newblock In {\em {Proc. 6th International Particle Accelerator Conf.
  (IPAC2015)}}, pages 138--140. JACoW, 2015.

\bibitem{Kramer:IPAC18-TUPML052}
F.~Kramer, P.~Goslawski, A.~Jankowiak, M.~Ries, M.~Ruprecht, and
  A.~Sch\"alicke.
\newblock Characterization of the second stable orbit generated by transverse
  resonance island buckets {(TRIBs)}.
\newblock In {\em Proc. 9th Int. Particle Accelerator Conf. (IPAC'18)}, page
  1656. JACoW Publishing, May 2018.

\bibitem{TRIBs-BESSY1:ipac19}
P.~Goslawski, F.~Andreas, F.~Armborst, T.~Atkinson, J.~Feikes, A.~Jankowiak,
  J.~Li, T.~Mertens, M.~Ries, A.~Schälicke, G.~Schiwietz, and G.~Wüstefeld.
\newblock {Two orbit operation at BESSY II during a user test week}.
\newblock In {\em {Proc. 10th International Particle Accelerator Conf.
  (IPAC2019)}}, pages 3419--3421. JACoW, 2019.

\bibitem{TRIBs-MAXIV:ipac19}
P.~F. Tavares, E.~Al-Dmour, Å. Andersson, J.~Breunlin, F.~Cullinan,
  E.~Mansten, S.~Molloy, D.~Olsson, D.K. Olsson, M.~Sjöström, and S.~Thorin.
\newblock {Status of the MAX IV accelerators}.
\newblock In {\em {Proc. 10th International Particle Accelerator Conf.
  (IPAC2019)}}, pages 1185--1190. JACoW, 2019.

\bibitem{NOCE17}
M.~Giovannozzi, N.~Carmignani, and A.~Franchi.
\newblock {Could synchrotron light sources benefit from the experience at
  {CERN} with beams split in horizontal phase space?}
\newblock In {\em {Proc. Nonlinear Dynamics and Collective Effects NOCE 2017 in
  Particle Beam Physics, Arcidosso, Italy}}. World Scientific 2017, ISBN:
  978-981-3279-61-2, 2017.

\bibitem{Piminov:PAC09-TH6PFP093}
P.~A. Piminov, E.~B. Levichev, and D.~N. Shatilov.
\newblock Nonlinear beam dynamics with strong damping and space charge in the
  clic damping ring.
\newblock In {\em Proc. 23rd Particle Accelerator Conf. (PAC'09)}, pages
  3925--3927. JACoW Publishing, May 2009.

\bibitem{madx}
{MAD - Methodical Accelerator Design}.
\newblock \url{https://mad.web.cern.ch/mad/}.

\bibitem{Persson:IPAC21-WEPAB028}
T.~H.~B. Persson, H.~Burkhardt, L.~Deniau, A.~Latina, and P.~K. Skowronski.
\newblock {MAD-X} for future accelerators.
\newblock In {\em Proc. 12th Int. Particle Accelerator Conf. (IPAC'21)}, page
  2664. JACoW Publishing, May 2021.

\bibitem{Terebilo:PAC01-RPAH314}
A.~Terebilo.
\newblock Accelerator modeling with matlab accelerator toolbox.
\newblock In {\em Proc. 19th Particle Accelerator Conf. (PAC'01)}, pages
  3203--3205. JACoW Publishing, Jun. 2001.

\bibitem{Nash:IPAC15-MOPWA014}
B.~Nash et~al.
\newblock New functionality for beam dynamics in accelerator toolbox (at).
\newblock In {\em Proc. 6th Int. Particle Accelerator Conf. (IPAC'15)}, pages
  113--116. JACoW Publishing, May 2015.
\newblock https://doi.org/10.18429/JACoW-IPAC2015-MOPWA014.

\bibitem{4327987}
R.~Chasman, G.~K. Green, and E.~M. Rowe.
\newblock Preliminary design of a dedicated synchrotron radiation facility.
\newblock {\em IEEE Transactions on Nuclear Science}, 22(3):1765--1767, 1975.

\bibitem{Helmut-diffusion}
H.~Burkhardt.
\newblock {Monte Carlo generation of the energy spectrum of synchrotron
  radiation}.
\newblock Technical report, {CERN}, Geneva, June 2007.

\bibitem{giovannozzi2021novel}
M.~Giovannozzi, L.~Huang, A.~Huschauer, and A.~Franchi.
\newblock A novel non-adiabatic approach to transition crossing in a circular
  hadron accelerator.
\newblock arXiv:2109.00387 [physics.acc-ph], 2021.

\bibitem{Andrea-CouplingESRF}
A.~Franchi, L.~Farvacque, J.~Chavanne, F.~Ewald, B.~Nash, K.~Scheidt, and
  R.~Tomás.
\newblock Vertical emittance reduction and preservation in electron storage
  rings via resonance driving terms correction.
\newblock {\em Phys. Rev. ST Accel. Beams}, 14:34002, Mar 2011.

\bibitem{PhysRevE.50.R4298}
E.~Todesco.
\newblock Analysis of resonant structures of four-dimensional symplectic
  mappings, using normal forms.
\newblock {\em Phys. Rev. E}, 50:R4298--R4301, Dec 1994.

\bibitem{Nakamura:EPAC90}
S~Nakamura, O~Asai, N~Awaji, A~Chiba, T~Iida, S~Kawazu, R~Kitano, M~Kodaira,
  K~Kondo, M~Mizota, H~Nishizawa, M~Ohno, M~Shiota, M~Takanaka, T~Tomimasu, and
  Y~Yamamoto.
\newblock Present status of the 1 gev synchrotron radiation source at sortec.
\newblock In {\em Proc. 2nd European Particle Accelerator Conf. (EPAC'90)},
  pages 472--475. JACoW Publishing, Jun. 1990.

\bibitem{Chen:EPAC96-MOP115G}
J.~Chen, K.~T. Hsu, K.~K. Lin, T.~S. Ueng, and W.~T. Weng.
\newblock Topping up experiments at srrc.
\newblock In {\em Proc. 5th European Particle Accelerator Conf. (EPAC'96)}.
  JACoW Publishing, Jun. 1996.

\bibitem{Spring-8-06}
H.~Tanaka, M.~Adachi, T.~Aoki, T.~Asaka, A.~Baron, S.~Dat\'e, K.~Fukami,
  Y.~Furukawa, H.~Hanaki, N.~Hosoda, T.~Ishikawa, H.~Kimura, K.~Kobayashi,
  T.~Kobayashi, S.~Kohara, N.~Kumagai, M.~Masaki, T.~Masuda, S.~Matsui,
  A.~Mizuno, T.~Nakamura, T.~Nakatani, T.~Noda, T.~Ohata, H.~Ohkuma,
  T.~Ohshima, M.~Oishi, S.~Sasaki, J.~Schimizu, M.~Shoji, K.~Soutome,
  M.~Suzuki, S.~Suzuki, Y.~Suzuki, S.~Takano, M.~Takao, T.~Takashima,
  H.~Takebe, A.~Takeuchi, K.~Tamura, R.~Tanaka, Y.~Tanaka, T.~Taniuchi,
  Y.~Taniuchi, K.~Tsumaki, A.~Yamashita, K.~Yanagida, Y.~Yoda, H.~Yonehara,
  T.~Yorita, M.~Yoshioka, and M.~Takata.
\newblock Stable top-up operation at spring-8.
\newblock {\em J. Synchrotron Rad.}, 13:378, 2006.

\bibitem{PhysRevSTAB.10.123501}
Kentaro Harada, Yukinori Kobayashi, Tsukasa Miyajima, and Shinya Nagahashi.
\newblock New injection scheme using a pulsed quadrupole magnet in electron
  storage rings.
\newblock {\em Phys. Rev. ST Accel. Beams}, 10:123501, Dec 2007.

\bibitem{Ohkuma:EPAC08-MOZCG01}
H.~Ohkuma.
\newblock Top-up operation in light sources.
\newblock In {\em Proc. 11th European Particle Accelerator Conf. (EPAC'08)},
  pages 36--40. JACoW Publishing, Jun. 2008.

\bibitem{PEAKE2008143}
D.J. Peake, M.J. Boland, G.S. LeBlanc, and R.P. Rassool.
\newblock Measurement of the real time fill-pattern at the australian
  synchrotron.
\newblock {\em Nucl. Instrum. Methods Phys. Res. A}, 589(2):143--149, 2008.

\bibitem{PhysRevSTAB.13.020705}
Hiroyuki Takaki, Norio Nakamura, Yukinori Kobayashi, Kentaro Harada, Tsukasa
  Miyajima, Akira Ueda, Shinya Nagahashi, Miho Shimada, Takashi Obina, and
  Tohru Honda.
\newblock Beam injection with a pulsed sextupole magnet in an electron storage
  ring.
\newblock {\em Phys. Rev. ST Accel. Beams}, 13:020705, Feb 2010.

\bibitem{PhysRevSTAB.15.050705}
S.~C. Leemann.
\newblock Pulsed sextupole injection for sweden's new light source max iv.
\newblock {\em Phys. Rev. ST Accel. Beams}, 15:050705, May 2012.

\bibitem{VSR-tds}
VV. AA.
\newblock {Technical Design Study BESSY VSR}.
\newblock Technical report, Helmholtz-Zentrum Berlin für Materialien und
  Energie GmbH, Berlin, June 2015.

\bibitem{Nature-BESSY-II}
K.~Holldack, C.~Sch{\"u}ssler-Langeheine, P.~Goslawski, N.~Pontius, T.~Kachel,
  F.~Armborst, M.~Ries, A.~Sch{\"a}licke, M.~Scheer, W.~Frentrup, and
  J.~Bahrdt.
\newblock Flipping the helicity of x-rays from an undulator at unprecedented
  speed.
\newblock {\em Communications Physics}, 3(1):61, 2020.

\end{thebibliography}
\end{document}